\documentclass[a4paper,12pt]{report}

\newcommand{\beqn}{\begin{eqnarray}}
\newcommand{\beq}{\begin{equation}}

\newcommand{\eeqn}{\end{eqnarray}}
\newcommand{\eeq}{\end{equation}}
\newcommand{\nn}{\nonumber}

\newcommand{\lsim}{\raise0.3ex
\hbox{$\;<$\kern-0.75em\raise-1.1ex\hbox{$\sim\;$}}}
\newcommand{\gsim}{\raise0.3ex
\hbox{$\;>$\kern-0.75em\raise-1.1ex\hbox{$\sim\;$}}}

\usepackage{amsmath}
\usepackage[dvips]{graphicx}
\usepackage{amssymb}
\usepackage[mathcal]{euscript}
\usepackage{setspace}

\numberwithin{equation}{section}
\begin{document}


\title{Supersymmetry and Phenomenology of Heterotic and Type I
Superstring Models}

\author{David J Clements}

\maketitle

\begin{abstract}
This work is the discussion of heterotic and type I string
phenomenology. The heterotic string model is based on the
free--fermionic formalism.  In this scenario, one has the first
case where non--Abelian VEV's, as opposed to singlet VEV's are
required for the cancellation of the Fayet--Iliopoulos term.  It
is noted that non--Abelian fields are the only fields that can
give rise to the satisfaction of the $D$--flat constraints in this
model.  In addition, there is an enhancement to the observable and
hidden $SU(3)$ gauge groups to $SU(4)$.

The type I models discussed are based on
$T^6/(\mathbb{Z}_2\times\mathbb{Z}_2^\text{s})$ and
$T^6/(\mathbb{Z}_2\times\mathbb{Z}_2\times\mathbb{Z}_2^{\text{s}})$
compactifications.  The first example is a simpler case of the
second model. It has $N=2$ supersymmetry and includes a rank
reduction of the $D5$ gauge groups as a result of using a freely
acting Kaluza Klein shift $\mathbb{Z}_2^\text{s}$.  The second
case, is a freely acting Kaluza Klein shift of the
$T^6/(\mathbb{Z}_2\times\mathbb{Z}_2)$ orbifold $N=1$ model.  As
such one has a choice of sign $\epsilon=\pm 1$ that arises in the
model from terms not related to the principle orbits by $S$ and
$T$ transformations.  This allows the breaking of supersymmetry
with the introduction of antibranes.  In addition, one has four
models for each choice of sign. However, I discuss the problems
inherent in the $\epsilon=-1$ case with respect to particle
interpretation.

I also discuss the $T^6/(\mathbb{Z}_2\times\mathbb{Z}_2)$ model
when background magnetic fields are introduced for the
$\epsilon=-1$ case.  In this scenario, one keeps the twisted terms
in the transverse annulus which in turn does not allow for
consistent cancellation of tadpoles.  This is caused by a residual
term that is proportional to the magnetic field.  In this case I
consider the magnetization of the first two tori.  This leads to
tadpole complications for the $g$ and $f$ twisted sectors, but
allows the $h$ twisted sector to behave normally. I discuss the
issues of tachyonic excitations that arise from the low lying
states in the direct channel.
\end{abstract}

\tableofcontents

\listoffigures

\listoftables



\def\beq{\begin{equation}}
\def\eeq{\end{equation}}
\def\beqn{\begin{eqnarray}}
\def\eeqn{\end{eqnarray}}

\def\no{\noindent }
\def\nolabel{\nonumber }
\def\ie{{\it i.e.}}
\def\eg{{\it e.g.}}
\def\half{{\textstyle{1\over 2}}}
\def\third{{\textstyle {1\over3}}}
\def\quarter{{\textstyle {1\over4}}}
\def\sixth{{\textstyle {1\over6}}}
\def\m{{\tt -}}
\def\p{{\tt +}}

\def\Tr{{\rm Tr}\, }
\def\tr{{\rm tr}\, }

\def\slash#1{#1\hskip-6pt/\hskip6pt}
\def\slk{\slash{k}}
\def\GeV{\,{\rm GeV}}
\def\TeV{\,{\rm TeV}}
\def\y{\,{\rm y}}
\def\SM{Standard--Model }
\def\SUSY{supersymmetry }
\def\SSSM{supersymmetric standard model}
\def\vev#1{\left\langle #1\right\rangle}
\def\l{\langle}
\def\r{\rangle}
\def\o#1{\frac{1}{#1}}

\def\Htw{{\tilde H}}
\def\chibar{{\overline{\chi}}}
\def\qbar{{\overline{q}}}
\def\ibar{{\overline{\imath}}}
\def\jbar{{\overline{\jmath}}}
\def\Hbar{{\overline{H}}}
\def\Qbar{{\overline{Q}}}
\def\abar{{\overline{a}}}
\def\alphabar{{\overline{\alpha}}}
\def\betabar{{\overline{\beta}}}
\def\tautwo{{ \tau_2 }}
\def\thetatwo{{ \vartheta_2 }}
\def\thetathree{{ \vartheta_3 }}
\def\thetafour{{ \vartheta_4 }}
\def\ttwo{{\vartheta_2}}
\def\tthree{{\vartheta_3}}
\def\tfour{{\vartheta_4}}
\def\ti{{\vartheta_i}}
\def\tj{{\vartheta_j}}
\def\tk{{\vartheta_k}}
\def\calF{{\cal F}}
\def\smallmatrix#1#2#3#4{{ {{#1}~{#2}\choose{#3}~{#4}} }}
\def\ab{{\alpha\beta}}
\def\Minv{{ (M^{-1}_\ab)_{ij} }}
\def\bone{{\bf 1}}
\def\ii{{(i)}}
\def\V{{\bf V}}
\def\N{{\bf N}}

\def\b{{\bf b}}
\def\S{{\bf S}}
\def\X{{\bf X}}
\def\I{{\bf I}}
\def\mb{{\mathbf b}}
\def\mS{{\mathbf S}}
\def\mX{{\mathbf X}}
\def\mI{{\mathbf I}}
\def\balpha{{\mathbf \alpha}}
\def\bbeta{{\mathbf \beta}}
\def\bgamma{{\mathbf \gamma}}
\def\bxi{{\mathbf \xi}}

\def\t#1#2{{ \Theta\left\lbrack \matrix{ {#1}\cr {#2}\cr }\right\rbrack }}
\def\C#1#2{{ C\left\lbrack \matrix{ {#1}\cr {#2}\cr }\right\rbrack }}
\def\tp#1#2{{ \Theta'\left\lbrack \matrix{ {#1}\cr {#2}\cr }\right\rbrack }}
\def\tpp#1#2{{ \Theta''\left\lbrack \matrix{ {#1}\cr {#2}\cr }\right\rbrack }}
\def\l{\langle}
\def\r{\rangle}

\def\beq{\begin{equation}}
\def\eeq{\end{equation}}
\def\beqn{\begin{eqnarray}}
\def\eeqn{\end{eqnarray}}

\def\no{\noindent }
\def\nolabel{\nonumber }
\def\ie{{\it i.e.}}
\def\eg{{\it e.g.}}
\def\half{{\textstyle{1\over 2}}}
\def\third{{\textstyle {1\over3}}}
\def\quarter{{\textstyle {1\over4}}}
\def\sixth{{\textstyle {1\over6}}}
\def\tenth{{\textstyle {1\over 10}}}
\def\hund{{\textstyle {1\over 100}}}
\def\m{{\tt -}}
\def\p{{\tt +}}

\def\Tr{{\rm Tr}\, }
\def\tr{{\rm tr}\, }

\def\slash#1{#1\hskip-6pt/\hskip6pt}
\def\slk{\slash{k}}
\def\GeV{\,{\rm GeV}}
\def\TeV{\,{\rm TeV}}
\def\y{\,{\rm y}}
\def\SM{Standard--Model }
\def\SUSY{supersymmetry }
\def\SSSM{supersymmetric standard model}
\def\vev#1{\left\langle #1\right\rangle}
\def\l{\langle}
\def\r{\rangle}
\def\o#1{\frac{1}{#1}}

\def\Htw{{\tilde H}}
\def\chibar{{\bar{\chi}}}
\def\qbar{{\bar{q}}}
\def\ibar{{\bar{\imath}}}
\def\jbar{{\bar{\jmath}}}
\def\Hbar{{\bar{H}}}
\def\Qbar{{\bar{Q}}}
\def\abar{{\bar{a}}}
\def\alphabar{{\bar{\alpha}}}
\def\betabar{{\bar{\beta}}}
\def\tautwo{{ \tau_2 }}
\def\thetatwo{{ \vartheta_2 }}
\def\thetathree{{ \vartheta_3 }}
\def\thetafour{{ \vartheta_4 }}
\def\ttwo{{\vartheta_2}}
\def\tthree{{\vartheta_3}}
\def\tfour{{\vartheta_4}}
\def\ti{{\vartheta_i}}
\def\tj{{\vartheta_j}}
\def\tk{{\vartheta_k}}
\def\calF{{\cal F}}
\def\smallmatrix#1#2#3#4{{ {{#1}~{#2}\choose{#3}~{#4}} }}
\def\ab{{\alpha\beta}}
\def\Minv{{ (M^{-1}_\ab)_{ij} }}
\def\bone{{\bf 1}}
\def\bo{{\bf 1}}
\def\ii{{(i)}}
\def\V{{\bf V}}
\def\N{{\bf N}}

\def\bfb{{\bf b}}
\def\bfS{{\bf S}}
\def\bfX{{\bf X}}
\def\bfI{{\bf I}}
\def\ma{{\mathbf a}}
\def\mb{{\mathbf b}}
\def\mS{{\mathbf S}}
\def\mX{{\mathbf X}}
\def\mI{{\mathbf I}}
\def\malpha{{\mathbf \alpha}}
\def\mbeta{{\mathbf \beta}}
\def\mgamma{{\mathbf \gamma}}
\def\mzeta{{\mathbf \zeta}}
\def\mxi{{\mathbf \xi}}

\def\t#1#2{{ \Theta\left\lbrack \matrix{ {#1}\cr {#2}\cr }\right\rbrack }}
\def\C#1#2{{ C\left\lbrack \matrix{ {#1}\cr {#2}\cr }\right\rbrack }}
\def\tp#1#2{{ \Theta'\left\lbrack \matrix{ {#1}\cr {#2}\cr }\right\rbrack }}
\def\tpp#1#2{{ \Theta''\left\lbrack \matrix{ {#1}\cr {#2}\cr }\right\rbrack }}
\def\l{\langle}
\def\r{\rangle}

\def\x#1{\phi_{#1}}
\def\bx#1{{\bar{\phi}}_{#1}}

\def\cl#1{{\cal L}_{#1}}
\def\bcl#1{\bar{\cal L}_{#1}}

\def\bt{{\bar 3}}
\def\h#1{h_{#1}}
\def\Q#1{Q_{#1}}
\def\L#1{L_{#1}}

\def\N#1{N_{#1}}
\def\bN#1{{\bar{N}}_{#1}}

\def\S#1{S_{#1}}
\def\Ss#1#2{S_{#1}^{#2}}
\def\bS#1{{\bar S}_{#1}}
\def\Sb#1{{\bar S}_{#1}}
\def\bSs#1#2{{\bar{S}}_{#1}^{#2}}

\def\D#1{D_{#1}}
\def\Ds#1#2{D_{#1}^{#2}}
\def\bD#1{{\bar{D}}_{#1}}
\def\bDs#1#2{{\bar{D}}_{#1}^{#2}}

\def\p#1{\phi_{#1}}
\def\bp#1{{\bar{\phi}}_{#1}}

\def\P#1{\Phi_{#1}}
\def\bP#1{{\bar{\Phi}}_{#1}}
\def\X#1{\Phi_{#1}}
\def\bX#1{{\bar{\Phi}}_{#1}}
\def\Ps#1#2{\Phi_{#1}^{#2}}
\def\bPs#1#2{{\bar{\Phi}}_{#1}^{#2}}
\def\ps#1#2{\phi_{#1}^{#2}}
\def\bps#1#2{{\bar{\phi}}_{#1}^{#2}}
\def\php{\phantom{+}}

\def\H#1{H_{#1}}
\def\bH#1{{\bar{H}}_{#1}}
\def\Hb#1{{\bar{H}}_{#1}}

\def\xH#1#2{H^{#1}_{#2}}
\def\bxH#1#2{{\bar{H}}^{#1}_{#2}}

\def\UA{U(1)_{\rm A}}
\def\QA{Q^{(\rm A)}}
\def\mssm{SU(3)_C\times SU(2)_L\times U(1)_Y}

\def\inbar{\,\vrule height1.5ex width.4pt depth0pt}

\def\IC{\relax\hbox{$\inbar\kern-.3em{\rm C}$}}
\def\IQ{\relax\hbox{$\inbar\kern-.3em{\rm Q}$}}
\def\IR{\relax{\rm I\kern-.18em R}}
 \font\cmss=cmss10 \font\cmsss=cmss10 at 7pt
\def\IZ{\relax\ifmmode\mathchoice
 {\hbox{\cmss Z\kern-.4em Z}}{\hbox{\cmss Z\kern-.4em Z}}
 {\lower.9pt\hbox{\cmsss Z\kern-.4em Z}}
 {\lower1.2pt\hbox{\cmsss Z\kern-.4em Z}}\else{\cmss Z\kern-.4em Z}\fi}


\def\AEF{A.E. Faraggi}
\def\NPB#1#2#3{{\it Nucl.\ Phys.}\/ {\bf B#1} (#2) #3}
\def\PLB#1#2#3{{\it Phys.\ Lett.}\/ {\bf B#1} (#2) #3}
\def\PRD#1#2#3{{\it Phys.\ Rev.}\/ {\bf D#1} (#2) #3}
\def\PRL#1#2#3{{\it Phys.\ Rev.\ Lett.}\/ {\bf #1} (#2) #3}
\def\PRT#1#2#3{{\it Phys.\ Rep.}\/ {\bf#1} (#2) #3}
\def\MODA#1#2#3{{\it Mod.\ Phys.\ Lett.}\/ {\bf A#1} (#2) #3}
\def\IJMP#1#2#3{{\it Int.\ J.\ Mod.\ Phys.}\/ {\bf A#1} (#2) #3}
\def\nuvc#1#2#3{{\it Nuovo Cimento}\/ {\bf #1A} (#2) #3}
\def\RPP#1#2#3{{\it Rept.\ Prog.\ Phys.}\/ {\bf #1} (#2) #3}
\def\JHEP#1#2#3{ {\it JHEP } {\bf #1} (#2)  #3}
\def\etal{{\it et al\/}}

\hyphenation{su-per-sym-met-ric non-su-per-sym-met-ric}
\hyphenation{space-time-super-sym-met-ric} \hyphenation{mod-u-lar
mod-u-lar--in-var-i-ant}

\def\P#1{\Phi_{#1}}
\def\Pp#1{\Phi^{'}_{#1}}
\def\Pb#1{{\bar{\Phi}}_{#1}}
\def\bP#1{{\bar{\Phi}}_{#1}}
\def\Pbp#1{{\bar{\Phi}}^{'}_{#1}}
\def\Ppb#1{{\bar{\Phi}}^{'}_{#1}}
\def\Ppx#1{\Phi^{(')}_{#1}}
\def\Pbpx#1{\bar{\Phi}^{(')}_{#1}}

\def\p#1{\phi_{#1}}
\def\pp#1{\phi^{'}_{#1}}
\def\pb#1{{\bar{\phi}}_{#1}}
\def\bp#1{{\bar{\phi}}_{#1}}
\def\pbp#1{{\bar{\phi}}^{'}_{#1}}
\def\ppb#1{{\bar{\phi}}^{'}_{#1}}
\def\ppx#1{\phi^{(')}_{#1}}
\def\pbpx#1{\bar{\phi}^{(')}_{#1}}

\def\pbx#1{{\bar{\phi}_{#1 +}}}
\def\px#1{{     {\phi}_{#1 +}}}
\def\pbn#1{{\bar{\phi}_{#1 -}}}
\def\pn#1{{     {\phi}_{#1 -}}}
\def\Hp{H^{a}_{+}}
\def\Hm{H^{a}_{-}}
\def\Hbp{\bar{H}_{+}}
\def\Hbm{\H^{b}_{-}}
\def\cH#1{{\cal H}_{#1}}
\def\cHb#1{{\bar{\cal H}}_{#1}}
\def\cL#1{{\cal L}_{#1}}

\def\Db#1{{\bar{D}}_{#1}}
\def\Dm#1{{\mathbf D}_{#1}}
\def\Dmv#1{{\mathbf D}^{v}_{#1}}

\setcounter{footnote}{0}

\chapter{Introduction}

The standard model has provided a wealthy insight into low energy
physics and interactions of fundamental particles and their
respective force mediators.  In particular, it has been shown to
be consistent with virtually all physics down to scales
approaching $10^{-16}$cm. However, despite its success, the
``Standard Model of Elementary Particles'' could not support a
theory in which the electromagnetic, weak, strong and gravity
forces could be unified.

And so a new framework would be needed whereby the well known low
energy behavior of particles is captured, and that the gauge
symmetries could be unified at a higher scale. Along the path of
understanding this framework, it would be hoped that other
unsolved questions would also be answered. This motivation would
provoke, and indeed require, an unprecedented level of insight
into the mathematical and conceptual framework that represents the
underlying principles of nature.

The standard model entails the symmetry of particles as the group
$SU(3)\times SU(2)_L\times U(1)_Y$.  The group structure describes
the force mediators that set the multiplet families of matter. The
matter multiplets organize as fifteen multiplets of
spin-$\frac{1}{2}$ fermions in three generations of five. The
Standard model force mediator particles are supplemented with
gravity as a general relativistic theory.

There is a distinct conceptual underpinning of determinism versus
uncertainty in the case of gravity and the force mediators of the
standard model. In addition, there are a relatively large number of
free parameters that determine the dynamics within the Lagrangian.
From this, one must ask a question of the origin of the arrangements
of multiplets and what mechanism might determine the particular
values of the parameters used in the standard model structure.

Applying the highly successful method of field quantization (as used
for the standard model forces) to gravitation has arguably proven to
be the most complicated mathematical undertaking in theoretical
physics. A conventional attempt to from a union between the quantum
formalism and gravity yields a nonrenormalizable theory. This
suggests that if gravity is to be unified consistently with the
concept of uncertainty, there are physical mechanisms at work that
exist at higher probing scales that are beyond the scope of standard
model physics.

String theory is at present the best candidate for a unified
framework for fundamental forces. Of the many attractive features
it provides, it solves certain divergence problems that occur in
equivalent point particle field theory diagrams because of the
extended nature of a string.  String theory includes gravitation
defined within its own framework as a spin-2 closed string.
Indeed, it is capable of describing physics of gravity well beyond
the point where details of processes no longer have a consistent
interpretation (when, for example, approaching a singular point in
general relativity, which is not possible at the resolution of
standard model physics).

By contrast to the rather large number of parameters that reside
in the Lagrangian of the standard model, String theory has one
freedom which involves the string tension.  This presents a
striking first move toward a final theory, since such a theory (or
theories at scales approaching it) would necessarily involve fewer
or no freely fixable parameters.  It has been argued that a final
theory (that would necessarily include a unified description of
fundamental forces) might show itself by means of self consistent
and self explanatory mathematical structure.

While great strides are being made to develop the mathematical
ideas and tools necessary to explore strings, future generations
of accelerators will prompt the direction of such research with
the production of related data.  Although, there is still a gap in
cooperation between theory development and corroborative data at
scales beyond the standard model.

The large number of additional dimensions (by comparison to the
four that exist in the standard model case) allows significant
freedom to include string states with differing polarizations in
the spacetime and internal directions. The requirement of
additional dimensions leads to interesting compactification
scenarios that provide new quantum numbers and moduli.  For
example, in addition to the conventional Kaluza Klein quantum
number (which generically manifests itself in periodic or confined
particle quantum systems) resulting from compactification on a
circle, one also has towers of states that correspond to winding
modes.  Whereby the extended nature of the string allows it to
wrap around a compact dimension.

With the introduction of left and right moving oscillators that
involve left and right momentum components (that are symmetric and
antisymmetric under interchange of winding and Kaluza Klein
states) one has a simple duality relation for two theories defined
at two different radii.  The two radii are inversely related.  The
idea of this duality, combined with other operations that identify
different string theories has become a pivotal point in
contemporary research of string theory.

The picture that has emerged so far is that there are different
perturbative string theories that are related and exist in ten
dimensions, in addition to eleven dimensional supergravity.
Collectively, these theories are linked by duality relations form
perturbative limits of an underlying theory that doesn't have a
clearly defined formulation. This more fundamental theory, is
traditionally called M--theory.

In the context of M--theory, the true fundamental theory of
nature, the theory should realize itself in a nonperturbative
description.  However, at the current level of understanding, what
is known about this illusive theory are its perturbative limits,
which are the $SO(32)$ and $E_8\times E_8$ heterotic, type I, IIA
and IIB along with eleven dimensional supergravity theories.
Therefore, one should regard these limits as tools that allow the
ability to probe the properties of the fundamental vacuum. But it
may well be that none of these limits is capable of characterizing
the vacuum completely.  It is likely that all
perturbative limits will need to be used to isolate the true
M--theory vacuum.  As such, it may well be that different
perturbative string limits may provide more useful means to study
different properties of the true nonperturbative vacuum.

There is yet no dynamical mechanism that provides instructions on
how to select, and for what reason, a particular vacua.  Indeed,
while there are attempts in progress, such a mechanism may well
lead to, or otherwise outline the form of the underlying
M--theory.

The additional area of freedom that exists to explore appealing
models resides in the concept of compactification.  Since current
accelerators are not able to probe the scales at which these
hidden dimensions would allow their properties to be tested, there
is at present a wide range of possibilities for their structure
that are being considered. In this way, we can probe those
properties that pertain to the observed experimental and
cosmological data by using low energy effective field theory. The
study of the perturbative limits in both their low and high energy
behavior is the focus of intense study in the strings community.
Much progress has been made in recent years in the basic
understanding and intuitive meaning of the many new concepts and
ideas related to compactification that have unfolded as a result.

The study of the $\mathbb{Z}_2\times \mathbb{Z}_2$ modulated
spectrum, which is an orbifold structure upon which all models in
the work are considered, provides very appealing phenomenology. In
the heterotic limit, the models that are based on this
compactification have given rise to the most realistic superstring
models to date.  They contain the three generation structure with
each of the standard model generations contained in one of the
twisted sectors.  One would conclude that this group structure
provides some reason as to the manifest nature of the three
generation family as being dependant on the underlying geometry.

In addition, models based on the $\mathbb{Z}_2\times \mathbb{Z}_2$
orbifold have the $SO(10)$ embedding of the standard model
spectrum. As such, in these models $U(1)_Y$ has the standard
$SO(10)$ normalization with $k_Y=\frac{5}{3}$. In fact, the
$E_8\times E_8$ \cite{heterotic} heterotic construction is the
only perturbative limit that contains the standard $SO(10)$
embedding of the standard model. The reason for this is that it
contains the spinorial $\mathbf{16}$ representation in the
perturbative massless spectrum.  In this respect it may well be
that other perturbative string limits may provide more useful
means to study different properties of the true nonperturbative
vacuum.

In terms of vacuum selection, while a dynamical mechanism for
choosing one which is preferred is not known, it is appropriate to
suggest that the realistic vacuum resides in the vicinity of the
$\mathbb{Z}_2\times \mathbb{Z}_2$ class of models.  In addition,
with the standard model compatibility that the $E_8\times E_8$
model has, it is certainly suggested that a true vacuum would
favor this limit. The full study of phenomenological aspects of
the $\mathbb{Z}_2\times \mathbb{Z}_2$ orbifold is therefore of
significant benefit.  It would then follow that the study of other
limits based on the $\mathbb{Z}_2\times \mathbb{Z}_2$
compactification scheme would also be of benefit.

Much of string physics, with the exception of recent work being
done with time varying backgrounds, takes place on backgrounds
that are fixed. This is largely because of the geometrical
dependency that extended objects have when propagating on or
interacting with a boundary that changes with time.

There is yet no known general approach to tackling the different
limits with dynamical backgrounds.  In order to make steps in this
direction, it is very informative to understand such limits in the
context of their behavior under different fixed backgrounds.
Attempts to provide a greater picture in this way may well
highlight some mechanism that is present which allows such string
models to be defined consistently. Such a mechanism might be
extrapolated to show a more general framework of string scenarios
with dynamical backgrounds.

It would be a very strong step forward if such a potential
mechanism were to enlighten, or perhaps solve the problem of
stabilizing the "unfrozen moduli" that exist in string theory (for
example, in the case of the magnetized type I theory etc$\ldots$).
In looking at increasingly complicated backgrounds, such as
allowing the presence of magnetic monopoles in the type I setting,
or antisymmetric $\mathbf{B}$ field have the effect of adding to
the number of free moduli.

In the type I picture, the effect of magnetic monopoles admit an
alternative interpretation in terms of rotated branes in the type
IIA orientifold. The properties of the spectral content in this
scenario, with respect to three-generation like properties, have
been studied in \cite{CSU}. However, the type I framework offers
this description in terms of background magnetic fields
\cite{ACNY,FT}.

The six--dimensional case of the $T^4/\mathbb{Z}_2$ with two
homogeneous magnetic fields in each $T^2$ (resulting from
$T^4/\mathbb{Z}_2 \rightarrow \left(T^2(H_1)\times
T^2(H_2)\right)/\mathbb{Z}_2$ after introducing the two magnetic
fields $H_1$ and $H_2$) was demonstrated in \cite{CAAS}. This was
a particular case in which all tachyon's and possible terms that
could give rise to divergences and inconsistencies are absent for
the case that the two background field values are the same.

In extending this case to the study of the interesting class of
$\mathbb{Z}_2\times\mathbb{Z}_2$ models in four dimensions, such
background fields present difficulties. Tachyonic instabilities
present themselves without regard to the number of tori one wishes
to magnetize (where $T^6/\mathbb{Z}_2\times\mathbb{Z}_2$ has a
torus structure $T^2\times T^2\times T^2$). This is due to the
very nature of the $\mathbb{Z}_2\times \mathbb{Z}_2$ orbifold
generators which means that the compact directions have
arrangements of orbifold operations that include $(+,-,-)$,
$(-,+,-)$ and $(-,-,+)$.  In addition, the magnetically shifted
spectral states will have a dependence on the magnetic fields that
can break supersymmetry.

This shift happens in a twofold way.  The contributions from world
sheet fermions give an addition to the mass squared term as a
magnetic moment term that is sensitive to the helicity of the state
in the directions of the tori that carry a magnetic field. Secondly,
one has a tower of states that arise from the zero mode
contributions for untwisted bosonic excitations as another
contribution.  The magnetic moment couplings will be sensitive to
the sign of both the overall magnetic charge and the spin of the
state. In this way, it is thus possible to introduce a mass
splitting in supersymmetric multiplets.

So it is appreciated that if one is to assert that a magnetically
deformed type I stable vacua exists, then there must exist an
additional mechanism to prevent these excitations.

Moreover, the twisted structure of type I strings on such
manifolds (and indeed in the simpler $T^4/\mathbb{Z}_2$ case for
general choices of background fields) gives rise to divergences at
massless level in the tube (or transverse) channel. These
divergences take the form of residual terms that do not depend on
the measure parameter $\tau_2$, which would normally cancel as a
tadpole condition (in the absence of background magnetic fields).
For the $T^4/\mathbb{Z}_2$ example, however, these are easily
resolved with an appropriate field choice.  This is not possible
in the $\mathbb{Z}_2\times\mathbb{Z}_2$ case, due to the structure
of the generators (demonstrated in chapter \ref{ch:MagDeform} with
two magnetic fields). These effects have not been studied with the
choice of saturating the compact directions with magnetic fields.

These are then examples of how the study of increasingly
complicated background scenarios cause further questions on
realizing an underlying structure of string vacua that is capable
of simulating the real world.

In the work presented here, phenomenological aspects of the
heterotic $E_8\times E_8$ NAHE based free fermionic and $SO(32)$
type I models are considered.   In the light of the motivations
for research already given, this work is intended to highlight
particular points and phenomenology of these theories.

In the first chapter, the work is centered on the heterotic free
fermionic models which discusses a particular extension of the
NAHE set. The choice of extended vectors $\{\alpha,\beta,\gamma\}$
with appropriate GSO coefficients give rise to a low lying
spectrum that will allow a satisfaction of the $D$--flat
constraints only for the use of VEVs of non--Abelian fields. This
is the first time that non--Abelian VEVs have been required for
$D$--flat consistency as opposed to other possible
phenomenological reasons.

In chapter \ref{ch:OpenDescendents}, the effect of discrete Wilson
lines is discussed in type I strings compactified on
$T^6/(\mathbb{Z}_2\times\mathbb{Z}_2^\text{s})$ and
$T^6/(\mathbb{Z}_2\times\mathbb{Z}_2\times\mathbb{Z}_2^\text{s})$
manifolds.  In the first case, I discuss a toy model which has one
$\mathbb{Z}_2$ orbifold generator accompanied by a freely acting
shift ($\mathbb{Z}_2^\text{s}$).  The second case, which is
essentially a generalization of this relies on the
$\mathbb{Z}_2\times\mathbb{Z}_2$ orbifold structure.

As described previously, the NAHE heterotic models are underlined
by a geometric manifold which is an orbifold compactification of
the type $\mathbb{Z}_2\times \mathbb{Z}_2$. This structure is
common to all the three generation free fermionic models.

The type I models show the effects on the open and closed spectrum
from the projection of the torus with a freely acting momentum
shift.  It will be seen that while the first model is
straightforwardly consistent, the second model presents some
subtleties.  Specifically, the symmetrization of states in the
open and closed sectors.

Projections involving shifts that are defined as part of the
$\mathbb{Z}_2\times \mathbb{Z}_2$ structure have been studied in
\cite{AADS}.  In this case, the shifts were not freely acting since
they involve projectors of the form $(\delta,-\delta,-)$ for
example.  Here, a shift operation, denoted by $\delta$, acts
simultaneously with an orbifold operation.  These models remove
twisted terms in the torus that are independent orbits (terms that
are not related to the principle orbits
$\{(o,o),(o,g),(o,f),(o,h)\}$ by the generators of modular
invariance, which is demonstrated in appendix \ref{app:Boundaries}).
The presence of such terms, as will be seen in the second model,
allow a sign freedom ($\epsilon=\pm 1$). This freedom will allow for
the introduction of eight possible subclasses of models. However, I
show that the subclasses belonging to $\epsilon=-1$, has a problem
with particle interpretation.

I will demonstrate the mechanisms of supersymmetry breaking that
result in the four subclasses of models belonging to
$\epsilon=+1$. The sign $\epsilon$ is defined in terms of three
signs $\epsilon_i=\pm 1$ for $i=1,2,3$ where
$\epsilon=\epsilon_1\epsilon_2\epsilon_3$ and the models are
classified by $(\epsilon_1,\epsilon_2,\epsilon_3)$. The only fully
supersymmetric model in all eight that are possible from
$\epsilon=\pm 1$ is the $(+,+,+)$ model.

In the third chapter, I demonstrate the appearance of residual
terms in the $\mathbb{Z}_2\times \mathbb{Z}_2$ open spectrum with
two background magnetic fields in a model with discrete torsion.
These terms give rise to tadpoles in the twisted sector that do
not cancel.  In the light of this model, it is appropriate to
discuss the details of consistency conditions in type I strings.
Additionally, the models in chapter \ref{ch:OpenDescendents} also
show interesting deviations from the satisfaction of these
consistency requirements.

In the ten dimensional case, the orientifold projection of the
type IIB superstring torus amplitude, which is of course the ten
dimensional type I superstring, gives rise to
\begin{eqnarray}
{\cal T}&=&\frac{1}{2}\int_{\cal F}
\frac{d^2\tau}{\tau_2^2}\frac{1}{\tau_2^4}\frac{|V_8-S_8|^2(\tau)}{\eta (\tau)}\nn\\
{\cal K}&=&\frac{1}{2}\int_{0}^\infty
\frac{d\tau_2}{\tau_2^2}\frac{1}{\tau_2^4}\frac{(V_8-S_8)(2i\tau_2)}{\eta(2i\tau_2)}
\end{eqnarray}
where the second case is the Klein bottle amplitude, and in both
cases their respective dependencies on the measures $\tau$ and
$2i\tau_2$ are shown. The details of the construction are found in
chapter \ref{ch:OpenDescendents}. After an $S$ transformation, the
Klein can be written in terms of a closed string propagating
between two non--dynamical objects known as $O9$--planes.  This is
illustrated in the diagrams of appendix (\ref{app:TransDiagrams})
for the Klein, annulus and Mobius couplings.  However, this closed
string coupling has the interpretation of two one--point functions
at the $O9$--planes with a propagator of
\begin{eqnarray}\label{eqn:prop}
\frac{1}{p^2+m^2}.
\end{eqnarray}
The factor of $\frac{1}{\tau_2^4}$ can be converted into a
momentum integral according to equation (\ref{eqn:Guass}).  So at
zero momentum (and after the appropriate rescaling of
$2i\tau_2=it=\frac{i}{l}$) the Klein amplitude has the form
\begin{eqnarray}
\tilde{{\cal K}}=\frac{2^5}{2}\int_{0}^\infty
dl\frac{(V_8-S_8)(il)}{\eta (il)}\sim \frac{2^5}{2}\int_{0}^\infty
dl ~e^{-lm^2}=\frac{1}{m^2}.
\end{eqnarray}
This of course carries a divergence, comparing with equation
(\ref{eqn:prop}) at zero momentum.

So, part of the rationale for having an open sector is to allow
cancellation of this tadpole divergence term.  The transverse
annulus and Mobius amplitudes are
\begin{eqnarray}
\tilde{\cal A}&=&\frac{2^{-5}N^2}{2}\int_{0}^\infty
dl~\frac{(V_8-S_8)(il)}{\eta (il)}\nn\\
\tilde{\cal M}&=&\frac{2\epsilon N}{2}\int_{0}^\infty
dl~\frac{(\hat{V}_8-\hat{S}_8)(il)}{\hat{\eta} (il)}.
\end{eqnarray}
The Chan--Paton charges $N$ count the stacks of dynamical objects
known as $D9$--branes to which the open string ends couple (in the
direct channel).  In the present discussion, one has closed
strings propagating between branes. This then shows the
contributions from $\tilde{ \cal K}+\tilde{ \cal A}+\tilde{ \cal
M}$ as proportional to
\begin{eqnarray}\label{eqn:Div}
\frac{2^{-5}}{2}(2^5+\epsilon N)^2\int_{0}^\infty dl~e^{-lm^2}.
\end{eqnarray}
The low energy effective action that describes these $O9$-- and
$D9$-- brane Neveu--Schwarz couplings is
\begin{eqnarray}\label{eqn:NSaction}
S_{NS}\sim (2^5+\epsilon N)^2\int d^{10}x
e^{-\phi}\sqrt{-\text{det}(G_{\mu \nu})}
\end{eqnarray}
for dilaton field $\phi$ and ten dimensional metric $G_{\mu \nu}$.
So one has a cancellation of the divergence (\ref{eqn:Div}) for
the the choice of $\epsilon=-1$.  Moreover this confines $N=32$,
which of course allows the ten dimensional type I gauge group of
$SO(32)$. The other choice of $\epsilon=+1$ leads to a gauge group
of $USp(32)$.  These gauge groups can be seen from the direct
channel expansion of the open sector as the number of vector
bosons (coming from $V_8$) that act as the generators with
\begin{eqnarray}
{\cal A}+{\cal M}\sim\frac{N(N+\epsilon)}{2}(V_8-S_8).
\end{eqnarray}

Moving to describe the Ramond sector couplings, one has to take
into account the potentials that arise from the bi--spinor
expansion
\begin{eqnarray}
\Phi_\alpha\bar{\Phi}_\beta=C_0 \Gamma_{\alpha \beta}^0+C_{\mu
\nu}\Gamma_{\alpha \beta}^{\mu \nu}+C_{\mu \nu \rho
\kappa}\Gamma_{\alpha \beta}^{\mu \nu \rho \kappa}.
\end{eqnarray}
The low energy effective action is of the form
\begin{eqnarray}\label{eqn:Raction}
S_{R}\sim (2^5+\epsilon N)^2\mu_{(9)}\int C_{10}.
\end{eqnarray}
In this case one has both $O9$--planes and the $D9$ branes
carrying the Ramond charge $\mu_{(9)}$.  There must therefore be
an overall vanishing of the Ramond contributions.  In the case of
the action (\ref{eqn:NSaction}), one has an interpretation in
terms if an increased vacuum energy if $\epsilon=+1$. However, for
contributions arising from (\ref{eqn:Raction}) there are charges
with associated field lines.  In a noncompact space, one can
simply have the vanishing of the $C_{p+1}$ forms at infinity. In a
compact space, as will be the case in the models discussed in this
work, the fields lines must begin and end somewhere, which can
only be accomplished if there as many sources as sinks. In this
case, there are exactly $32$ sources as $D9$--branes and an
$O9$--plane. The relative charges carried by a $D9$--brane and an
$O9$--plane is of course $32$.

In the magnetic chapter, it will be shown that the residual terms
appearing in the transverse twisted sectors of the annulus cannot
lead to consistent cancellation of both the Ramond and the
Neveu--Schwarz sector tadpoles.  This will therefore lead to an
inconsistent model.  While this is true for the case with discrete
torsion, the model is completely consistent for the case without
discrete torsion \cite{MLGP}

I now turn to the relevant discussions of the heterotic and type I
string phenomenology.

\setcounter{footnote}{0}

\chapter{$D$--flatness in a 4D Heterotic Model Requiring Non--Abelian Fields}

In this section I look at the necessity of non--Abelian $D$--flat
constraints of a four dimensional toy model in the free fermionic
construction for the one loop amplitude \cite{cff}. Unlike other
models before it, it will be seen that the spectrum does not allow
any solutions to the $D$--flat constraint equations for the use of
VEVs of fields that are non--Abelian singlets.

The spectral content and phenomenology of these string models are
determined by choices of boundary condition vectors and the one
loop GSO phases.  This model is underlined by the NAHE set which
contains a set of boundary condition vectors that forms a base for
a very large class of phenomenologically interesting models.

Although string consistency requirements impose conditions on the
boundary conditions, there is freedom within the restrictions of
consistency for models with greatly different phenomenology. In
this toy model, the choice of boundary condition vectors plus an
additional change to one of the GSO phases gives rise to:
\begin{itemize}

\item An enhancement in the observable and hidden sector gauge
groups.

\item Solution of the $D$--flat constraint equations exclusively through
non--Abelian VEVs.

\end{itemize}

The string model discussed here belongs to a particular class of
models know as Left--Right Symmetric models (LRS). These are free
fermionic models with an observable sector gauge group that
contains $SU(2)$ terms as $SU(2)_L\times SU(2)_R$.

I begin by reviewing the construction tools necessary for model
building.

\section{The Free Fermionic Construction}\label{sec:FermionicConstruction}

The free fermionic construction relies on an action that allows
world sheet bosonic and fermionic fields to act freely of one
another with no interaction.  The heterotic action arises from
taking left moving superstring degrees of freedom and right moving
bosonic degrees of freedom and compactifying directly to four
spacetime dimensions.  In the fermionic formulation, one then
applies fermionization to all internal bosonic fields.  One finds
that a complex fermionic field $\psi$ and abosonic field $H$
written as $e^{iH}$ satisfy the same operator product exapnsions.
From each bosonic field $H$, two fermionic fields can be written
as
\begin{eqnarray}
\psi\sim e^{iH},\quad {\psi^\ast}\sim e^{-iH}
\end{eqnarray}

The action for this construction is then defined by
\begin{eqnarray}
\int d^2\sigma\bigg[\partial_\alpha X^\mu_L
\partial^\alpha X_{\mu,L}+\partial_\alpha X^\mu_R
\partial^\alpha X_{\mu,R}
-i\bar{\psi}^\mu_L\gamma^\alpha\partial_\alpha\psi_{\mu,L} \nn\\
-i\bar{\psi}^I_L\gamma^\alpha
\partial_\alpha\psi_{I,L}-i\bar{\psi}^I_R\gamma^\alpha
\partial_\alpha\psi_{I,R}\bigg],
\end{eqnarray}
where $\mu=1,2$ are the spacetime indices and $I$ are the internal
ones. The labels $L$ and $R$ imply left and right moving fermions.

The spectrum is then composed of left moving coordinates
$X^\mu_L$, $\psi^\mu_L$ for transverse coordinates and eighteen
internal real fermions, which in the notation used in this model,
are denoted as six sets of three by the index $i=1,\dots,6$. The
right moving sector has transverse bosonic coordinates $X^\mu_R$
and forty four ($j=1,\ldots 44$) internal real fermionic
coordinates as
\begin{eqnarray}\label{eqn:HeteroticLeftRightCoord}
\text{left moving }\left\{ \begin{array}{l} X^\mu(z) \\ \Psi^\mu(z) \\
\chi_i(z), y_i(z), \omega_i(z) \\
\end{array} \right.
\text{ and right moving }\left\{ \begin{array}{l} \bar{X}^\mu(\bar{z}) \\
\bar{\phi}_j(\bar{z})
\end{array} \right..
\end{eqnarray}

The total of sixty four fermions are grouped as complex fermions
\begin{eqnarray}
\psi=\frac{1}{\sqrt{2}}\left(\eta_j+i\eta_{k}\right),\quad
\psi^\ast=\frac{1}{\sqrt{2}}\left(\eta_j-i\eta_{k}\right),
\end{eqnarray}
the particular grouping will of course rely on fermions having the
same boundary condition where $\eta_j$ and $\eta_k$ are real.

A boundary condition vector defined as
\begin{eqnarray}\label{eqn:BCvec}
\mathbf{a}=\{\mathbf{a}_L|\mathbf{a}_R\}=\{a_1,\ldots,a_{10}|a_{11},\ldots,a_{22}\}
\end{eqnarray}
with components $a_i$ will impose the condition on the
$i^\text{th}$ complex fermion as
\begin{eqnarray}\label{eqn:boundaryconditions}
\Psi^i(\tau,\sigma+2\pi)=-e^{-\pi i a_i}\Psi^i(\tau,\sigma).
\end{eqnarray}
The labels $L$ and $R$ show the parts that act on the left and
right moving fermions. One obtains $NS$ (Neveu--Schwarz) and $R$
(Ramond) boundary condition when the entries $a_i$ are 0 or 1
respectively.

Model building begins with a prescription of vectors
\begin{eqnarray}\label{eqn:initalset}
\{\mathbf{b}_1,\ldots,\mathbf{b}_n\}
\end{eqnarray}
that underly a general model. These vectors span a finite additive
group
\begin{eqnarray}\label{eqn:AssGroup}
\Xi = \sum_{i=1}^n m_i\mathbf{b}_i
\end{eqnarray}
for $m_i=0,\ldots ,N_i-1$ such that $N_i{\bf b}_i=0\text{ (mod
}2)$.

In terms of boundary condition vectors $\alpha$, $\beta\in \Xi$
(with components $\alpha_l$ and $\beta_l$) the partition function
will be written as
\begin{eqnarray}\label{eqn:PartitionFunction}
Z=\int \frac{d^2\tau}{[{\rm
Im}(\tau)]^2}Z_B(\tau,\bar{\tau})\sum_{\text{$\alpha,\beta$}} C
\left(
\begin{array}{c} \alpha\\ \beta
\end{array}\right)\prod_{l=1}^{10}Z_l \left[ \begin{array}{c} \alpha_l \\ \beta_l
\end{array}
\right]\prod_{l=11}^{22}\bar{Z}_k \left[ \begin{array}{c} \alpha_l \\
\beta_l
\end{array}
\right].
\end{eqnarray}
The phases $C\left( \begin{array}{c} \alpha \\
\beta
\end{array} \right)$ are chosen to allow the partition function to
be modular invariant.  Modular invariance is necessary to ensure
that the partition function counts only distinct tori.  A more
formal discussion of modular invariance is left until the type I
construction in the following chapter.

The phases that correspond to a given set of basis vectors, as
will be clarified later, can introduce a freedom in the form of a
choice of sign.

The invariance of $Z$ will introduce rules or conditions on the
basis vectors and phases introduced in
(\ref{eqn:PartitionFunction}). In addition, a GSO projection based
on these vectors and phases will allow a projection of states of a
given model to describe the spectral content. The calculation of
these rules can be found in \cite{KLT} and are summarized below.

Basis vectors that are consistent with modular invariance satisfy
\begin{eqnarray}
N_{ij}\mathbf{b}_i.\mathbf{b}_j&=&0\text{
(mod }4).
\end{eqnarray}
where $N_{ij}$ is the least common multiple of the integers $N_i$
and $N_j$.  If one sets $i=j$, this condition holds for $N_i$ odd.
In the case that $N_i$ is even, one has
\begin{eqnarray}
N_i {\bf b}_i.{\bf b}_i&=&0\text{ (mod }8).
\end{eqnarray}
The operation ($.$) counts the left and right
components of two parameters $\mathbf{a}$ and $\mathbf{b}$ as
\begin{eqnarray}
\mathbf{a}.\mathbf{b}=\bigg(\sum_{{\rm Left}}-\sum_{{\rm
right}}\bigg)a_jb_j.
\end{eqnarray}

The GSO projection is defined as
\begin{eqnarray}\label{eqn:GSO}
e^{i\pi{\bf b}_j.F_\alpha}|s>_\alpha = \delta_\alpha C \left(
\begin{array}{c} \alpha \\ \mathbf{b}_j \end{array} \right)^\ast |s>_\alpha
\end{eqnarray}
with $\alpha \in \Xi$ and $|s>_\alpha$ defines a state $|s>$ in
the sector $\alpha$ ($\ast$ denotes complex conjugation), and
\begin{eqnarray}
\delta_{\alpha}=e^{i\pi \alpha(\psi^\mu)}=
\left\{\begin{array}{cc} -1 & \alpha(\psi^\mu)=1 \\
+1 & \alpha(\psi^\mu)=0
\end{array}\right..
\end{eqnarray}

The phases that appear in the partition function, and thus the GSO
projection, will have a sign freedom (in some sectors). Such
choices can result in vastly different spectral content.  In the
case studied here, the choice of a different sign results, in
particular, in an enhancement of the observable and hidden gauge
groups.

The phases depend on the vectors according to
\begin{eqnarray}\label{eqn:PhaseEquations}
C\left(\begin{array}{c} \mathbf{b}_i \\
\mathbf{b}_j
\end{array}\right)=\delta_{\mathbf{b}_j}exp\left({\frac{2i \pi
m_i}{N_i}}\right)exp\left({\frac{i\pi
\mathbf{b}_i.\mathbf{b}_j}{2}}\right)
\end{eqnarray}
where $m_i$ and $N_i$ are understood by equation
(\ref{eqn:AssGroup}).

The world sheet fermion number operator $F$ acts on the complex
modes $\psi$ and $\psi^\ast$ as a sign, so that $F\psi=+\psi$ and
$F\psi^\ast=-\psi$.

The massless spectral content will generally have states that are
pure vacuum which is degenerate.  Such configurations have
spinorial representations $|\pm>$. As such, one has $F|+>=0|+>$
and $F|->=-|->$ for these states.

The study of the phenomenology of these models involves the
massless states from each sector.  For each element $\alpha \in
\Xi$, the oscillators allowed in the massless spectrum of the
sector $\alpha$ are confined through the equations
\begin{eqnarray}
M_L^2=0=-\frac{1}{2}+\frac{\alpha_L.\alpha_L}{8}+N_L,\nn\\
M_R^2=0=-1+\frac{\alpha_R.\alpha_R}{8}+N_R.
\end{eqnarray}
Here, the terms $N_{L,R}$ are the left and right moving fermionic
number operators and $\alpha_{L,R}$ are the components of $\alpha$
that act on the corresponding left and right fermions.

Thus, a model will be fully specified by the basis vectors and a
choices for the phases, where such choices exists.

For an example of the choice that can arise, if one takes a vector
$S$ that has the boundary conditions on the fermions
$\{\Psi^\mu,\chi_{1,2},\chi_{3,4},\chi_{5,6}\}=1$ and all others
0, then $N_s=2$.  So $m_s\in\{0,1\}$, and these two possibilities
lead to different phases if one projects using the sector ${\bf
b}_j=\bf 1$, with reference to equation
(\ref{eqn:PhaseEquations}). As models become more complicated, so
do the choices of the phases.

The following notation for the left and right sectors will be used
for the total of sixty four fermions,
\begin{eqnarray}\label{eqn:fermions1}
\psi^\mu_{1,2}(\chi_1y_1\omega_1)(\chi_2y_2\omega_2)(\chi_3y_3\omega_3)
(\chi_4y_4\omega_4)(\chi_5y_5\omega_5)(\chi_6y_6\omega_6)
\end{eqnarray}
for the left sector, which carries the fermion spacetime field.
In the right sector, one has
\begin{eqnarray}\label{eqn:fermions2}
\bar{y}_1\bar{\omega}_1\bar{y}_2\bar{\omega}_2\bar{y}_3\bar{\omega}_3
\bar{y}_4\bar{\omega}_4\bar{y}_5\bar{\omega}_5\bar{y}_6\bar{\omega}_6
\bar{\Psi}_{1\ldots
5}\bar{\eta}_1\bar{\eta}_2\bar{\eta}_3\bar{\phi}_{1\ldots 8}.
\end{eqnarray}
In all models, the fields $\bar{\Psi}_{1\ldots 5},
\bar{\eta}_1,\bar{\eta}_2,\bar{\eta}_3$ and $\bar{\phi}_{1\ldots
8}$ are complex.  The pairings of the remaining internal fields
will depend on the choice of boundary condition vectors.


\setcounter{footnote}{0}
\section{$F$-- and $D$-- Term Supersymmetry Breaking}


It will be helpful to review some of details of breaking
supersymmetry through $F$-- and $D$-- terms. The Fayet--Iliopoulos
(FI) term is an addition to the Lagrangian in the form $\xi D$ for
the auxiliary field $D$ \cite{dsw1,dsw2}. The variation with
respect to this field then allows the inclusion of the factor
$\xi$ in the equation of motion.  For the $D$ terms, supersymmetry
can be spontaneously broken via $\langle D_A\rangle\neq 0$ if
there are no fields $\chi_k$ with charges $Q_A^k$ such that
$\sum_k Q^k_A\vert\langle \chi_k\rangle\vert=-\xi$.  Similarly,
supersymmetry is spontaneously broken via $\langle F\rangle\neq
0$.

The terms that result from variation with respect to the auxiliary
fields $D$ and $F$ are
\begin{eqnarray}
D_A&=&
\sum_k Q_A^k\vert\chi_k\vert^2+\xi,\label{da}\\
D_\alpha&=&
\sum_k Q_\alpha^k\vert\chi_k\vert^2\quad (\alpha\ne A),\label{dalpha}\\
F_i&=&-\frac{\partial W}{\partial\eta_i},
\end{eqnarray}
where
\begin{eqnarray}
&&\langle D_A\rangle=\langle D_\alpha\rangle=\langle
F_i\rangle=0\label{dterms},\\
&&\text{ }\xi={\frac{g^2({\rm Tr} Q_A)}{192\pi^2}}M_{\rm Pl}^2
\label{dxi}.
\end{eqnarray}
The fields $\chi_k$ are those which acquire VEVs of order
$\sqrt\xi$ and the $\eta_i$ fields are scalars. $W$ is the super
potential, the terms $Q_A^k$ and $Q_\alpha^k$ denote the anomalous
and non--anomalous charges and $M_{\rm Pl}\approx2\times 10^{18}$
GeV denotes the reduced Planck mass. The solution ({\it i.e.}\ the
choice of fields with non--vanishing VEVs) to the set of equations
(\ref{dterms})--(\ref{dalpha}), though nontrivial, is not unique.
Therefore in a typical model there exist a moduli space of
solutions to the $F$ and $D$ flatness constraints.

Very interesting aspects of the phenomenology of models involves
the analysis and classification of flat directions. The methods
for the analysis of $D$--flatness in string models, as the reader
will appreciate from the discussion in section \ref{sec:SDFC},
have been made quite systematic \cite{systematic,mshsm}. It has in
general been assumed in the past that in a given string model
there should exist a solution to the $F$ and $D$ flatness
constraints.  The simpler type of solutions to the flatness
constraints utilize only fields that are singlets of all the
non--Abelian groups in a given model (type I solutions).

In the following section, the string model discussed does not
contain a type I solution.  That is, the VEVs required to ensure
that there are solutions to the flat directions must be from a set
of fields that are not solely composed of singlet VEVs. Moreover,
a solution using only singlet VEVs does not exist. This particular
model was the first instance in which non-Abelian fields in the
spectrum are required to have non vanishing VEVs if one insists on
conforming to the constraints $\langle D_a \rangle=\langle
D_\alpha \rangle=0$.

\section{The string model}

The free fermionic string model discussed here is constructed by
specifying a set of one loop GSO projection coefficients
\cite{fff,ff1} that correspond to the basis vectors
$\{1,\mathbf{S},\mathbf{b}_1,\mathbf{b}_2,\mathbf{b}_3\}$ and
$\{\mathbf{\alpha},\mathbf{\beta},\mathbf{\gamma}\}$.  These
vectors then form the set that was defined in
(\ref{eqn:initalset}).

The first set of vectors corresponds to the NAHE set
\cite{nahe,nahe1}, which are common to all the semi--realistic
free fermionic models. The second consists of three additional
boundary condition basis vectors. The rules for extracting the
superpotential terms were derived in ref. \cite{kln}.

For the NAHE set, the boundary conditions on the left and right
moving coordinates are defined in table \ref{tab:NAHEset}.
\begin{eqnarray}
&\begin{tabular}{c|c|ccc|c|ccc|c} & $\psi^\mu$ & $\chi^{12}$ &
$\chi^{34}$ & $\chi^{56}$ & $\bar{\psi}^{1,\ldots,5}$ &
$\bar{\eta}^{1}$ & $\bar{\eta}^{2}$ & $\bar{\eta}^{3}$ &
$\bar{\phi}^{1,\ldots,8}$ \\
\hline \hline
            $1$ & 1 & 1 & 1 & 1 & 1 & 1 & 1 & 1 & 1 \\
   $\mathbf{S}$ & 1 & 1 & 1 & 1 & 0 & 0 & 0 & 0 & 0 \\
 $\mathbf{b}_1$ & 1 & 1 & 0 & 0 & 1 & 1 & 0 & 0 & 0 \\
 $\mathbf{b}_2$ & 1 & 0 & 1 & 0 & 1 & 0 & 1 & 0 & 0 \\
 $\mathbf{b}_3$ & 1 & 0 & 0 & 1 & 1 & 0 & 0 & 1 & 0 \\
\end{tabular}
\nonumber\\
   ~  &  ~ \nonumber\\
   ~  &  ~ \nonumber\\
&\begin{tabular}{c|c|c|c} & $y^3{y}^6$ $y^4{\bar y}^4$ $y^5{\bar
y}^5$ ${\bar y}^3{\bar y}^6$ & $y^1{\omega}^5$ $y^2{\bar y}^2$
$\omega^6{\bar\omega}^6$ ${\bar y}^1{\bar\omega}^5$ &
$\omega^2{\omega}^4$ $\omega^1{\bar\omega}^1$
$\omega^3{\bar\omega}^3$ ${\bar\omega}^2{\bar\omega}^4$ \\
\hline \hline
$1$            & 1 ~~~ 1 ~~~ 1 ~~~ 1  & 1 ~~~ 1 ~~~ 1 ~~~ 1  &
1 ~~~ 1 ~~~ 1 ~~~ 1 \\
$\mathbf{S}$   & 0 ~~~ 0 ~~~ 0 ~~~ 0  & 0 ~~~ 0 ~~~ 0 ~~~ 0  &
0 ~~~ 0 ~~~ 0 ~~~ 0 \\
$\mathbf{b}_1$ & 1 ~~~ 1 ~~~ 1 ~~~ 1  & 0 ~~~ 0 ~~~ 0 ~~~ 0  &
0 ~~~ 0 ~~~ 0 ~~~ 0 \\
$\mathbf{b}_2$ & 0 ~~~ 0 ~~~ 0 ~~~ 0  & 1 ~~~ 1 ~~~ 1 ~~~ 1  &
0 ~~~ 0 ~~~ 0 ~~~ 0 \\
$\mathbf{b}_3$ & 0 ~~~ 0 ~~~ 0 ~~~ 0  & 0 ~~~ 0 ~~~ 0 ~~~ 0  &
1 ~~~ 1 ~~~ 1 ~~~ 1 \\
\end{tabular}\label{tab:NAHEset}
\end{eqnarray}

The boundary conditions of the three basis vectors which extend
the NAHE set are shown in Table \ref{model1}.  This table as well
as table \ref{tab:NAHEset} above show the pairings of left and
right moving real fermions from the set $\{y_i,\omega_i|{\bar
y}_i,{\bar\omega}_i\}$, for $i=1,\ldots,6$. These fermions are
paired to form complex left or right moving fermions. The
particular pairings will of course vary with different models that
assign alternate boundary conditions to these fermions.

The generalized GSO coefficients determining the physical massless
states (through he GSO projection) of Model 1 appear in matrix
\ref{phasesmodel1}. \vskip 0.4truecm
\begin{eqnarray}
&\begin{tabular}{c|c|ccc|ccccc|c|cccc} & $\psi^\mu$ & $\chi^{12}$
& $\chi^{34}$ & $\chi^{56}$ & $\bar{\psi}^{1}$ & $\bar{\psi}^{2}$
& $\bar{\psi}^{3}$ & $\bar{\psi}^{4}$ & $\bar{\psi}^{5}$ &
$\bar{\eta}^{1,2,3}$ & $\bar{\phi}^{1}$ & $\bar{\phi}^{2,3,4}$ &
$\bar{\phi}^{5,6,7}$ & $\bar{\phi}^{8}$\\
\hline \hline
$\alpha$ & 0 & 0 & 0 & 0 & 1 & 1 & 1 & 0 & 0 & 0 & 1 & 1 & 0 & 0 \\
$\beta$  & 0 & 0 & 0 & 0 & 1 & 1 & 1 & 0 & 0 & 0 & 1 & 1 & 0 & 0 \\
$\gamma$ & 0 & 0 & 0 & 0 & $\frac{1}{2}$ & $\frac{1}{2}$ & $\frac{1}{2}$ & 0 &
0 & $\frac{1}{2}$ & 0 & $\frac{1}{2}$ & $\frac{1}{2}$ & 0 \\
\end{tabular}
\nonumber\\
   ~  &  ~ \nonumber\\
   ~  &  ~ \nonumber\\
&\begin{tabular}{c|c|c|c} & $y^3{y}^6$ $y^4{\bar y}^4$ $y^5{\bar
y}^5$ ${\bar y}^3{\bar y}^6$ & $y^1{\omega}^5$ $y^2{\bar y}^2$
$\omega^6{\bar\omega}^6$ ${\bar y}^1{\bar\omega}^5$ &
$\omega^2{\omega}^4$ $\omega^1{\bar\omega}^1$
$\omega^3{\bar\omega}^3$ ${\bar\omega}^2{\bar\omega}^4$ \\
\hline \hline
$\alpha$ & 1 ~~~ 1 ~~~ 1 ~~~ 0  & 1 ~~~ 1 ~~~ 1 ~~~ 0  &
1 ~~~ 1 ~~~ 1 ~~~ 0 \\
$\beta$  & 0 ~~~ 1 ~~~ 0 ~~~ 1  & 0 ~~~ 1 ~~~ 0 ~~~ 1  &
1 ~~~ 0 ~~~ 0 ~~~ 0 \\
$\gamma$ & 0 ~~~ 0 ~~~ 1 ~~~ 1  & 1 ~~~ 0 ~~~ 0 ~~~ 0  &
0 ~~~ 1 ~~~ 0 ~~~ 1 \\
\end{tabular}\label{model1}
\end{eqnarray}

LRS Model 1 Generalized GSO Coefficients:
\begin{equation}
{\bordermatrix{
          &{\bf 1}&\mS & &{\mb_1}&{\mb_2}&{\mb_3}& &{\malpha}&{\mbeta}
&{\mgamma}\cr
       {\bf 1}&~~1&~~1 & & -1   &  -1 & -1  & & ~~1     & ~~1   & ~~i   \cr
           \mS&~~1&~~1 & &~~1   & ~~1 &~~1  & &  -1     &  -1   &  -1   \cr
          &   &    & &      &     &     & &         &       &       \cr
       {\mb_1}& -1& -1 & & -1   &  -1 & -1  & &  -1     &  -1   & ~~i   \cr
       {\mb_2}& -1& -1 & & -1   &  -1 & -1  & &  -1     &  -1   & ~~i   \cr
       {\mb_3}& -1& -1 & & -1   &  -1 & -1  & &  -1     &  -1   & ~~i   \cr
          &   &    & &      &     &     & &         &       &       \cr
     {\malpha}&~~1& -1 & &~~1   & ~~1 &~~1  & & ~~1     & ~~1   & ~~1   \cr
      {\mbeta}&~~1& -1 & & -1   &  -1 &~~1  & &  -1     &  -1   &  -1   \cr
     {\mgamma}&~~1& -1 & &~~1   &  -1 &~~1  & &  -1     &  -1   & ~~1   \cr}}
\label{phasesmodel1}
\end{equation}

After the NAHE set, one has $N=1$ supersymmetry with a gauge group
of $SO(10)\times E_8 \times SO(6)^3$ and a total of 48 spinorial,
with $16$ of $SO(10)$, each $16$ correspondingly arises from each
sector $b_1$, $b_2$ and $b_3$. The $SO(10)$ comes from the
fermions $\bar{\Psi}^{1,\ldots,5}$.

In matrix (\ref{phasesmodel1}) only the entries above the diagonal
are independent and those below and on the diagonal are fixed by
the modular invariance constraints.  The relation between off
diagonal components can be seen from
\begin{eqnarray}
C\left(\begin{array}{c} \mathbf{b}_i \\
\mathbf{b}_j
\end{array}\right)=exp\left(\frac{i\pi \mathbf{b}_i.\mathbf{b}_j}{2}\right)
C\left(\begin{array}{c} \mathbf{b}_j \\
\mathbf{b}_i
\end{array}\right)^\ast.\nn
\end{eqnarray}
Blank lines are inserted to emphasize the division of the free
phases between the different sectors of the realistic free
fermionic models. Thus, the first two lines involve only the GSO
phases of
\begin{eqnarray}
c\left(\begin{array}{c}\{\mathbf{1},\mathbf{S}\} \\
\mathbf{a}_i \end{array}\right).\nn
\end{eqnarray}
The set $\{\mathbf{1},\mathbf{S}\}$ generates the $N=4$ model with
$\mS$ being the space--time supersymmetry generator.

Note that the boundary condition basis vectors that generate the
string model are those of Model 3 \cite{lrsmodels}. The two models
differ by a GSO phase with
\begin{eqnarray}\label{eqn:Phase1}
c\left(\begin{array}{c} \mathbf{b}_3 \\
\beta \end{array}\right)=-1
\end{eqnarray}
in this model, and
\begin{eqnarray}
c\left(\begin{array}{c} \mathbf{b}_3 \\
\beta \end{array}\right)=+1
\end{eqnarray}
in Model 3. As is elaborated on below, the consequence of this GSO
phase change is that the gauge symmetry is enhanced, with one of
the Abelian generators being absorbed into the enhanced
non--Abelian gauge symmetry. Consequently, the number of Abelian
group factors is reduced, which simplifies somewhat the analysis
of the $D$--flat directions.

The final gauge group of the string model arises as follows: In
the observable sector the NS boundary conditions produce gauge
group generators for
\begin{eqnarray}
SU(3)_C\times SU(2)_L\times SU(2)_R\times U(1)_C\times
U(1)_{1,2,3}\times U(1)_{4,5,6}.
\end{eqnarray}
Thus, the $SO(10)$ symmetry, which arises from the fermion group
$\bar{\Psi}^{1,\ldots,5}$ is broken to $SU(3)\times SU(2)_L\times
SU(2)_R\times U(1)_C$, where,
\begin{eqnarray}
U(1)_C={\rm Tr}\, U(3)_C~\Rightarrow~Q_C=
\sum_{i=1}^3Q({\bar\psi}^i). \label{u1c}
\end{eqnarray}
The three symmetries denoted by $U(1)_{j}$ $(j=1,2,3)$ arise from
the world--sheet currents ${\bar\eta}^j{\bar\eta}^{j^*}$. These
three $U(1)$ symmetries are present in all the three generation
free fermionic models which use the NAHE set. Additional $U(1)$
symmetries, denoted by $U(1)_{j}$ $(j=4,5,6)$, arise by pairing
two real fermions from the sets $\{{\bar y}^{3,\cdots,6}\}$,
$\{{\bar y}^{1,2},{\bar\omega}^{5,6}\}$, and
$\{{\bar\omega}^{1,\cdots,4}\}$.

The final observable gauge group depends on the number of such
pairings. {From} both tables \ref{model1} and \ref{tab:NAHEset},
it can be seen that pairs can be formed out of $\bar{y}^{3,6}$,
$\bar{y}^1\bar{\omega}^5$ and $\bar{\omega}^{2,4}$, since they
have identical boundary conditions for all basis vectors.  In this
model these are the pairings which generate the three additional
$U(1)$ symmetries, denoted by $U(1)_{4,5,6}$.

In the hidden sector, which arises from the complex world--sheet
fermions ${\bar\phi}^{1\cdots8}$, the NS boundary conditions
produce the generators of \beq SU(3)_{H_1}\times U(1)_{H_1}\times
U(1)_{7^\prime}\times SU(3)_{H_2}\times U(1)_{H_2}\times
U(1)_{8^\prime}\, . \eeq  Here, the $U(1)$'s are rotated with the
$U(1)_{7',8'}$ combinations of world--sheet charges given later,
and $U(1)_{H_1}$ and $U(1)_{H_2}$ correspond to the combinations
\begin{eqnarray}
Q_{H_1}&=&Q({\bar\phi^1})-Q({\bar\phi^2})-Q({\bar\phi^3})+Q({\bar\phi^4})-
\sum_{i=5}^7Q({\bar\phi}^i)+Q({\bar\phi})^8,\label{qh1}\\
Q_{H_2}&=&\sum_{i=1}^4Q({\bar\phi}^i) -Q({\bar\phi^5})+
\sum_{i=6}^8Q({\bar\phi}^i). \label{qh2}
\end{eqnarray}

The sector $\mzeta\equiv1+\mb_1+\mb_2+\mb_3$ produces the
representations $(3,1)_{-2,0}\oplus({\bar3},1)_{2,0}$ and
$(1,3)_{0,2}\oplus(1,{\bar3})_{0,-2}$ of $SU(3)_{H_1}\times
U(1)_{H_1}$ and $SU(3)_{H_2}\times U(1)_{H_2}$. Thus, the $E_8$
symmetry reduces to $SU(4)_{H_1}\times SU(4)_{H_2}\times U(1)^2$.
The additional $U(1)$'s in $SU(4)_{H_{1,2}}$ are given by the
combinations in equations (\ref{qh1}) and (\ref{qh2}),
respectively. The remaining $U(1)$ symmetries in the hidden
sector, $U(1)_{7^\prime}$ and $U(1)_{8^\prime}$, correspond to the
combinations of world--sheet charges
\begin{eqnarray}
Q_{7^\prime}&=&Q({\bar\phi^1})-Q({\bar\phi^8}),\label{q7prime}\\
Q_{8^\prime}&=&Q({\bar\phi^1})-\sum_{i=2}^4Q({\bar\phi}^i) +
\sum_{i=5}^7Q({\bar\phi}^i)+Q({\bar\phi^8}). \label{q8prime}
\end{eqnarray}

In addition to the NS and $\mzeta$ sector the string model
contains a combination of non--NAHE basis vectors with $\mX_L\cdot
\mX_L=0$, which therefore may give rise to additional space--time
vector bosons. The vector combination is given by
$\mX\equiv\mzeta+2\mgamma$, where
$\mzeta\equiv1+\mb_1+\mb_2+\mb_3$. This combination arises only
from the NAHE set basis vectors plus $2\mgamma$, with $\gamma$
inducing the left--right symmetry breaking pattern $SO(6)\times
SO(4)\rightarrow SU(3)\times U(1)\times SU(2)_L\times SU(2)_R$.
This vector combination is therefore generic for the pattern of
symmetry breaking $SO(10)\rightarrow SU(3)_C\times U(1)_C\times
SU(2)_L\times SU(2)_R$, in NAHE based models.

The sector $\mX$ gives rise to six additional space--time vector
bosons which are charged with respect to the world--sheet $U(1)$
currents, and transform as $3\oplus{\bar 3}$ under $SU(3)_C$.
These additional gauge bosons enhance the $SU(3)_C\times
U(1)_{C^\prime}$ symmetry to  $SU(4)_C$, where $U(1)_{C^\prime}$
is given by the combination of world--sheet charges,
\begin{eqnarray}
Q_{C^\prime}=Q({\bar\psi^1})-Q({\bar\psi^2})-Q({\bar\psi^3})-
\sum_{i=1}^3Q({\bar\eta}^i)+Q({\bar\phi}^1)-Q({\bar\phi}^8).
\label{u1su4}
\end{eqnarray}
The remaining orthogonal $U(1)$ combinations are
\begin{eqnarray}
Q_{1^\prime}&=&Q(\bar{\eta}^1)-Q(\bar{\eta}^2),\nn\\
Q_{2^\prime}&=&Q(\bar{\eta}^1)+Q(\bar{\eta}^2)-2Q(\bar{\eta}^3),\nn\\
Q_{3^\prime}&=&3Q_C-(Q(\bar{\eta}^1)+Q(\bar{\eta}^2)+Q(\bar{\eta}^3)),\nn\\
Q_{7^{\prime\prime}}&=&Q_C+3(Q(\bar{\eta}^1)+Q(\bar{\eta}^2)+Q(\bar{\eta}^3))
+5Q_{7^\prime}.
\label{u1com}
\end{eqnarray}
and $Q_{4,5,6,8^\prime}$ are unchanged. Thus, the full massless
spectrum transforms under the final gauge group, $SU(4)_C\times
SU(2)_L\times SU(2)_R\times
U(1)_{1^\prime,2^\prime,3^\prime}\times U(1)_{4,5,6}\times
SU(4)_{H_1}\times SU(4)_{H_2}\times
U(1)_{7^{\prime\prime},8^{\prime}}$.

In addition to the graviton, dilaton, antisymmetric sector and
spin--1 gauge bosons, the NS sector gives three pairs of $SO(10)$
singlets with $U(1)_{1,2,3}$ charges; and three singlets of the
entire gauge group.

The states from the sectors $\mb_j\oplus \mb_j+\mX~(j=1,2,3)$
produce the three light generations. The states from these sectors
and their decomposition under the entire gauge group are shown in
Table 1 at the end of the Appendicies\footnote{The reader should
note that all charges in this table have been scaled up by a
factor of four.}. The leptons (and quarks) are singlets of the
color $SU(4)_{H_1,H_2}$ gauge groups and the $U(1)_{8^{\prime}}$
symmetry of equation (\ref{q8prime}) becomes a gauged leptophobic
symmetry.

The remaining massless states in the model and their quantum
numbers are also given in Table 1.

I next turn to the definition of the weak--hypercharge in this LRS
model. Due to the enhanced symmetry there are several
possibilities to define a weak--hypercharge combination which is
still family universal and reproduces the correct charge
assignment for the Standard Model fermions. One option is to
define the weak--hypercharge with the standard $SO(10)$ embedding,
as in equation (\ref{U1Y}),
\begin{eqnarray}\label{U1Y}
U(1)_Y={\frac{1}{3}}\,U(1)_C+{\frac{1}{2}}\,U(1)_L.
\end{eqnarray}
This is identical to the weak--hypercharge definition in
$SU(3)\times SU(2)\times U(1)_Y$  free fermionic models, which do
not have enhanced symmetries, as for example in Model 3 of ref.\
\cite{lrsmodels}. Alternatively, one can define the
weak--hypercharge to be the combination
\begin{eqnarray}\label{U1Y2}
U(1)_Y={\frac{1}{{10}}}\left(U(1)_{3^\prime}+
{\frac{1}{3}}U_{7^{\prime\prime}}\right)+{\frac{1}{2}}U(1)_L
\end{eqnarray}
where $U(1)_{3^\prime}$ and $U(1)_{7^{\prime\prime}}$ are given in
(\ref{u1com}). This combination still reproduces the correct
charge assignment for the Standard Model states. The reason being
that the states from the sectors $\mb_i$ $i=1,2,3$ which are
identified with the Standard Model states, are not charged with
respect to the additional Cartan sub--generators that form the
modified weak hypercharge definition. In some models it is found
that such alternative definitions allow all massless exotic states
to be integrally charged.

However, the study of this model is motivated by the possible
existence of $D$--flat directions that use non--Abelian VEVs.  It
is not intended to be a semi--realistic candidate. It is therefore
concluded that the model admits a sensible weak--hypercharge
definition, which for the purpose of this work is sufficient.

\section{Anomalous $U(1)$}\label{anomalousu1}

The string model contains an anomalous $U(1)$ symmetry. The
anomalous $U(1)$ is a combination of $U(1)_4$, $U(1)_5$ and
$U(1)_6$, which are generated by the world--sheet complex fermions
${\bar y}^3{\bar y}^6$, ${\bar y}^1{\bar\omega}^5$ and
${\bar\omega}^2{\bar\omega}^4$, respectively.  So it is seen that $U_{4,5,6}$ arise from the internal ``compactified'' degrees of freedom.

It is convenient to rotate $U(1)_4$, $U(1)_5$ and $U(1)_6$ so that
only one these symmetries is anomalous. The anomalous $U(1)_A$
combination in this model is given by
\beq U_A\equiv U_4+U_5+U_6, \label{anomau1infny}
\eeq
with ${\rm Tr}Q_A=-72$.  The two orthogonal linear combinations,
\beqn
U_{4^\prime} &=& U_4-U_5 \\
U_{5^\prime} &=& U_4+U_5-2 U_6 \nolabel
\eeqn are then both traceless.

Since ${\rm Tr}Q_A<0$, the sign for the Fayet--Iliopoulos term
($\xi$) is negative, as can be seen from equation (\ref{dxi}).
Requiring $D$--flatness then implies that there must exist a field
(or a combination of fields) with positive total $U(1)_A$ charge
that receive a VEV (or VEVs) to cancel the $U(1)_A$ $D$--term. The
equation that enforces this for $D$--flatness is shown in
(\ref{da}).

Looking at the massless spectrum of the model, given in Table 1 of
the appendix (\emph{Model 1 Fields}), it is immediately understood
that the model does not contain any non--Abelian singlet fields
with such a positive charge. Therefore, if $D$--flatness exists,
some non--Abelian fields must acquire a VEV.  The spectrum shows
that the only states that carry positive $U(1)_A$ charge are the
$SU(2)_L$ and $SU(2)_R$ ($\mathcal{L}_{L_k,R_k}$) doublets from
the three sectors $\mb_k + \mzeta + 2 \mgamma \equiv
\bo+\mb_i+\mb_j+ 2 \mgamma$, $(i,j,k=1,2,3)$ with $i$, $j$, $k$
all distinct.  These states are shown in Table 1 in the sectors
$\mathbf{b}_k \oplus \mathbf{b}_k+\zeta+2\gamma$. The same result
holds also in Model 3 of ref.\ \cite{lrsmodels}, in which there is
no colour gauge enhancement from the sector $\mzeta+2\mgamma$.

I next turn to discuss the possibility of $D$--flatness in this
model.

\section{Solutions of the $D$--Flat Constraint Equations}\label{sec:SDFC}

Table 1 in the appendices lists all of the massless states that
appear in this LRS string model. There are a total of 68 fields,
38 of which may be used to form 19 sets of vector--like pairs of
fields (which have opposite $U(1)$ charges). Of these 19
vector--like pairs, 13 pairs are singlets under all non--Abelian
gauge groups, while three pairs are $\bar{\mathbf 4}$/$\mathbf 4$
($\bar{D_i}/D_i$, $i=1,2,3$) sets under $SU(4)_C$ and two pairs
are $\mathbf 6$/$\mathbf 6$'s ($\bar{\cal{H}}_i$/${\cal H}_i$,
$i=1,2$) sets under $SU(4)_{H_2}$. The 30 non--vector--like fields
are all non--Abelian reps. That is, all singlets occur in
vector--like pairs.

The anomalous charge trace of $U(1)_A$ is negative for this model.
Thus, the anomaly can only be cancelled by fields with positive
anomalous charge. In this model, as can be seen from Table 1, {\it
none} of the non--Abelian singlets carry anomalous charge $\QA$.
The only fields with positive anomalous charge are three $SU(2)_L$
doublets, $\cL{L1,L2,L3}$ and three $SU(2)_R$ doublets,
$\cL{R1,R2,R3}$.

To systematically study $D$--flatness for this model, first one
must generate a complete basis of directions $D$--flat for all
non--anomalous Abelian symmetries. These basis directions are
provided in Table 2 at the end of the appendices.

The process begins from a set of simultaneous equations of the
form
\begin{eqnarray}
\mathbf{D}.\mathbf{x}=\mathbf{b}
\end{eqnarray}
where the matrix $\mathbf{D}$ has components $D_{ij}=Q^{(i)}_j$
which are the charges of the state $\phi_j$ with the norm squared
of its VEV as the components of the vector $\mathbf{x}$,
$x_j=|\vev{\phi_j}|^2$.  The vector $\mathbf{b}$ has zero entries
except for the row $i=A$ which corresponds to the net anomalous
charge, and has value $-\xi$.  This is consistent with the
equations (\ref{da}) and (\ref{dalpha}).

{From} here, one uses the reduced matrix $\mathbf{D'}$ which is
simply the original matrix with the $i=A$ row removed.  One then
applies the SVD method\footnote{This method is based on the
``Singular Value Decomposition'' method, which was demonstrated
for the systematic resolution of $D$--flat constraint equations by
G. Cleaver.  The method itself is reviewed in appendix
\ref{app:SVD}.} to systematically provide a set of basis
directions. These are then rotated to give the basis directions in
the form
\begin{eqnarray}\label{1}
\mathbf{x}_{1}=\left(\begin{array}{c}|<{\cal L}_{R1}>|^2 \\ 0 \\ 0 \\
\vdots \\
\text{common fields} \\ 0 \\ \vdots \\
\end{array}\right),\quad
\mathbf{x}_{2}=\left(\begin{array}{c} 0 \\ |<{\cal L}_{L1}>|^2 \\ 0 \\
\vdots \\
\text{common fields} \\ 0 \\ \vdots \\
\end{array}\right),\ldots
\end{eqnarray}
so that each contains at least one unique field VEV and a group of
field VEVs that are common to all directions.

For a given row in Table 2, the first column entry denotes the
name of the $D$--flat basis direction ($\mathbf{x}_i$). The next
row specifies the anomalous charge of the basis direction. The
following seven entries specify the ratios of the norms of the
VEVs of the fields common to these directions. The first five of
these fields have vector--like partners. For these, a negative
norm indicates the vector--partner acquires the VEV, rather than
the field specified at the top of the respective column. The last
two of these seven fields are not vector--like. Thus, the norm
must be non--negative for each of these for a flat direction
formed from a linear combination of basis directions to be
physical. The next to last entry specifies the norm of the VEV of
the field unique to a given basis direction, while the identity of
the unique field is given by the last entry.

These $D$--flat directions have been labelled as $\mathbf{x}_1$
through $\mathbf{x}_{41}$. The first eight $D$--flat directions
($\mathbf{x}_{1}$ to $\mathbf{x}_{8}$) carry a positive net
anomalous charge. The next fourteen ($\mathbf{x}_{9}$ to
$\mathbf{x}_{22}$) carry a negative net anomalous charge, while
the remaining nineteen ($\mathbf{x}_{23}$ to $\mathbf{x}_{41}$)
lack a net anomalous charge. There are two classes of basis
vectors lacking anomalous charge. The first class contains six
directions for which the unique field is non--vector--like. These
directions also contain VEVs for $\cH{1}$ $(\cHb{1})$ and/or
$\cH{2}$ $(\cHb{2})$. The second class contains thirteen basis
directions wherein the unique fields are vector--like and which do
not contain $\bH{2'}$ and/or $\bH{4'}$. Thus, these thirteen
directions are themselves vector--like and are denoted as such by
a superscript ``$v$''. For each vector--like basis direction,
$\mathbf{x}^v$ there is a corresponding $-\mathbf{x}^v$ direction,
wherein the fields in $\mathbf{x}^v$ are replaced by their
respective vector--like partners.

None of the positive $\QA$ directions are good in themselves
because one or both of $|\vev{\bH{2'}}|^2$ and $|\vev{\bH{4'}}|^2$
are non--zero and negative while $|\vev{\bH{2'}}|^2$ and
$|\vev{\bH{4'}}|^2$ are not vector--like representations. In
particular, the $\QA= 12$ directions have either
    $|\vev{\bH{2'}}|^2= -2$ and
    $|\vev{\bH{4'}}|^2=  0$ or
    $|\vev{\bH{2'}}|^2=  0$ and
    $|\vev{\bH{4'}}|^2= -2$,
    while the $\QA= 24$ directions all have
$|\vev{\bH{2'}}|^2= |\vev{\bH{4'}}|^2= -4$. In this basis one also
finds that the $|\vev{\bH{2'}}|^2$ and $|\vev{\bH{4'}}|^2$ charges
of all of the $\QA=0$ directions are zero or negative. So the
$|\vev{\bH{2'}}|^2$ and $|\vev{\bH{4'}}|^2$ negative charges on
the positive $\QA$ directions cannot be made zero or positive by
adding $\QA=0$ directions to $\QA >0$ directions. In contrast, all
of the $\QA= -12$ directions have either
    $|\vev{\bH{2'}}|^2= 2$ and
    $|\vev{\bH{4'}}|^2= 0$ or
    $|\vev{\bH{2'}}|^2= 0$ and
    $|\vev{\bH{4'}}|^2= 2$;
    the $\QA= -24$ directions all have either
    $|\vev{\bH{2'}}|^2= |\vev{\bH{4'}}|^2= 4$ or
    $|\vev{\bH{2'}}|^2= |\vev{\bH{4'}}|^2= 2$; while the
$\QA= -48$ directions all have $|\vev{\bH{2'}}|^2=
|\vev{\bH{4'}}|^2= 4$ . Therefore, physical $D$--flat directions
must necessarily be formed from linear combinations of $\QA> 0$
and $\QA<0$ directions such that the net $\QA$,
$|\vev{\bH{2'}}|^2$, and $|\vev{\bH{4'}}|^2$ are all positive.
Physical $D$--flat directions may also contain $\QA= 0$ components
that keep $|\vev{\bH{2'}}|^2$, $|\vev{\bH{4'}}|^2 \ge 0$.

The specific values of $\QA$, $|\vev{\bH{2'}}|^2$, and
$|\vev{\bH{4'}}|^2$ in the basis directions indicate that the
roots of all physical flat directions must contain either
$\mathbf{x}_{19}$ or $\mathbf{x}_{20}$ and combinations of basis
vectors $\mathbf{x}_{1}$, $\mathbf{x}_{2}$, $\mathbf{x}_{3}$, and
$\mathbf{x}_{4}$ of the form
\begin{eqnarray}\label{sol1a}
n_1\mathbf{x}_{1}+n_2\mathbf{x}_{2}+n_3\mathbf{x}_{3}+
n_4\mathbf{x}_{4}+n_{19}\mathbf{x}_{19}+n_{20}\mathbf{x}_{20}.
\end{eqnarray}
The non--negative integers $n_1$, $n_2,$ $n_3$, $n_4$, $n_{19}$,
$n_{20}$ satisfy the constraints
\begin{eqnarray}
n_1+n_2+n_3+n_4-2n_{19}-2n_{20} &>& 0, \label{sol1b1}\\
-n_1-n_2+2n_{19}+2n_{20} &\ge& 0, \label{sol1b2}\\
-n_3-n_4+2n_{19}+2n_{20} &\ge& 0. \label{sol1b3}
\end{eqnarray}
which provide the positivity requirements, in agreement with the
Fayet--Iliopoulos term $\xi$, and the positive semi--definiteness
of the norm squared of the field VEVs for $\bar{H}_{4'}$ and
$\bar{H}_{2'}$ so that they represent fields with physical VEVs.

For example, one of the simplest $D$--flat solutions for all
Abelian gauge groups is $n_1= n_2=n_3=n_4=2$, $n_{19}= n_{20}= 1$.
This direction is simply $|\vev{\cL{L1}}|^2 = |\vev{\cL{L2}}|^2 =
|\vev{\cL{L3}}|^2 =|\vev{\cL{R1}}|^2 = |\vev{\cL{R2}}|^2 =
|\vev{\cL{R3}}|^2 $. The corresponding fields are three exotic
$SU(2)_L$ doublets, $\cL{L1}$, $\cL{L2}$, and $\cL{L3}$, and three
exotic $SU(2)_R$ doublets, $\cL{R1}$, $\cL{R2}$, and $\cL{R3}$.
These six fields are singlets under all other non--Abelian groups.

For this model any $D$--flat direction must contain $SU(2)_L$ or
$SU(2)_R$ fields. Thus, let us examine more closely $SU(2)$
$D$--flat constraints. The only $SU(2)$ fields in this model are
doublet representations, which are generically denoted $L_i$.
Thus, the related three $SU(2)$ $D$--terms,
\begin{eqnarray}\label{dtgen}
D_{a=1,2,3}^{SU(2)}&\equiv& \sum_m L_i^{\dagger}
T^{SU(2)}_{a=1,2,3} L_i\,\, ,
\end{eqnarray}
contain matrix generators $T^{SU(2)}_a$ that take on the values of
the three Pauli matrices,
\begin{eqnarray}
\sigma_x=\left(\begin{array}{cc}
0 & 1 \\
1 & 0 \\
\end{array} \right ), \,\,
\sigma_y = \left (
\begin{array}{cc}
0 & -i \\
i &  0 \\
\end{array} \right ), \,\,
\sigma_z = \left (
\begin{array}{cc}
1 & 0 \\
0 & -1 \\
\end{array} \right ),
\label{pauli}
\end{eqnarray}
respectively.

As discussed in \cite{cfnw}, each component of the vector
$\vec{D}^{SU(2)}$ is the total ``spin expectation value'' in the
given direction of the internal space, summed over all $SU(2)$
doublet fields of the gauge group. Thus, for all of the
$\vev{D^{SU(2)}_a}$ to vanish, the $SU(2)$ VEVs must be chosen
such that the total $\hat{x}, \hat{y},$ and $\hat{z}$ expectation
values are zero.

I take the explicit representation of a generic $SU(2)$ doublet
\cite{cfnw} $L(\theta, \phi)$ as
\begin{eqnarray} L(\theta, \phi) \equiv A
\left (
\begin{array}{c}
\cos{\frac{\theta}{2}} \, e^{-i \frac{\phi}{2}} \\
\sin{\frac{\theta}{2}} \, e^{+i \frac{\phi}{2}} \\
\end{array} \right )\, ,
\label{spinor}
\end{eqnarray}
where $A$ is the overall amplitude of the VEV. The range of
physical angles, $\theta = 0 \rightarrow \pi$ and $\phi = 0
\rightarrow 2\pi$ provide for the most general possible doublet

The contribution of  $L(\theta, \phi)$ to each $SU(2)$ $D$--term
is,
\begin{eqnarray}
D_{1}^{SU(2)}(L) &\equiv&  L^{\dagger}\,
\left (
\begin{array}{cc}
0 & 1 \\
1 & 0 \\
\end{array} \right )\,
L =  |A|^2 \sin\, \theta \,\, \cos\, \phi \label{dl1}\\
D_{2}^{SU(2)}(L) &\equiv&  L^{\dagger}\,  \left (
\begin{array}{cc}
0 & - i \\
i &  0 \\
\end{array} \right )\,
L =  |A|^2 \sin\, \theta \,\, \sin\, \phi \label{dl2}\\
D_{3}^{SU(2)}(L) &\equiv&  L^{\dagger}\,  \left (
\begin{array}{cc}
1 &   0 \\
0 &  -1 \\
\end{array} \right )\,  L =  |A|^2 \cos\, \theta\,\, . \label{dl3}
\end{eqnarray}
{From} this one can see that the VEVs of three $SU(2)$ doublets
$\cL{i=1,2,3}$ with equal norms $|A_1|^2=|A_2|^2=|A_3|^2\equiv
|A|^2$ can, indeed, produce an $SU(2)$ $D$--flat direction with
the choice of angles, $\theta_1 = 0$, $\theta_2 = \theta_3 =
2\pi/3$, $\phi_2 = 0$, $\phi_3 = \pi$. For these angles, the total
$\hat x$, $\hat y$, and $\hat z$ expectation values are all zero.

This flat direction gives a specific example of what will occur
for every flat direction of this model: non--Abelian VEVs (for at
least $SU(2)_L$ or $SU(2)_R$ doublets) are a necessary if one
requires $D$--flatness, or a solution to the equations of motion
(\ref{da}) and (\ref{dalpha})).

I now consider the status of $F$--flatness at the FI scale.  By
$U(1)$ gauge invariance (by refering to table 1), the direction
$|\vev{\cL{L1}}|^2 = |\vev{\cL{L2}}|^2 =|\vev{\cL{L3}}|^2
=|\vev{\cL{R1}}|^2 = |\vev{\cL{R2}}|^2 = |\vev{\cL{R3}}|^2 $ was
found to comply with $F$--flatness. This is understood since one
of the constraints to forming superpotential terms is that each
should have vanishing total $U(1)$ charge. {From} this,
$F$-flatness remains the case to all finite order in the
superpotential. While this is true for the simplest flat direction
as is used in this example, it is expected that dangerous
$F$--terms will arise for all but a few of the more complicated
directions.

\section{Discussion}

This is a toy model with specific choices of GSO phases and
boundary condition vectors (consistent with modular invariance)
that give rise to a spectrum which supports $D$-- and $F$--
flatness for the use of non--Abelian terms.  Of course, this is
certainly the case for the simple direction where the only fields
that take on VEVs are the exotic $SU(2)_{L,R}$ doublets ${\cal
L}_{L_i,R_i}$, for $i=1,2,3$ where $|\vev{\cL{L1}}|^2 =
|\vev{\cL{L2}}|^2 =|\vev{\cL{L3}}|^2 =|\vev{\cL{R1}}|^2 =
|\vev{\cL{R2}}|^2 = |\vev{\cL{R3}}|^2 $.  As was mentioned before,
this expected not to be the case for most of the other directions
that are solutions of equations (\ref{sol1b1}), (\ref{sol1b2}) and
(\ref{sol1b3}).

As the reader will appreciate, the number of massless sectors that
derive from the NAHE plus $\{\alpha,\beta,\gamma\}$ sectors is
large.  Generally there is no mechanism, other than computing and
examining a model, for determining the appropriate vectors and GSO
phases to obtain the precisely desired phenomenology.  However, it
is through the collective study of different models that one might
begin to understand a mechanism behind the freedom to choose the
vectors and GSO phases.

This work is not intended to elaborate on such a mechanism.
Presented here is simply a four dimensional scenario whereby the
choices (\ref{tab:NAHEset}), (\ref{model1}) and
(\ref{phasesmodel1}) lead to new and interesting spectral, and
most importantly, $D$--flatness aspects.

It was pointed out before that this model could not be admitted as a
possible realistic candidate.  The non-Abelian fields $\mathcal{L}$,
that were used here for $D$--flatness, carry fractional electric
charge. In addition to this, all Higgs doublets from the
Neveu-Schwarz sector are projected out by the choice of GSO phases.
This makes it problematic to obtain realistic fermion mass spectrum.

This was the first instance in the study of the realistic free
fermionic models in which non--Abelian VEVs are enforced by the
requirement of $D$--flatness rather than by other possible
phenomenological considerations.

\setcounter{footnote}{0}

\chapter{Open Descendants of Type I $T^6/(\mathbb{Z}_2\times\mathbb{Z}_2
^\text{s})$ and
$T^6/(\mathbb{Z}_2\times\mathbb{Z}_2\times\mathbb{Z}_2^\text{s})$
models}\label{ch:OpenDescendents}

\setcounter{footnote}{0}

In this and the next chapter, I will discuss phenomenological
aspects of type I models in four spacetime dimensions whereby the
internal coordinates are compactified on orbifolds \cite{cfo}.
Unlike toroidal compactification, orbifolds are the
compactification of string coordinates on singular manifolds.  The
presence of orbifold identifications of these coordinates will
lead to a much richer spectrum than compactification on an
equivalent dimension torus.

The first chapter will involve a discussion of the open
descendants and their spectral content with compactification on
$T^6/(\mathbb{Z}_2\times \mathbb{Z}_2^\text{s})$ and
$T^6/(\mathbb{Z}_2\times\mathbb{Z}_2\times \mathbb{Z}_2^\text{s})$
orbifolds.  The additional $\mathbb{Z}_2^\text{s}$ group is a
$\mathbb{Z}_2$ shift, which in both cases will be freely acting in
that it will not generate any additional fixed points with respect
to the orbifold fixed points.

In the $T^6/(\mathbb{Z}_2\times \mathbb{Z}_2\times
\mathbb{Z}_2^\text{s})$ case, the freely acting shift will allow
for the inclusion of twisted sectors that appear in the
$T^6/(\mathbb{Z}_2\times \mathbb{Z}_2)$ case, this is illustrated
in appendix \ref{app:Boundaries}.

The construction of these models uses the Hamiltonian formulation,
and details of the language used will now follow \cite{CAAS}.


\section{Type I Construction}\label{sec:TypeIConstruction}


The discussions that will follow on type I phenomenology involve
amplitudes that have zero Euler character. The Euler character
$\chi=2-2h-b-c$ defines an invariant number of a surface with $h$
handles, $b$ boundaries and $c$ cross caps. The parent amplitude
is the torus ($h=1$, $b=0$, $c=0$).  The remaining surfaces are
the Klein bottle ($h=0$, $b=0$, $c=2$), annulus ($h=0$, $b=2$,
$c=0$) and Mobius strip ($h=0$, $b=1$, $c=1$). The torus is formed
by first taking the complex coordinates
\begin{eqnarray}\label{eqn:TorusIdent}
z=\sigma_1+\tau \sigma_2,\quad \bar{z}=\sigma_1+\bar{\tau}
\sigma_2
\end{eqnarray}
for the complex parameter $\tau=\tau_1+i\tau_2$, which defines the
shape of the torus, and the two coordinates $\sigma_1$ and
$\sigma_2$. The relation
\begin{eqnarray}\label{eqn:TorusCoords}
z\sim z+m+n\tau
\end{eqnarray}
periodically identifies the coordinates of each direction
$\sigma_1$ and $\sigma_2$ so that the shape of the torus can be
controlled by the parameter $\tau$.  However, counting over these
surfaces involves summing over some equivalent tori. These
equivalent surfaces are identified by the transformation
\begin{eqnarray}\label{eqn:Transform}
\tau \rightarrow \frac{a\tau+b}{c\tau+d}
\end{eqnarray}
which defines the modular group
\begin{eqnarray}
{\rm SL}(2,\mathbb{Z})/\mathbb{Z}_2={\rm PSL}(2,\mathbb{Z})
\end{eqnarray}
for integers $a,b,c,d$ and $ad-bc=1$.  The modulating group
$\mathbb{Z}_2$ ensures that tori which are equivalent under the
transformation $\{a,b,c,d\}\rightarrow \{-a,-b,-c,-d\}$ are not
doubly counted, since this would leave the transformation
(\ref{eqn:Transform}) unchanged.

The transformation (\ref{eqn:Transform}) is generated by $S$ and
$T$ operations which act on $\tau$ as
\begin{eqnarray}\label{eqn:ActionOfS&T}
&&T: \tau \rightarrow \tau+1\nn\\
&&S: \tau \rightarrow -\frac{1}{\tau}.
\end{eqnarray}

The fundamental region of integration for the parameters $\tau_1$
and $\tau_2$, which defines the shape of the torus sheet, is given
by
\begin{eqnarray}
{\cal F}=\{-\frac{1}{2}<\tau_1\leq \frac{1}{2},|\tau|\geq1\}.
\end{eqnarray}

The amplitudes that are to be discussed are represented in terms
of Jacobi theta functions and the Dedekind eta function.  I will
review their structure and transformation properties here.

The Dedekind eta function is defined as
\begin{eqnarray}\label{eqn:Eta}
\eta(\tau)=q^{\frac{1}{24}}\prod_{n=1}^{\infty}\big(1-q^n\big)
\end{eqnarray}
for
\begin{eqnarray}
q=e^{2\pi i \tau}.
\end{eqnarray}
It transforms under $S$ and $T$ as
\begin{eqnarray}
&&\eta(\tau+1)=e^{i\frac{\pi}{12}}\eta(\tau), \nn\\
&&\eta(-\frac{1}{\tau})=\sqrt{-i\tau}\eta(\tau)
\end{eqnarray}
respectively, which shows the function
\begin{eqnarray}\label{eqn:EtaFunction}
\frac{1}{\sqrt{\tau_2}|\eta(\tau)|^2}
\end{eqnarray}
has a modular invariant form.

The Jacobi theta functions are defined by the infinite product
\begin{eqnarray}\label{eqn:ThetaProExp}
\theta(z|\tau)\left[ \begin{array}{c} \gamma \\
\vartheta \\
\end{array}\right]=e^{2i\pi
\gamma(z+\vartheta)}q^{\frac{\gamma^2}{2}}\prod_{n=1}^\infty(1-q^n)
(1+q^{n+\gamma-\frac{1}{2}}e^{2i\pi(z+\vartheta)})
(1+q^{n-\gamma-\frac{1}{2}}e^{-2i\pi(z+\vartheta)}).\nn\\
\end{eqnarray}
which has an equivalent expression (which is more convenient to
identify particular massive modes in the spectrum)
\begin{eqnarray}
\theta \left[ \begin{array}{c}\gamma \\ \vartheta
\end{array}
\right](z|\tau)=\sum_{n}q^{\frac{1}{2}(n+\gamma)^2}e^{2\pi i
(n+\gamma)(z+\varphi)}
\end{eqnarray}
The elements $\gamma$ and $\varphi$ describe the periodic boundary
conditions $\gamma,\varphi=0$ or anti periodic with
$\gamma,\varphi=\frac{1}{2}$.

The periodicity values of 0 and $\frac{1}{2}$ define NS
(Neveu--Schwarz) and R (Ramond) boundary conditions for the upper
$\gamma$ characteristic.

The four separate theta functions arising from all possible
boundary conditions are defined as
\begin{eqnarray}\label{eqn:Theta's}
&&\theta_1(z|\tau) = \left[ \begin{array}{c} \frac{1}{2} \\
\frac{1}{2} \end{array} \right](z|\tau), \quad \theta_2(z|\tau) =
\left[ \begin{array}{c} \frac{1}{2} \\ 0 \end{array} \right](z|\tau),\nn\\
&&\theta_3(z|\tau) = \left[ \begin{array}{c} 0 \\ 0 \end{array}
\right](z|\tau),\quad \theta_4(z|\tau) = \left[
\begin{array}{c} 0 \\
\frac{1}{2} \end{array} \right](z|\tau).
\end{eqnarray}
However, it can easily be seen by (\ref{eqn:ThetaProExp}), that
$\theta_1$ vanishes identically for argument $z=0$.  The parameter
$z\neq 0$ will play a crucial role in the type I models with
magnetic deformations, in other cases that are studied it has a
value $z=0$.

The behavior of (\ref{eqn:ThetaProExp}) under $S$ and $T$
transformations is shown by
\begin{eqnarray}\label{eqn:STtransform}
\theta \left[ \begin{array}{c} \gamma \\ \vartheta
\end{array} \right]\bigg(\frac{z}{\tau}\bigg|-\frac{1}{\tau}\bigg)=
(-i\tau)^\frac{1}{2}e^{2\pi i
\gamma\varphi+i\pi\frac{z^2}{\tau}}\theta \left[
\begin{array}{c} \vartheta \\ -\gamma \end{array} \right](z|\tau)
\end{eqnarray}
and
\begin{eqnarray}
\theta \left[ \begin{array}{c} \gamma \\ \vartheta
\end{array} \right](z|\tau+1)= e^{-\pi i \gamma(\gamma-1)}\theta
\left[ \begin{array}{c} \gamma \\ \vartheta+\gamma-\frac{1}{2}
\end{array} \right](z|\tau)
\end{eqnarray}
respectively.

Now that the preliminary elements have been discussed, I turn to
the construction techniques of type I model building.

The technology used to describe the open and closed sectors of the
type I models \cite{CAAS} discussed in this chapter are based on
the partition function in the Hamiltonian formulation. The form of
the generator of translation, or Laurent zero mode, will depend on
the 10-dimensional superstring action
\begin{eqnarray}\label{eqn:Action}
S=-\frac{1}{4\pi \alpha'}\int d^2\zeta(\partial^\alpha X^\mu
\partial_\alpha X^\nu \eta_{\mu \nu}+i\bar{\psi}^\mu \gamma^\alpha
\partial_\alpha \psi_\mu).
\end{eqnarray}
The action contains two dimensional Majorana spinors and
corresponding bosonic superpartners $X^\mu$.

One obtains a description of closed string spectra in terms of
four sectors, the first two of which are the left-right pairings
of antiperiodic ($NS$) and periodic ($R$) fields $NS$--$NS$ and
$R$--$R$ respectively. These account for fields that behave as
spacetime bosons.  The two remaining pairings are $R$--$NS$ and
$NS$--$R$, and these describe spacetime fermions.

In the case of the open spectrum, the fields are the left moving
fields of the closed string.  In this case, the $R$ fields
describe spacetime fermions and those of the bosons correspond to
$NS$ fields.

The zero Laurent mode has the form
\begin{eqnarray}\label{eqn:LaurantModes}
L_o=\frac{1}{2}:\sum_l
\alpha_{-l}^i\alpha_{l,i}:+\frac{1}{2}:\sum_w
w\phi_{-w}^i\phi_{w,i}:+\Delta
\end{eqnarray}
Where the string spectrum is built by acting on the vacuum using
the creation operators from $X^\mu$ as $\alpha^\mu_{-l}$ and
$\Psi^\mu$ as $\phi^\mu_{-w}$. The world sheet fermion modes
($\phi^\mu_w$) are summed over the index $w$ which is integer for
the $R$ sector and half--odd integer for the $NS$.

The quantity $\Delta$ is a normal ordering constant, and acquires
contributions of $-\frac{1}{48}$ from the $NS$ sector and
$\frac{1}{24}$ from the $R$ of the fermionic coordinates.  The
bosonic coordinates provide $-\frac{1}{24}$. As such, the total in
$D$ spacetime dimensions is given by
\begin{eqnarray}\label{eqn:Delta}
\Delta=-\frac{1}{16}(D-2)
\end{eqnarray}
for the $NS$ sector and 0 for the $R$ sector.

The type I theory is the orientifold projection (or world sheet
parity projection) of the type IIB open and closed theory under
$\Omega$. The action of $\Omega$ on the world sheet modes is
defined as
\begin{eqnarray}\label{eqn:OmegaOnClosed}
\Omega : \alpha^\mu_n \leftrightarrow \tilde{\alpha}^\mu_n
\end{eqnarray}
for the left and right modes of the closed string and
\begin{eqnarray}
\Omega : \alpha^\mu_n \rightarrow (-1)^n\alpha^\mu_n
\end{eqnarray}
for the open string.

With the insertion of the world sheet parity projection in the
partition function as
\begin{eqnarray}\label{eqn:PF}
{\rm Tr}\frac{(1+\Omega)}{2}q^{L_o}\bar{q}^{\bar{L}_o},
\end{eqnarray}
the closed sector comprises the torus amplitude and the Klein
bottle. The torus being defined by the insertion of 1 and the
Klein by $\Omega$.

It will be seen that as a consequence of the orbifold operation,
that the orientifold planes that exist in the Klein are $O9$ and
$O5$ planes. Additionally, the action of $\Omega$ on the left and
rights states is simply seen to reduce the counting of states
according to
\begin{eqnarray}
\sum_{L,R}<L,R|\Omega q^{L_o-1}\bar{q}^{\bar{L}_o-1}|L,R>
=\sum_{R}<R,R|(\bar{q}q)^{L_o-1}|R,R>.
\end{eqnarray}
This identification will also have a non-trivial effect on the
lattices that involve quantum numbers in the form of Kaluza Klein
and winding states.  Closed string momentum is defined by
\begin{eqnarray}
&&p_{L}=\frac{m}{R}+\frac{nR}{\alpha^\prime},\nn\\
&&p_{R}=\frac{m}{R}-\frac{nR}{\alpha^\prime}.
\end{eqnarray}
So the left right identification of a state $|p_L,p_R>$
necessarily confines the counting to either pure winding of Kaluza
Klein towers.

The different topologies that result from the $\Omega$ projection
define different forms for the Tecichm\"{u}ller parameter
$\tau=\tau_1+i\tau_2$.

The open sector has contributions from the annulus and Mobius
amplitudes.  The modulus for the Mobius amplitude is not purely
imaginary, it transforms according to
\begin{eqnarray}\label{eqn:Ptransform}
P: \frac{1}{2}+\frac{i\tau_2}{2}\rightarrow
\frac{1}{2}+\frac{i}{2\tau_2}
\end{eqnarray}
It is straight forward to see that this can be written in terms of
the conventional $S$ and $T$ operators as
\begin{eqnarray}\label{eqn:Ptransform1}
P=TST^2S.
\end{eqnarray}
The action of $S$ and $T$ are defined by their actions on the
Teichm\"{u}ller parameter in (\ref{eqn:ActionOfS&T}).

Under the $P$ transform, the $\eta$ function behaves as
\begin{eqnarray}\label{eqn:Ptranseta}
P:\hat{\eta}(\frac{it}{2}+\frac{1}{2})\rightarrow
\hat{\eta}(\frac{i}{2t}+\frac{1}{2})
=\sqrt{t}\hat{\eta}(\frac{it}{2}+\frac{1}{2}).
\end{eqnarray}

The $S$ transformation has an interesting effect on the Klein,
annulus and Mobius amplitudes. In the direct channel of the
annulus and Mobius amplitudes one has an interpretation of an open
string coupling to endpoint boundaries. An $S$ transformation
allows a $\frac{\pi}{2}$ coordinate rotation that now defines
world sheet time running from one boundary to the other (as can be
seen by the tadpole diagrams illustrated in appendix
\ref{app:TransverseTadpoleDiagrams}). This is understood as a
closed string propagating in the bulk between the boundaries.

The consideration of diagrams under $S$ transformation will be a
convenient frame in which to extract tadpole cancellation
conditions.

It is constructive to demonstrate the representations of
(\ref{eqn:LaurantModes}) in terms of Jacobi theta functions, in
the presence and absence of orbifold operations.

In addition to the untwisted bosonic contribution in the absence
of an orbifold identification, which has the form
\begin{eqnarray}\label{eqn:EtaFunction1}
{\rm Tr}q^{\sum n\alpha_n \cdot \alpha_{-n}}=\prod_{n=1}^{\infty}
\frac{1}{1-q^n}=\frac{q^{\frac{1}{24}}}{\eta}
\end{eqnarray}
the other possible states are combinations of twisted strings with
or without orbifold insertions, and untwisted strings with the
action of an orbifold.

The effect of twisting and the presence of orbifold operations in
the trace of (\ref{eqn:EtaFunction1}) will provide rich spectral
content.  Twisted states are associated with strings, in the open
sector, with ends terminating on different boundaries such as $D9$
and $D5$ branes. The modes of these strings appear with the mode
shift $n\rightarrow n-\frac{1}{2}$.

The orbifold operation in the trace of the torus allows (by
enforcing modular invariance) a much extended spectrum, this will
be appreciated in models described later.  An orbifold can be
reached by subjecting smooth covering manifolds to discrete
identifications. This action in general leaves points that are
fixed on the original manifold.  For example, an orbifold
projection in a trace can be realized by the identification of a
compact coordinate
\begin{eqnarray}\label{eqn:OrbIdent}
X^\mu\sim -X^\mu
\end{eqnarray}
and so leaving only states that are even under this parity
operation.  This type of operation is a $\mathbb{Z}_2$
identification.  This action results from the effect on the string
modes as $\beta:\alpha^\mu_n\rightarrow -\alpha^\mu_n$, for
$\beta\in \mathbb{Z}_2$.

So to exhaust the remaining bosonic states, one has
\begin{eqnarray}\label{eqn:LatticeModesTheta's}
{\rm Tr}\beta q^{\sum n\alpha_n \cdot
\alpha_{-n}}&=&\prod_{n=1}^{\infty} \frac{1}{1+q^n}=
q^{\frac{1}{24}}\bigg(\frac{2\eta}{\theta_2}\bigg)^{\frac{1}{2}},\nn
\\
{\rm Tr} q^{\sum (n-\frac{1}{2})\alpha_n \cdot
\alpha_{-n}}&=&\prod_{n=1}^{\infty}\frac{1}{1-q^{(n-\frac{1}{2})}}=
q^{-\frac{1}{48}}\bigg(\frac{\eta}{\theta_4}\bigg)^{\frac{1}{2}},\nn \\
{\rm Tr} \beta q^{\sum (n-\frac{1}{2})\alpha_n \cdot
\alpha_{-n}}&=&\prod_{n=1}^{\infty}\frac{1}{1+q^{(n-\frac{1}{2})}}=
q^{-\frac{1}{48}}\bigg(\frac{\eta}{\theta_3}\bigg)^{\frac{1}{2}}.
\end{eqnarray}
The boundary conditions and their particular forms of the Jacobi
theta function can be seen from equations (\ref{eqn:Theta's}) and
(\ref{eqn:ThetaProExp}) respectively.

The effect of the orbifold identification (\ref{eqn:OrbIdent}) on
the torus coordinates (\ref{eqn:TorusCoords}), is to create the
four fixed points
\begin{eqnarray}\label{fixedtau}
z_i=\{0,1/2,\tau/2,(1+\tau)/2\}
\end{eqnarray}
which are the points left invariant under the $\mathbb{Z}_2$
action (in both directions of the torus)
\begin{eqnarray}
z_i\sim -z_i.
\end{eqnarray}

In the case of the fermionic states, the Pauli exclusion principle
requires only one fermion in each state, and so one has
\begin{eqnarray}
{\rm Tr}q^{\sum_w w\phi_{w}\cdot\phi_w}=\prod_w (1+q^w)^8
\end{eqnarray}
which applies to both the $NS$ and $R$ sectors with the simple
modification of $w\rightarrow w-\frac{1}{2}$ for the $NS$ fields.
Therefore, one has the full expression for $R$ states as
\begin{eqnarray}
{\rm Tr}q^{L_0}=16\frac{\prod_{n=1}^\infty
(1+q^n)^8}{\prod_{n=1}^\infty (1-q^n)^8}
\end{eqnarray}
where the factor of 16 takes into account the overall degeneracy
of the $R$ vacuum ($|\pm,\pm,\pm,\pm>\sim 2^4$ states).  In the
$NS$ case, the expression
\begin{eqnarray}
{\rm Tr}q^{L_0}=\frac{1}{q^\frac{1}{2}}\frac{\prod_{n=1}^\infty
(1+q^{n-\frac{1}{2}})^8}{\prod_{n=1}^\infty (1-q^n)^8}
\end{eqnarray}
shows an additional factor of $q^{-\frac{1}{2}}$ which arises from
equation (\ref{eqn:Delta}).

The $GSO$ projection
\begin{eqnarray}\label{eqn:GSO1}
{\rm Tr}\left(\frac{\big(1-(-1)^F\big)}{2}q^{L_0}\right)
=\frac{1}{2q^\frac{1}{2}}\frac{\prod_{n=1}^\infty
(1+q^{n-\frac{1}{2}})^8 -\prod_{n=1}^\infty
(1-q^{n-\frac{1}{2}})^8}{\prod_{n=1}^\infty (1-q^n)^8}
\end{eqnarray}
is exactly the eight dimensional representation of the vector
denoted by $V_8$, as defined more generally below for $SO(2n)$
representations.  This involves a restriction to odd numbers of
modes by use of the fermion number operator $F$.

In the lightcone gauge, one has eight transverse directions which
corresponds to an $SO(8)$ invariance.  The $SO(2n)$ Lie group
contains four conjugacy classes of representations which include
the scalar, vector and two spinors.  The representations of the
$NS$ fields that include the scalar and vector for $SO(2n)$ are
given by the "constant" theta functions $z=0$
\begin{eqnarray}\label{eqn:ThetaO}
O_{2n}(0|\tau)=\frac{\theta^n_3(0|\tau)+\theta^n_4(0|\tau)}{2\eta^n}
\end{eqnarray}
which is the scalar, and by reference to (\ref{eqn:GSO1}) shows it
to correspond to a restriction to even numbers of modes. This
expression thus contains the tachyon.  Similarly
\begin{eqnarray}\label{eqn:ThetaV}
V_{2n}(0|\tau)=\frac{\theta^n_3(0|\tau)-\theta^n_4(0|\tau)}{2\eta^n}
\end{eqnarray}
begins with lowest contribution $q^\frac{1}{2}$ for all
$n=1,2,\ldots$.

In the $R$ case, one has two characters of opposite chirality.
The characters
\begin{eqnarray}\label{eqn:ThetaS}
S_{2n}(0|\tau)=\frac{\theta^n_2(0|\tau)+i^{-n}\theta^n_1(0|\tau)}{2\eta^n}
\end{eqnarray}
and
\begin{eqnarray}\label{eqn:ThetaC}
C_{2n}(0|\tau)=\frac{\theta^n_3(0|\tau)-i^{-n}\theta^n_1(0|\tau)}{2\eta^n}
\end{eqnarray}
provide such a description.  In both cases the low lying modes
begin with the wights $q^{\frac{n}{8}}$. They are projections of
the spectrum by
\begin{eqnarray}
\frac{\big(1+\Gamma_9 (-1)^F\big)}{2}.
\end{eqnarray}

The characters used in the discussion of Mobius amplitudes involve
real \emph{hatted} characters.  A given character $\chi_i$ is
inherently complex, due to the additional $\frac{1}{2}$ piece in
the measure ($il+\frac{1}{2}$ in the transverse channel). One can
then define the real hatted character $\hat{\chi}_i$ which then
differs from $\chi_i$ by a phase.

The string models which include the presence of antibranes,
provide a subtle change to the spectrum for certain couplings. For
strings coupling to $D5$ branes, the possible arrangements are
$D5-D5$, $\bar{D}5-\bar{D}5$ and two copies of $\bar{D}5-D5$. The
former two have no effect on the projection, however, the latter
will introduce a sign change in the GSO projection that will
change the characters.

In the
$T^6/(\mathbb{Z}_2\times\mathbb{Z}_2\times\mathbb{Z}_2^\text{s})$
model, there will be a sign freedom in the parent amplitude, in
the first case I consider, such a freedom will not exist.  This
sign freedom will realize the presence of antibranes.

All models in the following sections will be compactifications in
four dimensions of the $SO(8)$ type I string to $SO(4)^2$ and
$SO(2)^4$ which begins as the orientifold projection of the parent
type IIB torus with fermion contributions
\begin{eqnarray}\label{eqn:Parent}
{\cal T}\sim |V_8-S_8|^2,
\end{eqnarray}
This form gives rise to open descendants that allow an $SO(32)$
gauge group. The type IIA theory is described by the torus
\begin{eqnarray}\label{eqn:Parent1}
{\cal T}\sim (V_8-S_8)(\bar{V}_8-\bar{C}_8)
\end{eqnarray}
which has the standard left--right handed relative chirality
difference.


\subsection{Discrete and Continuous Wilson Lines}


Now that models compactified on both orbifold and toroidal
topologies have been considered, it is lastly constructive to
detail the effect of Wilson lines on the open spectrum and brane
multiplets \cite{AADS,z2z2orient,z2z2orient3,z2z2orient4}. The
details that will enable the reader to understand the specifics of
constructing amplitudes is left until I discuss particular
examples in the following section. The purpose of the illustration
below is simply to highlight some generic effects that result from
the use of Wilson lines.

Introducing Wilson lines has a direct effect on branes that are
transverse to a compact direction, in particular the lowest mass
state of a string stretched between two branes is discussed in
appendix \ref{app:WilsonLines}.

The distinct possibilities of the discrete operations that affect
the compact coordinates are
\cite{vwaaf}:
\begin{eqnarray}\label{eqn:a1a2a3}
&&A_1: X_{\rm L,R} \to X_{\rm L,R} + \frac{\pi R}{2} \,,
\nonumber \\
&&A_2: X_{\rm L,R} \to X_{\rm L,R} + {\textstyle{\frac{1}{2}}}
\left( \pi R \pm {\frac{\pi \alpha '}{R}} \right) \,,
\nonumber \\
&&A_3: X_{\rm L,R} \to X_{\rm L,R} \pm {\frac{\pi \alpha'}{2R}}
\,.
\end{eqnarray}
for a circle of radius $2\pi R$.  The shift which will be of
interest in this work is the $A_1$ operator.  This will affect the
state $e^{ip.X}$ as
\begin{eqnarray}
A_1:e^{ip_LX+ip_R\tilde{X}}\rightarrow (-1)^m
e^{ip_LX+ip_R\tilde{X}}
\end{eqnarray}

As an example of the continuous case with the six dimensional
orbifold compactification ($T^4/\mathbb{Z}_2$) \cite{CAAS},
introducing Wilson lines in the last direction involves a
transverse channel amplitude of the annulus as
\begin{eqnarray}\label{eqn:TransAnnulusWilsonLines}
\tilde{\cal
A}=\frac{2^{-5}}{4}\bigg\{(Q_o+Q_v)\bigg[N^2vW_4+\frac{1}{v}\sum_m
(D+\frac{d}{2}e^{2i\pi \alpha m}+\frac{d}{2}e^{-2i\pi \alpha m})^2P_3P_m\bigg]\nn\\
+2N(D+d)(Q_o-Q_v)\bigg(\frac{2\eta}{\theta_2}\bigg)^2
+4(Q_s+Q_c)(N^2_g+D^2_g)\bigg(\frac{2\eta}{\theta_4}\bigg)^2\nn\\
-2N_gD_g(Q_s-Q_c)\bigg(\frac{2\eta}{\theta_3}\bigg)^2\bigg\}.
\end{eqnarray}
%
%
%
%
\begin{figure}[!ht]
\begin{center}
\includegraphics[scale=.6]{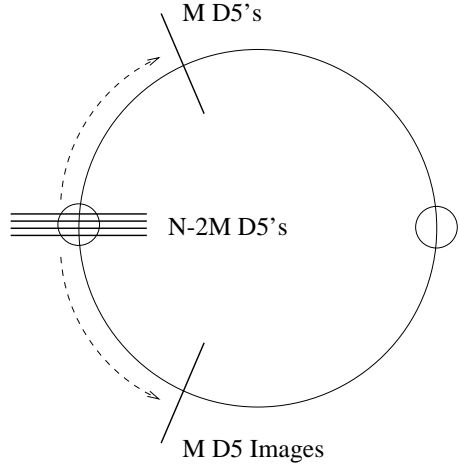}
\caption{Moving a brane off of the fixed
point}\label{dia:BraneMoving}
\end{center}
\end{figure}
%
%
The original $T^4/\mathbb{Z}_2$ amplitude is recovered by setting
$\alpha=0$.

Here a stack of $D5$'s (denoted as $d$) and their images are moved
a distance $2\pi \alpha R$ away from the fixed point they
originally occupied to the second one.  This is illustrated in
figure \ref{dia:BraneMoving}.

In the direct channel, Poisson resummation
\begin{eqnarray}\label{eqn:Resummation}
\sum_{n_i}e^{-\pi n^{\rm T}An+2i\pi b^{\rm }n}=
\frac{1}{\sqrt{{\rm det}(A)}}\sum_{m_i}e^{-\pi(m-b)^{\rm
T}A^{-1}(m-b)}
\end{eqnarray}
allows the lifting of states in mass, so that
\begin{eqnarray}
{\cal
A}=\frac{1}{4}\bigg\{(Q_o+Q_v)\bigg[N^2P_4+(D^2+\frac{d^2}{2})W_4
+\frac{d^2}{4}W_3(W_{n+2\alpha}+W_{n-2\alpha})\nn\\
+DdW_3(W_{n+\alpha}+W_{n-\alpha})\bigg]
+(Q_o-Q_v)(N^2_g+D^2_g)\bigg(\frac{2\eta}{\theta_2}\bigg)^2\nn\\
+2N_gD_g(Q_s-Q_c)\bigg(\frac{\eta}{\theta_3}\bigg)^2\bigg\}.
\end{eqnarray}
Here it is seen that for $\alpha=\frac{1}{2}$, lattice states
coupling to the $Dd$ configuration have $n+\frac{1}{2}$ windings.
In the general case that $\alpha\neq \frac{1}{2}$, one has a
generic gauge group braking of $U(16)_9\times U(16-2s)\times
USp(2s)$, where $d=2s$ which is required by consistency.  For the
interesting discrete case, (\ref{eqn:TransAnnulusWilsonLines})
allows $d$ $D5$ branes to move from one fixed point to the
opposite one. This has the effect of projecting the spectrum by
$\frac{1}{2}(1+(-1)^m)$.

The use of shifts in the context of $\mathbb{Z}_2\times
\mathbb{Z}_2$ (shift) orbifolds have been exhaustively studied in
\cite{AADS}. These models are descriptions of a partition function
that is projected by $\mathbb{Z}_2\times \mathbb{Z}_2$ generators
that incorporate the shift operators within the orbifold
generators themselves.  The shifts in these models act discretely.
The projections in the trace are defined as
\begin{eqnarray}\label{eqn:Gen'sOfNonFreelyActing}
\sigma_1(\delta_1,\delta_2,\delta_3)= \left(\begin{array}{rrr}
\delta_1 & -\delta_2 & -1 \\ -1 & \delta_2 & -\delta_3 \\
-\delta_1 & -1 & \delta_3
\end{array}\right),\quad
\sigma_2(\delta_1,\delta_2,\delta_3)= \left(\begin{array}{rrr}
\delta_1 & -1 & -1 \\ -1 & \delta_2 & -\delta_3 \\ -\delta_1 &
-\delta_2 & \delta_3
\end{array}\right).\nn\\
\end{eqnarray}
\begin{table}[!ht]
\begin{center}
\begin{tabular}{|l|c|c|c|}\hline
models & $D5_1$ & $D5_2$ & $D9$
\\
\hline \hline $p_3$ & ${N}=2$ & ${N}=2$ & ${N}=1$
\\
$w_2p_3$ & ${N}=2$ & ${N}=4$ & ${N}=2$
\\
$w_1w_2p_3$ & ${N}=4$ & ${N}=4$ & ${N}=4$
\\\hline
$p_{23}$ & ${N}=2$ & $-$ & ${N}=1$
\\
$w_1p_2$ & ${N}=4$ & $-$ & ${N}=2$
\\
$w_1p_2p_3$ & ${N}=4$ & $-$ & ${N}=2$
\\
$w_1p_2w_3$ & ${N}=4$ & $-$ & ${N}=4$
\\\hline
$p_{123}$ & $-$ & $-$ & ${N}=1$
\\
$p_1w_2w_3$ & $-$ & $-$ & ${N}=2$
\\
$w_1w_2w_3$ & $-$ & $-$ & ${N}=4$ \\\hline
\end{tabular}
\end{center}\caption{Model classes with corresponding brane supersymmetry}
\end{table}\label{tab:ShiftOrientModels}
It is then clear that these models necessarily involve
simultaneous actions of both shift and orbifold operations.  As
such these are not freely acting orbifolds, but they can be
interpreted as conventional $\mathbb{Z}_2\times \mathbb{Z}_2$
orbifolds with unconventional $\Omega$ projections.

In these models, one can see that the modulating group generators
given by equation (\ref{eqn:Gen'sOfNonFreelyActing}) will involve
terms of the form
\begin{eqnarray}\label{eqn:MatrixM}
\text{Tr}_{tw}\delta q^{L_0}\bar{q}^{\tilde{L}_o}.
\end{eqnarray}
This term is zero as the twisted trace involves a sum over fixed
points of the states.  As will be shown later by equation
(\ref{eqn:ShiftActionOnFixedPoints}), $\delta$ has the effect of
shifting the fixed point occupied by a state.  As such, twisted
states with shift insertions in the trace are zero.

In the freely acting shift case studied later, one can retain the
twisted terms corresponding to independent orbits (appendix
\ref{app:Boundaries}) which give rise to two classes of sub models
of the parent torus (each containing four sub models in the open
sector). These terms are not present in the models generated by
(\ref{eqn:Gen'sOfNonFreelyActing}) for the reason given above.

The various supersymmetries present on the branes for these
projections are displayed in table \ref{tab:ShiftOrientModels}.
{From} the table, the notations $p_i$ and $w_i$ are meant as a
Kaluza Klein shift $(-1)^{m_i}$ and winding shift $(-1)^{n_i}$ in
the direction of the $i^{\text{th}}$ torus.  Terms such as
$p_{ij}$ are shorthand for the product shift $(-1)^{m_i+m_j}$.  In
addition, $-$ denotes the absence of the corresponding brane.

In these cases, the presence of $D5$ brane types is determined by
the types of shift present.  If a shift is present in a line in
table \ref{tab:ShiftOrientModels}, the corresponding $D5$ brane is
absent for that direction. This is determined by the requirement
of tadpole cancellation. The presence of shifts will lift those
states in the Klein and Mobius that would normally contribute a
condition that would allow the $D5$ to remain. Therefore,
cancellation of the tadpole requires that the $D5$ (that would
otherwise contribute massless modes) be absent.

In the cases I turn to now the presence of shifts, which in this
case are freely acting, will not affect the numbers of $D5$'s in
the model.  However, the first model I consider has only one
orbifold element and in this case, for the same reasons of tadpole
cancellation, only one species of $D5$ exists.  The second model
has all possible types present.


\section{Open Descendants of a $T^6(\mathbb{Z}_2\times \mathbb{Z}_2^\text{s})$
Model with $N=2$ Supersymmetry}


To illustrate the effects of the freely acting shifts of the type
in equation (\ref{eqn:a1a2a3}) on the open descendants, I start
with a simpler example of a $\mathbb{Z}_2$ orbifold, with orbifold
operation denoted $g$ and an additional freely acting shift $h$.
The particular actions of $g$ and $h$ and their product on
coordinates and lattices are given in equation
(\ref{eqn:CompositeGenerators})\footnote{This model was analyzed
in collaboration with Carlo Angelantonj and Emilian Dudas.}.

The $\mathbb{Z}_2\times \mathbb{Z}_2^\text{s}$ generators comprise
a freely acting orbifold and shift.  In this case the effect of
the $\mathbb{Z}_2\times \mathbb{Z}_2^\text{s}$ is to break the
$SO(8)$ group to $SO(4)^2$ by virtue of the $\mathbb{Z}_2$
orbifold operation.  The shift of course has no effect in this
breaking.

While the structure of the compact directions is $T^6\rightarrow
T^2_{45}\times T^2_{67}\times T^2_{89}$, with subscripts referring
to the 2-tori directions, this model is (as far as the fermionic
excitations are concerned) effectively a compactification on
$(T^2\times (T^4/\mathbb{Z}_2))/\mathbb{Z}_2^\text{s}$.

The original type IIB theory is projected using
\begin{eqnarray}\label{eqn:CompositeGenerators}
\begin{array}{crrrrrr}
g=( & 1, & 1; & -1, & -1; & -1, & -1~~),\\
\nn\\
h=( & A_1, & 1; & A_1, & 1; & 1, & 1~~), \\
\nn\\
f=( & A_1, & 1; & -A_1, & -1; & -1, & -1~~).
\end{array}
\end{eqnarray}
for $A_1$ defined in (\ref{eqn:a1a2a3}).  The generators,
(\ref{eqn:CompositeGenerators}) illustrates the shift action on
only one of the coordinates of the relevant torus. The orbifolds
act on all coordinates within a given torus to provide four fixed
points in each that they act.

This is an interesting model that has a $\mathbb{Z}_2\times
\mathbb{Z}_2$ structure while preserving $N= 2$ supersymmetry in
the open sector.  The particular form of generators, that includes
one as a pure shift operator, implies that the independent orbit
diagrams (those not related by modular transformation, appendix
\ref{app:Boundaries}) do not contribute to the torus amplitude.

This takes away the consideration of a sub class of models
associated with a sign freedom.  These terms will appear in the
generalized version of this model in section
(\ref{sec:Z2xZ2xZ2Model1}).

The way the modulating group generators are written with composite
shift operators, has a twofold effect, firstly it will necessarily
force the number of distinct $D5$ types to become only one (in
this case, the $D5$ sitting in the first torus is the only one
present).

The character set is derived from the breaking of the original
SO(8) characters $O_8$, $V_8$, $C_8$ and $S_8$ to supersymmetric
representations involving $O_4$, $V_4$, $C_4$ and $S_4$.  Their
representations in terms of theta functions can be found in
section \ref{sec:TypeIConstruction}.

The supersymmetric world sheet fermion contributions are written
as
\begin{eqnarray}
&&Q_o=V_4O_4-C_4C_4 \quad Q_v=O_4V_4-S_4S_4 \nn\\
&&Q_s=O_4C_4-S_4O_4 \quad Q_c=V_4S_4-C_4V_4.
\end{eqnarray}

The function $\eta(\tau)$ is defined by equation (\ref{eqn:Eta}),
with the $SO(4)$ characters defined by equations
(\ref{eqn:ThetaO}), (\ref{eqn:ThetaV}), (\ref{eqn:ThetaS}) and
(\ref{eqn:ThetaC}).

In a given direction of a torus, one has a lattice of the form
\begin{eqnarray}\label{eqn:compact}
\Lambda_{m+a,n+b}=\frac{q^{\frac{\alpha\prime}{4}{\big{(}\frac{(m+a)}
{R}+\frac{(n+b)R}{\alpha\prime}\big{)}}^2}\bar{q}^{\frac{\alpha\prime}{4}
{\big{(}\frac{(m+a)}{R}-\frac{(n+b)R}{\alpha\prime}\big{)}}^2}}{\eta(q)
\eta(\bar{q})},
\end{eqnarray}
The form of which is demonstrated by equation (\ref{eqn:measure}).
In the models presented in this study, this notation will include
the lattice states of both directions. The labels $a$ and $b$ are
in anticipation of the effects of momentum and winding shifts,
which will of course only apply to the direction in which the
relevant shift acts.

To obtain modular invariance under $PSL(2,\mathbb{Z})$, as
required by the topology of the one loop string amplitude, one
must perform $S$ and $T$ transforms on the torus states. The
actions defined in (\ref{eqn:ActionOfS&T}) imply that the boundary
operations of states transform as
\begin{eqnarray}\label{eqn:tran}
&&S:(a,b)\rightarrow(b,a^{-1})\nn\\
&&T:(a,b)\rightarrow(a,a b).
\end{eqnarray}
Here, $a$ and $b$ label the boundary operations that pertain to
orbifold/twist's that are placed on world sheet fields. The full
orbit configuration of these operators is described for the
$\mathbb{Z}_2\times\mathbb{Z}_2$ case in figure \ref{app:Blocks}.
\begin{table}[!ht]
\begin{center}
\begin{tabular}{|c|c|c|}\hline
Lattice (in $\tilde{\cal A}$,${\cal K}$\text{ and }$\tilde{\cal
M}$) & $S(\tilde{\cal A})$ & $S({\cal K})$\text{ and
}$P(\tilde{\cal M})$
\\\hline
$P_m$ & $W_n$ &
$W_{2n}$\\
$P_{2m}$ & $W_n+W_{n+\frac{1}{2}}$ &
$W_{n}$\\
$(-1)^{m}P_m$ & $W_{n+\frac{1}{2}}$ &
$W_{2n+1}$\\
$(-1)^{m}P_{2m}$ & $W_{n+\frac{1}{2}}+W_{n}$ &
$W_{n+\frac{1}{2}}$\\
$W_n$ & $P_m$ &
$P_{m}$\\
$W_{2n}$ & $P_m+P_{m+\frac{1}{2}}$ & $P_{2m}$\\
$W_{2n+1}$ & $P_m-P_{m+\frac{1}{2}}$ & $(-1)^mP_{m}$
\\\hline
\end{tabular}
\caption{Lattice $S$ and $P$ transforms}\label{tab:Stransforms}
\end{center}
\end{table}

Table \ref{tab:Stransforms} summarizes some useful $S$ and $P$
transforms for lattice states that will be encountered in
determining the amplitude lattices in different channels. Here,
${\cal A}$, ${\cal K}$ and ${\cal M}$ are the annulus Klein and
Mobius contributions. The tilde denotes the transverse amplitude.
These results are obtained from equation (\ref{app:expL}) with the
appropriate boundary condition ($m=0$ or $n=0$), modular parameter
rescaling and equation (\ref{eqn:Resummation}) in combination with
$S$ transformation (\ref{eqn:ActionOfS&T}).

$P$ and $W$ are the restriction of $\Lambda_{m,n}$ to pure
Kaluza-Klein ($P$) or winding ($W$) modes. The more compact
notation of $P_e$ and $P_o$ will be used in the calculations (and
similarly for the winding sums) which are the restriction of the
counting to even or odd modes respectively.

$T$ transforms on the lattices behave as
\begin{eqnarray}\label{eqn:Ttransforms}
\begin{array}{lcl}
\Lambda_{m,n}\rightarrow\Lambda_{m,n}, & &
\Lambda_{m,n+\frac{1}{2}}\rightarrow(-1)^m\Lambda_{m,n+\frac{1}{2}}, \\
\Lambda_{m+\frac{1}{2},n}\rightarrow(-1)^n\Lambda_{m+\frac{1}{2},n},
& { \rm and } & \Lambda_{m+\frac{1}{2},n+\frac{1}{2}}\rightarrow
i(-1)^{m+n}\Lambda_{m+\frac{1}{2},n+\frac{1}{2}}.
\end{array}
\end{eqnarray}
These are easily seen by performing the $T$ operation defined in
(\ref{eqn:tran}) on the lattice of equation (\ref{eqn:compact})

The torus resulting from the projections
(\ref{eqn:CompositeGenerators}) is generated by enforcing modular
invariance under (\ref{eqn:tran}) on
\begin{eqnarray}\label{eqn:TraceT}
{\cal T}=\int \frac{d^2\tau}{\tau_2^2}\frac{1}{\tau_2^4}
\text{Tr}\frac{(1+g)}{2}\frac{(1+h)}{2}q^{L_o}\bar{q}^{\tilde{L}_o},
\end{eqnarray}
which of course will include the effect on lattice terms through
equation (\ref{eqn:measure}). The modular invariant torus
amplitude then simplifies to
\begin{eqnarray}
{\cal
T}=&\frac{1}{4}&\bigg\{\big[1+(-1)^{m_1+m_2}\big](\Lambda^1\Lambda^2+
\Lambda^1_{m,n+\frac{1}{2}}\Lambda^2_{m,n+\frac{1}{2}})\Lambda^3|Q_o+Q_v|^2\nn\\
&&+\big[1+(-1)^{m_1}\big]\Lambda^1|Q_o-Q_v|^2{\vline\frac{2\eta}{\theta_2}
\vline}^4\nn\\
&&+16(\Lambda^1+\Lambda^1_{m,n+\frac{1}{2}}){\vline\frac{\eta}{\theta_4}
\vline}^4|Q_s+Q_c|^2\nn\\
&&+16(\Lambda^1+(-1)^{m_1}\Lambda^1_{m,n+\frac{1}{2}}){\vline\frac{\eta}
{\theta_3}\vline}^4|Q_s-Q_c|^2\bigg\}.
\end{eqnarray}
World--sheet supersymmetry requires the actions on bosons and
fermions to be properly correlated, this is achieved by assigning
positive eigenvalues to $O_{2n}$ and $C_{2n}$ and negative ones to
$V_{2n}$ and $S_{2n}$, under the action of an orbifold element.

With the $A_1$ shift operator, the number of fixed points can be
seen to be halved. It acts on the fixed point coordinates, for
example, as
\begin{eqnarray}\label{eqn:ShiftActionOnFixedPoints}
(x_2,y_2;x_3,y_3)\rightarrow(x_2+\frac{1}{2},y_2;x_3+\frac{1}{2},y_3).
\end{eqnarray}
Where the labelling $(x_2,y_2;x_3,y_3)$ defines the collective
fixed point coordinates for the space $T^2_{67}\times T^2_{89}$,
with the values
$\{x_i,y_i|x_i\in\{0,\frac{1}{2}\},y_i\in\{0,\frac{1}{2}\}\}$.

The total number of fixed points without the shift operation is
$16=4\times 4$, which are detailed in table
(\ref{tab:fixedpoints}).
\begin{table}[!ht]
\begin{center}
\begin{tabular}{|llll|}\hline
${(0,0;0,0)}_1$ & ${(0,\frac{1}{2};0,0)}_2$ &
${(\frac{1}{2},0;0,0)}_3$ & ${(\frac{1}{2},\frac{1}{2};0,0)}_4$
\\ ${(0,0;0,\frac{1}{2})}_5$ & ${(0,0;\frac{1}{2},0)}_6$ &
${(0,0;\frac{1}{2},\frac{1}{2})}_7$ & ${(0,\frac{1}{2};0,\frac{1}{2})}_8$\\
${(0,\frac{1}{2};\frac{1}{2},0)}_9$ &
${(0,\frac{1}{2};\frac{1}{2},\frac{1}{2})}_{10}$ &
${(\frac{1}{2},0;0,\frac{1}{2})}_{11}$ &
${(\frac{1}{2},0;\frac{1}{2},0)}_{12}$\\
${(\frac{1}{2},0;\frac{1}{2},\frac{1}{2})}_{13}$ &
${(\frac{1}{2},\frac{1}{2};0,\frac{1}{2})}_{14}$ &
${(\frac{1}{2},\frac{1}{2};\frac{1}{2},0)}_{15}$ &
${(\frac{1}{2},\frac{1}{2};\frac{1}{2},\frac{1}{2})}_{16}$ \\
\hline
\end{tabular}
\caption{Unshifted fixed points}\label{tab:fixedpoints}
\end{center}
\end{table}

The origin of the lattice contributions of the torus amplitude is
\begin{eqnarray}
{\cal T}_0=|Q_o|^{2}+|Q_v|^2+8|Q_s|^2+8|Q_c|^2
\end{eqnarray}
and shows as expected 8 fixed points, reduced from 16.  The
independent coordinates of which are as in table
\ref{tab:shiftedFixedPoints}.
\begin{table}[!ht]
\begin{center}
\begin{tabular}{|llll|}\hline
${(0,0;0,0)}_1$ & ${(0,\frac{1}{2};0,0)}_2$ &
${(\frac{1}{2},0;0,0)}_3$ &
${(\frac{1}{2},\frac{1}{2};0,0)}_4$ \\
${(0,0;0,\frac{1}{2})}_5$ & ${(0,0;\frac{1}{2},\frac{1}{2})}_7$ &
${(0,\frac{1}{2};0,\frac{1}{2})}_8$ &
${(0,\frac{1}{2};\frac{1}{2},\frac{1}{2})}_{10}$ \\
\hline
\end{tabular}
\caption{Remaining fixed points}\label{tab:shiftedFixedPoints}
\end{center}
\end{table}

Vertex operators of states flowing in $\cal K$ and $\tilde{{\cal
A}}$ will acquire from the torus, by virtue of the action of the
shift in $T_{45}^2$ and $T_{67}^2$, a state projector
\begin{eqnarray}
V=\big[1+(-1)^{m_1+m_2}\big]V_{T_2\times (T^4/\mathbb{Z}_2)}.
\end{eqnarray}

For the Klein amplitude, $\Omega$ makes an effective
identification of the left and right modes.  As such, orbifold
elements in the trace of (\ref{eqn:TraceT}) acting on the world
sheet bosonic or fermionic oscillators are made ineffective by
$\Omega$. This is easily seen by a series expansion of such terms,
since the left and right modes contribute $(-1)^{k+\tilde{k}}$,
$k,\tilde{k}\in \mathbb{Z}$, the $\Omega$ identification then
neglects the orbifold presence.

In a similar fashion, this projection also reduces the lattice
modes to become either pure momentum or pure winding, this
situation is inverted with the assistance of an inserted orbifold
action $\alpha$:
\begin{eqnarray}
\Omega|p_L,p_R>&=&|p_R,p_L>\quad\Rightarrow\quad n=0,\nn\\
\Omega\alpha|p_L,p_R>&=&|-p_R,-p_L>\quad\Rightarrow\quad m=0.
\end{eqnarray}
$\Omega$ does not change the effect of twisting operations, since
these are realized as a shift in the oscillator modes. The Klein
amplitude thus takes the form,
\begin{eqnarray}
{\cal K}=&\frac{1}{8}&\bigg\{\bigg[\big(1+(-1)^{m_1+m_2}\big)P_1P_2P_3\nn\\
&&+\big(1+(-1)^{m_1}\big)P_1W_2W_3\bigg](Q_o+Q_v)\nn\\
&&+32(Q_s+Q_c)P_1{\bigg(\frac{\eta}{\theta_4}\bigg)}^2\bigg\}.
\end{eqnarray}

To derive the Klein amplitude in the transverse channel, one must
perform an $S$ transformation on the direct channel amplitude. The
measure and lattice resummations give rise to the factors of $2^2$
from the measure part and $\sqrt{2}$ from each internal direction.
These results can be seen more clearly by equation
(\ref{eqn:measure}). Moreover, table \ref{tab:Stransforms} shows
the effect on lattice state counting after $S$ transformation in
the Klein, annulus and Mobius amplitudes.

The $S$ transform of theta functions that are associated with
modes that are acted on by orbifold or twist operations are shown
by equation (\ref{eqn:STtransform}).  For the $SO(2n)$ characters,
the action of $S$ is represented as a matrix of the form
\begin{eqnarray}\label{eqn:Smatrix}
S_{2n}=\frac{1}{2}\left(\begin{array}{rrrr} 1 & 1 & 1 & 1 \\
1 & 1& -1 & -1 \\
1 & -1 & i^{-n} & -i^{-n} \\
1 & -1 & -i^{-n} & i^{-n} \\
\end{array}
\right),
\end{eqnarray}
and acts on the characters
$(O_{2n},V_{2n},S_{2n},C_{2n})^{\text{T}}$.  This can be seen by
deforming the individual theta functions of the $SO(2n)$
characters (\ref{eqn:ThetaO}), (\ref{eqn:ThetaV}),
(\ref{eqn:ThetaS}) and (\ref{eqn:ThetaC})

The direct Klein then has corresponding transverse amplitude
\begin{eqnarray}
\tilde{{\cal
K}}=&\frac{2^5}{8}&\bigg\{\bigg[(W_1^eW_2^e+W_1^oW_2^o)W_3^ev_1v_2v_3+
W_1P_2^eP_3^e\frac{v_1}{v_2v_3}\bigg](Q_o+Q_v)\nn\\
&&+2v_1W_1^e(Q_o-Q_v){\bigg(\frac{2\eta}{\theta_2}\bigg)}^2\bigg\}.
\end{eqnarray}

With the parity projection, the torus must be defined with an
additional factor of $\frac{1}{2}$.  It is seen that the torus has
the massive
$\Lambda_{n+\frac{1}{2}}^1\Lambda_{n+\frac{1}{2}}^2\Lambda^3$
lattice term in the untwisted sector that must symmetrize by
itself.  The presence of the operators $\Omega$ and $g$ allow this
to have the correct numerical factor to cancel the $\frac{1}{8}$
from the projections.  For all unshifted winding and momentum
quantum numbers set to zero, this state is simply
\begin{eqnarray}\label{eqn:StateExample}
|n+\frac{1}{2},n+\frac{1}{2}>
\end{eqnarray}
for the remaining non--zero lattice states that are counted in the
first two of the three tori. The spectrum must be invariant under
the operators $g$, and $\Omega$ ($\delta$ acts trivially on these
states), these have the actions
\begin{eqnarray}
g:(m,n) &\rightarrow &(-m,-n),\nn\\
\Omega:(m,n) &\rightarrow &(m,-n),\nn\\
g\Omega:(m,n) &\rightarrow &(-m,n).
\end{eqnarray}
So their action on (\ref{eqn:StateExample}) is
\begin{eqnarray}\label{eqn:StateDegeneracy}
4|n+\frac{1}{2},n+\frac{1}{2}>=|n+\frac{1}{2},n+\frac{1}{2}>_1\oplus|n+\frac{1}{2},-n-\frac{1}{2}>_g\nn\\
\oplus|-n-\frac{1}{2},-n-\frac{1}{2}>_\Omega\oplus|-n-\frac{1}{2},n+\frac{1}{2}
>_{g\Omega}
\end{eqnarray}
which can be seen by reference to (\ref{eqn:compact}).  The labels
$1,g,\ldots$ simply refer to the action that has been taken on a
given state. The degeneracy factor then gives the correct
counting.

Although the $O$-planes present are not indicated explicitly
within amplitudes, there presence and dimension are understood
from the toroidal volumes given by the $v_i$ terms.  $O9$--plane's
occupy the entire compact space and correspond to the term
$v_1v_2v_3$. The $O5$--plane only has a presence in the first of
the three 2-tori, and has a volume term $\frac{v_1}{v_2v_3}$.

In the transverse, or tube channel, the coefficients of the
characters must arrange themselves as perfect squares at the
origin of the lattices.  These coefficients contain all the
boundary terms that correspond to the $O9$ and $O5$ planes (in the
case of the Klein bottle). Appendix
\ref{app:TransverseTadpoleDiagrams} illustrates the diagrams
associated with a closed string propagating between two
boundaries, with world sheet time in the direction of propagation.
The perfect square thus reflects the invariance of the amplitude
under the interchange of the boundaries.  The cross terms of the
squares then give the mixing of different orientifold types. The
same is true in the transverse annulus for brane couplings.

The transverse Klein amplitude shows the perfect square structure
\begin{eqnarray}\label{eqn:KleinTransZero}
\tilde{{\cal
K}}_o&=&\frac{2^5}{8}v_1\bigg\{\big(\sqrt{v_2v_3}+\frac{1}{\sqrt{v_2v_3}}
\big)^2Q_o+\big(\sqrt{v_2v_3}-\frac{1}{\sqrt{v_2v_3}}\big)^2Q_v\bigg\}.
\end{eqnarray}

The transverse annulus is directly derived from the states that
flow in the torus.  The annulus in this channel involves closed
strings propagating between boundaries, or branes of the type
$D9$, represented by the factor $N$, and $D5_1$ in this case,
which is shown as $D$. As such, the structure of the states that
flow in the annulus reflect the boundary conditions of the brane
couplings. With the restriction to winding (Neumann boundary
conditions) or Kaluza Klein (Dirichlet boundary conditions) for
given boundaries, the transverse amplitude is
\begin{eqnarray}\label{eqn:CompositeTransverseAnnulus}
\tilde{{\cal
A}}&=&\frac{2^{-5}}{8}v_1\bigg\{\bigg[N^2v_2v_3(W_1W_2+W_1^{n+\frac{1}{2}}
W_2^{n+\frac{1}{2}})W_3\nn\\
&&+\frac{4D^2}{v_2v_3}W_1P^e_2P_3\bigg](Q_o+Q_v)+4NDW_1(Q_o-Q_v)
{\bigg(\frac{2\eta}{\theta_2}\bigg)}^2\bigg\}\nn\\
&&+\frac{2^{-3}}{8}v_1\bigg\{\bigg[R_N^2(W_1+W_1^{n+\frac{1}{2}})+2R_D^2W_1\bigg]
(Q_s+Q_c){\bigg(\frac{2\eta}{\theta_4}\bigg)}^2\nn\\
&&-4R_NR_DW_1(Q_s-Q_c){\bigg(\frac{\eta}{\theta_3}\bigg)}^2\bigg\}.
\end{eqnarray}

Some explanation of the numerical coefficients in the above
amplitude is necessary.  In the case of the untwisted terms, one
must satisfy the perfect square structure for the $D5$ and $D9$
terms, as shown more clearly in (\ref{eqn:TransAnnZero}).

The twisted terms are more subtle.  Since such terms highlight the
occupation of branes on the fixed points, their numerical
coefficients must therefore reflect this.  The breaking term $R_N$
corresponds to the effect of the orbifold on the $D9$--brane which
fills all compact and non-compact dimensions. Since it is wrapped
around all compact dimensions it therefore \textit{sees} all the
fixed points. The coefficient formula is
\begin{eqnarray}\label{eqn:FixedPointSummary}
\sqrt{\frac{{\rm total~number~of~fixed~points}}{{\rm
number~of~seen~fixed~points}}}.
\end{eqnarray}
$R_N$ thus has the coefficient
\begin{eqnarray}
\sqrt{\frac{16}{16}}\sqrt{v_1}.
\end{eqnarray}

With the volume $v_1$ being provided by the remaining compact
directions that are not acted on by the orbifold element
$(+,-,-)$.  The $D5$ breaking term, $R_D$, involves a brane which
wraps only the first tori, and is transverse to the remaining
ones.  Since the orbifold element $(+,-,-)$ provides sixteen fixed
points in the second and third tori, this term therefore has a
coefficient of 4, as it sits at the origin of the other tori, and
hence sees only that fixed point.

Now, under the identification of the fixed points, one can
categorize the types of brane that see certain fixed points.  All
brane types see the fixed point $(0,0;0,0)$ (the fixed point at
the origin of the tori $T^2_{67}$ and $T^2_{89}$). So one has the
perfect square
\begin{eqnarray}
\big(R_N\pm 4R_D\big)^2v_1
\end{eqnarray}
for the fixed point coordinates $(0,0;0,0)$, where the sign
depends on which character they couple to. For all other fixed
points, $R_D$ does not contribute to the counting since it only
sees $(0,0;0,0)$. So, the remaining seven fixed points are taken
into account by $R_N$ alone.  There is an overall factor of 2 that
reflects the degeneracy of the original sixteen fixed points
(where half of the sixteen are identified), which is also required
for proper particle interpretation in the direct channel.

These details provide the form for the lattice origin of the
transverse annulus as
\begin{eqnarray}\label{eqn:TransAnnZero}
\tilde{{\cal A}}_o&=&\frac{2^{-5}}{8}v_1\bigg\{\bigg(
N\sqrt{v_2v_3}+\frac{2D}{\sqrt{v_2v_3}}
\bigg)^2Q_o+\bigg(N\sqrt{v_2v_3}-\frac{2D}{\sqrt{v_2v_3}}\bigg)^2Q_v\nn\\
&&+2\bigg[(R_N-4R_D)^2+7R_N^2\bigg]Q_s\nn\\
&&+2\bigg[(R_N+4R_D)^2+7R_N^2\bigg]Q_c\bigg\}.
\end{eqnarray}

\begin{figure}[!ht]
\begin{center}
\includegraphics[scale=.6]{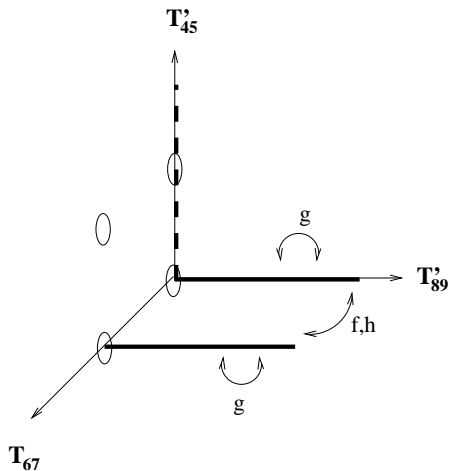}
\caption{Branes on the fixed points}\label{dia:Z2xZ2branes}
\end{center}
\end{figure}

Since the transverse annulus describes closed strings coupling to
boundaries, one can read off the boundary positions with reference
to the dilaton wave function
\begin{equation}\label{eqn:dilaton}
\phi(y_1,y_2)=\sum_{m_1,m_2}\bigg(cos\big(\frac{m_1y_1}{R_1}\big)
cos\big(\frac{m_2y_2}{R_2}\big)+sin\big(\frac{m_1y_1}{R_1}\big)
sin\big(\frac{m_2y_2}{R_2}\big)\bigg)\phi^{(m_1,m_2)}.\nn
\end{equation}
The dilaton is encoded in $Q_o$, and thus equation
(\ref{eqn:CompositeTransverseAnnulus}) gives the positions of the
branes.  After the use of $T$--dualities (as indicated by the primes
in a given direction) figure \ref{dia:Z2xZ2branes} shows the
positions of the $D9$ (now a $D5$ brane) shown as the dashed line
and the two stacks of rotated $D5$ branes sitting at the fixed
points.  By virtue of the shift, the $f$ and $h$ operations
interchange the brane stacks while $g$ leaves them unchanged.
%
%
%
%

The lattice origin of the Mobius amplitude is the product of
reflection coefficients from the transverse Klein and Annulus
amplitudes, with appropriate factors as shown for the $O9$--$D9$
coupling in (\ref{eqn:CouplingM}).  This is done for all the
common character sets in these two amplitudes, which includes only
those that are untwisted, since the transverse Klein amplitude has
only untwisted states. The resulting Mobius origin expression is
then dressed with the corresponding common lattice states of the
transverse Klein and annulus terms, such that the direct annulus
and Mobius symmetrize properly. These states are fully illustrated
in table \ref{tab:cc1}.  {From} the expressions
(\ref{eqn:KleinTransZero}) and (\ref{eqn:TransAnnZero}), the
lattice origin expression for the transverse Mobius is
\begin{eqnarray}
\tilde{\cal
M}_o=-\frac{1}{4}\bigg\{\left(\sqrt{v_1v_2v_3}+\sqrt{\frac{v_1}{v_2v_3}}\right)
\left(N\sqrt{v_1v_2v_3}+2D\sqrt{\frac{v_1}{v_2v_3}}\right)\hat{Q}_o\nn\\
+\left(\sqrt{v_1v_2v_3}-\sqrt{\frac{v_1}{v_2v_3}}\right)
\left(N\sqrt{v_1v_2v_3}-2D\sqrt{\frac{v_1}{v_2v_3}}\right)\hat{Q}_v
\bigg\}.
\end{eqnarray}
There is an ambiguity in the form of a sign in the Mobius, it is
chosen so as to allow a consistent tadpole cancellation. This
ambiguity comes from the square root of various coupling terms,
for example, the $D9$--$O9$ term is given by
\begin{eqnarray}\label{eqn:CouplingM}
\tilde{\cal
M}_{D9-O9}=\pm2\times\sqrt{\frac{2^5}{8}}\times\sqrt{\frac{2^{-5}N^2}{8}}.
\end{eqnarray}
This reflects the closed string coupling to an $O$--plane and a
$D$--brane, which includes a diagram factor of two for the
orientation of the boundaries.
\begin{table}
\begin{center}
\begin{tabular}{|c|c|}\hline
$D$--brane and $O$--plane couplings & Volumes \\ \hline $D9-D9$,
$D9-O9$,
$O9-O9$ & $v_1v_2v_3$ \\
$D5_1-D5_1$, $D5_1-O5_1$, $O5_1-O5_1$ & $\frac{v_1}{v_2v_3}$ \\
$D9-O5_l$, $O9-D5_l$ & $v_1$
\\ \hline \hline
Couplings from $\tilde{\cal A}$ and $\tilde{\cal K}$ & Lattice
States
\\\hline
$D9-D9$ & $(W^1W^2+W^1_{n+\frac{1}{2}}W^1_{n+\frac{1}{2}})W^3$
\\
$O9-O9$ & $(W^1_eW^2_e+W^1_oW^2_o)W^3_e$
\\
$D5_1-D5_1$ & $W^1P^2_eP^3$
\\
$O5_1-O5_1$ & $W^1P^2_eP^3_e$
\\ \hline \hline
$O-D$ Couplings in $\tilde{\cal M}$ & Lattice States
\\\hline
$D9-O9$ & $(W^1_eW^2_e+W^1_oW^2_o)W^3_e$
\\
$D9-O5_1$ & $W^1$
\\
$D5_1-O9$ & $W^1_e$
\\
$D5_1-O5_1$ & $W^1_eP^2_eP^3_e+(-1)^{m_2}W^1_oP^2_eP^3_e$
\\\hline
\end{tabular}
\end{center}
\caption{Lattice restrictions}\label{tab:cc1}
\end{table}

The coupling of the $D5$--brane and the $O5$--plane presents a
subtlety. Since the shift $h$ acts in one coordinate of the first
two tori, the direct channel Klein $O5-O5$ coupling has a
projected momentum lattice as $(1+(-1)^{m_1})P^1$. This leads to
the common lattice modes between the transverse annulus and Klein
amplitudes to have all winding modes.  This would naively lead to
inconsistent symmetrization with the direct channel annulus, since
the corresponding lattice in the annulus is $P^k$, and that of the
direct Mobius would be $2P^k_e$.  This can be rectified by
splitting the winding mode lattice to $W_e+W_o$ and introducing a
phase for the second momentum tower that exist with the odd
winding modes, as shown for the $D5_1$-$O5_1$ coupling in table
\ref{tab:cc1}. This then allows the proper symmetrization between
the direct annulus and Mobius for integer lattice modes.

With the remaining common states shown in table \ref{tab:cc1}, the
transverse Mobius results as
\begin{eqnarray}
\tilde{{\cal
M}}&=&-\frac{v_1}{4}\bigg\{\bigg[Nv_2v_3(W_1^eW_2^e+W_1^oW_2^o)W^e_3\nn\\
&&+\frac{2D}{v_2v_3}(W^e_1P^e_2+W^o_1(-1)^{m_2}P^e_2)P_3^e\bigg]
(\hat{Q}_o+\hat{Q}_v)\nn\\
&&+(NW_1+2DW^e_1)(\hat{Q}_o-\hat{Q}_v){\bigg(\frac{2\hat{\eta}}
{\hat{\theta}_2}\bigg)}^2\bigg\}.
\end{eqnarray}

The direct channel amplitude for the annulus is
\begin{eqnarray}
{\cal A}&=&\frac{1}{8}\bigg\{\bigg[N^2\big(1+(-1)^{m_1+m_2}\big)P_1P_2P_3\nn\\
&&+2D^2P_1(W_2+W_2^{n+\frac{1}{2}})W_3\bigg](Q_o+Q_v)\nn\\
&&+4NDP_1(Q_s+Q_c){\bigg(\frac{\eta}{\theta_4}\bigg)}^2+2\bigg[R_N^2P_1^e+
R_D^2P_1\bigg](Q_o-Q_v){\bigg(\frac{2\eta}{\theta_2}\bigg)}^2\nn\\
&&+4R_NR_DP_1(Q_s-Q_c){\bigg(\frac{\eta}{\theta_3}\bigg)}^2\bigg\}.
\end{eqnarray}

In deriving the direct channel Mobius it will be necessary to
perform $P$ transforms on the amplitude components.  Formally, the
$P$ operator is a combination of the already understood $S$ and
$T$ transforms as stated in equation (\ref{eqn:Ptransform1}). This
acts on the measure parameter according to equation
(\ref{eqn:Ptransform}), where a change of variables
$\tau_2=\frac{1}{2l}$ allows the same dependency on $il$ as the
Klein and annulus transverse amplitudes (with the additional
factor of $\frac{1}{2}$ in the Mobius).

The rescaling of the parameter $\tau_2$ for the $P$ transformation
of the Mobius is identical to that used for the Klein for the $S$
transformation.  As such the $P$ transformation on the Mobius
lattice terms is the same as an $S$ transformation on the Klein
lattice states.  The relevant lattice transformations are shown in
table \ref{tab:Stransforms}.

The $P$ transformation acts on the $SO(2n)$ characters as
\begin{eqnarray}\label{eqn:Pmatrix}
P_{2n}=\left(\begin{array}{cccc} c & s & 0 & 0 \\ s & -c & 0 & 0 \\
0 & 0 & \chi c & i\chi s \\ 0 & 0 & i\chi s & \chi c
\end{array}
\right)
\end{eqnarray}
for $s=sin(n\pi/4)$, $c=cos(n\pi/4)$ and
$\chi=e^{-i\frac{n\pi}{4}}$.

The direct channel Mobius is then
\begin{eqnarray}\label{eqn:DirectMobiusSimple}
{\cal M}&=&-\frac{1}{8}\bigg\{\big[N\big(1+(-1)^{m_1+m_2}\big)P_1P_2P_3\nn\\
&&+2D\big(P_1W_2+(-1)^{m_1}P_1W_2^{n+\frac{1}{2}}\big)W_3\big](\hat{Q}_o+
\hat{Q}_v)\nn\\
&&-2(NP^e_1+DP_1)(\hat{Q}_o-\hat{Q}_v){\bigg(\frac{2\hat{\eta}}{\hat{\theta}_2}
\bigg)}^2\bigg\}.
\end{eqnarray}

The massless modes in the amplitude $\tilde{{\cal K}}+\tilde{{\cal
A}}+\tilde{{\cal M}}$, which correspond to ${\mathcal{O}}(q^0)$
terms, give rise to a divergence. Such terms correspond to tadpole
diagrams, and consistency with the cancellation of these
contributions forces a constraint on the gauge group.  The
construction so far then yields the following tadpole conditions
\begin{eqnarray}
N=32,~2D=32,~R_N=0,~R_D=0.\nn
\end{eqnarray}
The Chan--Paton factors of the twisted sector ($R_N$ and $R_D$)
are identically zero as there are no complementary sectors in the
Mobius or Klein to allow otherwise.

The required Chan-Paton parameterization is then
\begin{eqnarray}
N=n+\bar{n},~D=d+\bar{d},~R_N=i(n-\bar{n}),~R_D=i(d-\bar{d}).\nn
\end{eqnarray}
This Chan-Paton charge arrangement shows that the vector multiplet
(contained in $Q_o$) is not present in the Mobius.  However, the
annulus does, and so the vector is oriented.  As such the
multiplicities have a unitary interpretation as shown above.

With these representations, one finds that the open sector has the
gauge group (from the character $Q_o$)
\begin{eqnarray}
U(16)_9\times U(8)_5\nn,
\end{eqnarray}
and shows the appropriate massless couplings in the open sector as
\begin{eqnarray}
{\cal
A}_o+M_o&=&(n\bar{n}+d\bar{d})Q_o+\frac{1}{2}\big(n(n-1)+\bar{n}
(\bar{n}-1)+d(d-1)+\bar{d}(\bar{d}-1)\big)Q_v\nn\\
&&+(n\bar{d}+\bar{n}d)Q_s+(nd+\bar{n}\bar{d})Q_c.
\end{eqnarray}
This demonstrates the rank reduction of the group associated with
the $D5$ brane.  In the generic $T^4/\mathbb{Z}_2$ model, one has
a gauge group of $U(16)_9\times U(16)_5$.

It is now a simple matter to extract the interesting spectral
content.  The open descendants contain hypermultiplets in the
representations $(120\oplus\overline{120},1)$ and
$(1,28\oplus\overline{28})$ from $Q_v$ as
\begin{eqnarray}
Q_v^h\sim O_2O_2(O_2V_2+V_2O_2)-(S_2S_2+C_2C_2)(S_2S_2+C_2C_2)\nn
\end{eqnarray}
and $(16,\overline{8})$ from $Q_s$ as
\begin{eqnarray}
Q_s^h\sim O_2O_2(C_2S_2+S_2C_2)-(S_2S_2+C_2C_2)O_2O_2.\nn
\end{eqnarray}
%
%


\section{$T^{6}/(\mathbb{Z}_{2}\times\mathbb{Z}_2\times\mathbb{Z}_2^\text{s})$ model}\label{sec:Z2xZ2xZ2Model1}


This model is in essence, a generalization of the previous one.
The orbifold group is enhanced from $\mathbb{Z}_2$ to
$\mathbb{Z}_2\times \mathbb{Z}_2$. In addition, the momentum shift
now extends to have an effect on modes in all three tori.

The projection that realizes the torus amplitude of this model has
the form
\begin{eqnarray}\label{eqn:TorusProjection}
\frac{1}{8}(1+g)(1+f)(1+\delta)\nn
\end{eqnarray}
The $\mathbb{Z}_2 \times \mathbb{Z}_2 \times
\mathbb{Z}_2^\text{s}$ elements are
\begin{eqnarray}
\begin{array}{crrrrrr}
g=( & 1, & 1; & -1, & -1; & -1, & -1~~),\\
\nn\\
f=( & -1, & -1; & 1, & 1; & -1, & -1~~), \\
\nn\\
h=( & -1, & -1; & -1, & -1; & 1, & 1~~), \\
\nn\\
\delta=( & A_1, & 1; & A_1, & 1; & A_1, & 1~~).
\end{array}
\end{eqnarray}
With an underlying $\mathbb{Z}_2 \times \mathbb{Z}_2$ group, there
will be terms in the torus amplitude that are not connected by $S$
and $T$ transforms to the principle orbits $(o,o), (o,g), (o,f)$
and $(o,h)$, as better illustrated in appendix
\ref{app:Boundaries}. These terms will be realized in terms of
modular orbits that are twisted sectors with different orbifold
insertions, such as $(f,g)$. As such, an ambiguity will be present
in the form of a sign freedom. This will give rise to models with
($\epsilon=-1$) or without ($\epsilon=+1$) discrete torsion, and
necessitate the study of different classes of models within a
choice of sign (as shown in \cite{CAAS} for the $\mathbb{Z}_2
\times \mathbb{Z}_2$ case without shifts).

The torus amplitude that results from the projection of the
spectrum from (\ref{eqn:TorusProjection}) has $N=2$ supersymmetry
in both amplitudes (with or without discrete torsion).  However,
the supersymmetric multiplet structure in each is quite different.

The torus is
\begin{eqnarray}\label{eqn:Torus2}
{\cal
T}&=&\frac{1}{8}\bigg{\{}|T_{oo}|^2\big{[}{\Lambda^1}_{m,n}{\Lambda^2}_{m,n}
{\Lambda^3}_{m,n}\nn\\
&&+{\Lambda^1}_{m,n+\frac{1}{2}}{\Lambda^2}_{m,n+\frac{1}{2}}
{\Lambda^3}_{m,n+\frac{1}{2}}\big{]}\big{(}1+(-1)^{m_1+m_2+m_3}\big{)}\nn\\
&&+|T_{ok}|^2{\Lambda^k}_{m,n}\big(1+(-1)^{m_k}\big){\vline\frac{2\eta}
{\theta_2}\vline}^4\nn\\
&&+16|T_{ko}|^2\big{(}{\Lambda^k}_{m,n}+{\Lambda^k}_{m,n+\frac{1}{2}}\big{)}
{\vline\frac{\eta}{\theta_4}\vline}^4\nn\\
&&+16|T_{kk}|^2\big{(}{\Lambda^k}_{m,n}+(-1)^{m_k}
{\Lambda^k}_{m,n+\frac{1}{2}}\big{)}
{\vline\frac{\eta}{\theta_3}\vline}^4 \nn\\
&&+\epsilon(|T_{gh}|^2+|T_{gf}|^2+|T_{fg}|^2+|T_{fh}|^2+|T_{hg}|^2+|T_{hf}|^2)
{\vline\frac{8{\eta}^3}{\theta_2\theta_3\theta_4}\vline}^2\bigg{\}}.
\end{eqnarray}
The values $k$,$m$ and $l$ take the values $\{1,2,3\}$ for the
bosonic lattice states, in correspondence with the generators
$g\sim 1,f\sim 2$ and $h\sim 3$.  The fermionic terms such as
$T_{kl}$ keep the labelling $l\in \{g,f,h\}$.

The origin of the lattice contributions has the form,
\begin{eqnarray}\label{eqn:TorusZero}
{\cal
T}_o&=&\big(|\tau_{oo}|^2+|\tau_{og}|^2+|\tau_{of}|^2+|\tau_{oh}|^2\big)\nn\\
&&+4(\epsilon+1)\big(|\tau_{go}|^2+|\tau_{gg}|^2+|\tau_{gf}|^2+|\tau_{gh}|^2\big)\nn\\
&&+4(1-\epsilon)\big(\tau_{go}\bar{\tau}_{gg}+\tau_{gg}\bar{\tau}_{go}
+\tau_{gf}\bar{\tau}_{gh}+\tau_{gh}\bar{\tau}_{gf}\big)\nn\\
&&+4(\epsilon+1)\big(|\tau_{fo}|^2+|\tau_{fg}|^2+|\tau_{ff}|^2+|\tau_{fh}|^2\big)\nn\\
&&+4(1-\epsilon)\big(\tau_{fo}\bar{\tau}_{ff}+\tau_{fg}\bar{\tau}_{fh}
+\tau_{ff}\bar{\tau}_{fo}+\tau_{fh}\bar{\tau}_{fg}\big)\nn\\
&&+4(\epsilon+1)\big(|\tau_{ho}|^2+|\tau_{hg}|^2+|\tau_{hf}|^2+|\tau_{hh}|^2\big)\nn\\
&&+4(1-\epsilon)\big(\tau_{ho}\bar{\tau}_{hh}+\tau_{hg}\bar{\tau}_{hf}
+\tau_{hf}\bar{\tau}_{hg}+\tau_{hh}\bar{\tau}_{ho}\big).
\end{eqnarray}
For either $\epsilon=\pm 1$, as in the last example, the presence
of the shift reduces the independent fixed points from sixteen to
eight.  The original $\mathbb{Z}_2\times \mathbb{Z}_2$ model has
52 hypermultiplets with 3 vector multiplets for $\epsilon=-1$ and
51 vector multiplets with 4 hypermultiplets for $\epsilon=+1$. The
untwisted sector contains the supergravity multiplet, four
hypermultiplets and 3 vector multiplets.  Each of the three
twisted sectors contains the remaining 48 hypermultiplets for
$\epsilon=+1$ and 48 vector multiplets for $\epsilon=-1$.  The
Hodge numbers for these models are $(51,3)$ and $(3,51)$.

In this model, one has eight fixed points with the action of the
freely acting shift.  As such, equation (\ref{eqn:TorusZero})
shows a model with the same multiplet structure in the untwisted
sector but a reduction to 24 hypermultiplets for $\epsilon=+1$,
and 24 vector multiplets for $\epsilon=-1$.  As such, the models
provide the Hodge numbers $(27,3)$ and $(3,27)$.

Again, a massive term
$\Lambda_{n+\frac{1}{2}}^1\Lambda_{n+\frac{1}{2}}^2\Lambda_{n+\frac{1}{2}}^3$
appears that must symmetrize by itself.  In precisely the same
manor as the argument given for (\ref{eqn:StateDegeneracy}), one
has states that correspond to the action of the operators $1$,
$g$, $f$, $fg$, $\Omega$, $g\Omega$, $f\Omega$ and $fg\Omega$.
These eight operators in addition to the factor of 2 from the
shift projection cancels precisely the $\frac{1}{16}$ (which
includes the factor of $\frac{1}{2}$ form the orientifold
projection).

The torus amplitude (\ref{eqn:Torus2}) clearly shows the two
separately connected parts, with the sign freedom $\epsilon$
associated with those orbits not related to the principle ones.
Discrete torsion is obtained by taking $\epsilon=-1$.  As shown,
the resulting spectral content for cases with and without torsion
are quite different.

This will also be the case for the open sector.  In addition,
there will be the possibility of SUSY breaking by the possible
presence of anti-branes, which will be discussed in section
(\ref{sec:OpenSpectrum}).

With an underlying $\mathbb{Z}_2\times \mathbb{Z}_2$ orbifold
group, the $SO(8)$ characters break to $SO(2)^4$ characters that
are defined by the following supersymmetric contributions:
\begin{eqnarray}
\tau_{oo}&=&V_2O_2O_2O_2+O_2V_2V_2V_2-S_2S_2S_2S_2-C_2C_2C_2C_2\nn\\
\tau_{og}&=&O_2V_2O_2O_2+V_2O_2V_2V_2-C_2C_2S_2S_2-S_2S_2C_2C_2\nn\\
\tau_{oh}&=&O_2O_2O_2V_2+V_2V_2V_2O_2-C_2S_2S_2C_2-S_2C_2C_2S_2\nn\\
\tau_{of}&=&O_2O_2V_2O_2+V_2V_2O_2V_2-C_2S_2C_2S_2-S_2C_2S_2C_2\nn\\
\nn\\
\tau_{go}&=&V_2O_2S_2C_2+O_2V_2C_2S_2-S_2S_2V_2O_2-C_2C_2O_2V_2\nn\\
\tau_{gg}&=&O_2V_2S_2C_2+V_2O_2C_2S_2-S_2S_2O_2V_2-C_2C_2V_2O_2\nn\\
\tau_{gh}&=&O_2O_2S_2S_2+V_2V_2C_2C_2-C_2S_2V_2V_2-S_2C_2O_2O_2\nn\\
\tau_{gf}&=&O_2O_2C_2C_2+V_2V_2S_2S_2-S_2C_2V_2V_2-C_2S_2O_2O_2\nn\\
\nn\\
\tau_{ho}&=&V_2S_2C_2O_2+O_2C_2S_2V_2-C_2O_2V_2C_2-S_2V_2O_2S_2\nn\\
\tau_{hg}&=&O_2C_2C_2O_2+V_2S_2S_2V_2-C_2O_2O_2S_2-S_2V_2V_2C_2\nn\\
\tau_{hh}&=&O_2S_2C_2V_2+V_2C_2S_2O_2-S_2O_2V_2S_2-C_2V_2O_2C_2\nn\\
\tau_{hf}&=&O_2S_2S_2O_2+V_2C_2C_2V_2-C_2V_2V_2S_2-S_2O_2O_2C_2\nn
\end{eqnarray}
\begin{eqnarray}\label{eqn:char}
\tau_{fo}&=&V_2S_2O_2C_2+O_2C_2V_2S_2-S_2V_2S_2O_2-C_2O_2C_2V_2\nn\\
\tau_{fg}&=&O_2C_2O_2C_2+V_2S_2V_2S_2-C_2O_2S_2O_2-S_2V_2C_2V_2\nn\\
\tau_{fh}&=&O_2S_2O_2S_2+V_2C_2V_2C_2-C_2V_2S_2V_2-S_2O_2C_2O_2\nn\\
\tau_{ff}&=&O_2S_2V_2C_2+V_2C_2O_2S_2-C_2V_2C_2O_2-S_2O_2S_2V_2.
\end{eqnarray}
Where one combines these into the character sums as
\begin{eqnarray}
&&T_{\gamma o}=\tau_{\gamma o}+\tau_{\gamma g}+\tau_{\gamma h
}+\tau_{\gamma f} \quad \quad
T_{\gamma g}=\tau_{\gamma o}+\tau_{\gamma g}-\tau_{\gamma h}-\tau_{\gamma f}\nn\\ \nn\\
&&T_{\gamma h}=\tau_{\gamma o}-\tau_{\gamma g}+\tau_{\gamma
h}-\tau_{\gamma f} \quad \quad T_{\gamma f}=\tau_{\gamma o}-\tau_{
\gamma g}-\tau_{\gamma h}+\tau_{\gamma f}.
\end{eqnarray}
Which for the sake of clarification, $\gamma \in \{0,1,2,3\}$
where $o\sim 0$ (the $\mathbb{Z}_2\times \mathbb{Z}_2$ identity),
with the normal relations for $g$, $f$ and $h$ in terms of the
$\mathbb{Z}_2\times \mathbb{Z}_2$ generators.

In the discussions that follow, where ever a sum occurs in
character set such as $T_{kl}$, it is taken that the condition $k
\neq l$ applies.  For $T_{kl}$, the index $k$ is the action of the
corresponding $k$ orbifold element, while $l$ is a twist.

The case considered in the previous section was the projection of
the partition function by $\mathbb{Z}_2\times
\mathbb{Z}_2^\text{s}$. Consequently, the counting of fixed points
(multiplicity factor) for the twisted sector in the torus, in
particular the $\Lambda_{m,n+\frac{1}{2}}$ massive states, is
preserved as eight after the orientifold projection. This was
realized by the factor of $\frac{1}{8}$ from the projection which
includes $\frac{1}{2}$ from the orientifold projection. In the
$\mathbb{Z}_2\times \mathbb{Z}_2\times \mathbb{Z}_2^\text{s}$
model, the freely acting shift acts as an additional modulating
group outside the $\mathbb{Z}_2\times \mathbb{Z}_2$ projection.
This requires an extra factor of one half in the trace.  As such,
the $n+\frac{1}{2}$ massive states in the twisted sector of the
torus have half the degeneracy they require for consistent
interpretation as states that exist at the eight fixed points.
Therefore, the Klein must also add an equal number of
$n+\frac{1}{2}$ states to compensate the half from the orientifold
projection
\begin{eqnarray}
\frac{1}{2}\big(8_{\cal T}\Lambda_{m,n+\frac{1}{2}}+8_{\cal
K}W_{n+\frac{1}{2}}\big),
\end{eqnarray}
with eight from the torus $8_{\cal T}$, and eight from the Klein
$8_{\cal K}$.  However, the naive insertion of such a
$W_{n+\frac{1}{2}}$ term leads to inconsistent factorization in
the transverse Klein amplitude.  This inconsistency arises due to
the $S$ transformation mapping $W_{n+\frac{1}{2}}$ states to
$(-1)^mP$.  In this case, the $O5_l-O5_k$ couplings would have a
factor of two.  A similar phenomenon arises in a six dimensional
example in \cite{CAAS}.  Here the authors consider
$T^4/\mathbb{Z}_2$ with an unconventional orientifold projection
$\xi\Omega$, for the phase $\xi^2=1$. This model defines a direct
Klein amplitude
\begin{eqnarray}
{\cal K}&=&\frac{1}{4}\bigg[(Q_o+Q_v)\bigg(\sum_m (-1)^m
\frac{q^{(\frac{\alpha^\prime}{2})m^{\rm T} g^{-1} m
}}{\eta^4}+\sum_n (-1)^n \frac{q^{(\frac{1}{2\alpha^\prime})n^{\rm
T} g n }}{\eta^4}\bigg)\nn\\
&&+2\times(n_++n_-)(Q_s+Q_c)\bigg(\frac{\eta}{\theta_4}\bigg)^4\bigg].
\end{eqnarray}
As such, in the transverse channel amplitude, the twisted sector
cannot be derived by factorization from the untwisted states that
are now entirely massive, by virtue of a redefinition of the
orientifold projection.  This then requires equal but opposite
eigenvalue assignments to the twisted states that effectively
render the counting zero with $n_+=8$ and $n_-=-8$.

The Klein amplitude for the $T^6/\mathbb{Z}_2\times
\mathbb{Z}_2\times \mathbb{Z}_2^\text{s}$ model is then given by
\begin{eqnarray}\label{eqn:DirectKlein}
{\cal
K}&=&\frac{1}{16}\bigg{\{}\bigg(P^{1}P^{2}P^{3}\big{(}1+(-1)^{m_1+m_2+m_3}\big{)}+
\big{(}1+(-1)^{m_1}\big{)}P^{1}W^{2}W^{3}\nn\\
&&+\big{(}1+(-1)^{m_2}\big{)}W^{1}P^{2}W^{3}+\big{(}1+(-1)^{m_3}\big{)}
W^{1}W^{2}P^{3}\bigg)T_{oo}\nn\\
&&+2\times16\epsilon_k\left(P^k+\epsilon
W^k\right)\bigg{(}\frac{\eta}{\theta_4}\bigg{)}^2T_{ko}+2\times(8-8)
W^k_{n+\frac{1}{2}}\bigg{(}\frac{\eta}{\theta_4}\bigg{)}^2
T_{ko}\bigg{\}}
\end{eqnarray}
where the parameter $\epsilon$ satisfies
\begin{eqnarray}\label{eqn:Epsilon}
\epsilon=\epsilon_1\epsilon_2\epsilon_3.
\end{eqnarray}
The arrangements of $\epsilon_j$ in equation
(\ref{eqn:DirectKlein}) are such that at the origin of the
lattices, the left--right symmetric states in the torus
(\ref{eqn:TorusZero}) can be seen to vanish for $\epsilon=-1$
(which is provided for by the term $P^k+\epsilon W^k$. Moreover,
the signs $\epsilon_j$ should arrange into the perfect square
structure (\ref{eqn:KleinPS}).

The transverse Klein amplitude is then
\begin{eqnarray}
\tilde{{\cal K}}=
\frac{2^5}{16}\bigg{\{}\bigg(v_1v_2v_2(W^1_eW^2_eW^3_e+W^1_oW^2_oW^3_o)
+\frac{v_k}{2v_lv_m}W^kP^l_eP^m_e\bigg)T_{oo}
\nn\\
\nn\\+2\epsilon_k\bigg[v_kW^k_e+\epsilon\frac{P^k_e}{v_k}\bigg]
{\biggl(\frac{2\eta}{\theta_2}\biggr)}^2T_{ok}\bigg{\}}.
\end{eqnarray}
The usual symmetrized summation convention is used for $k$,$l$ and
$m$.

The transverse Klein amplitude at the origin is
\begin{eqnarray}\label{eqn:KleinZeroModes}
\tilde{{\cal
K}}_o=\frac{2^5}{16}\bigg{\{}\bigg(v_1v_2v_2+\frac{v_k}{2v_lv_m}\bigg)
T_{oo}+2\epsilon_k\bigg(v_k+\epsilon\frac{1}{v_k}\bigg)T_{ok}\bigg{\}}
\end{eqnarray}
which has an expanded form
\begin{eqnarray}\label{eqn:KleinPS}
\tilde{{\cal K}_o}=\frac{2^5}{16}&&\bigg{\{}{\bigg{(}
\sqrt{v_1v_2v_3}+\epsilon_1\sqrt{\frac{v_1}{v_2v_3}}+
\epsilon_2\sqrt{\frac{v_2}{v_1v_3}}+
\epsilon_3\sqrt{\frac{v_3}{v_1v_2}}\bigg{)}}^2\tau_{oo}\nn\\
&&+{\bigg{(}\sqrt{v_1v_2v_3}+\epsilon_1\sqrt{\frac{v_1}{v_2v_3}}-
\epsilon_2\sqrt{\frac{v_2}{v_1v_3}}-
\epsilon_3\sqrt{\frac{v_3}{v_1v_2}}\bigg{)}}^2\tau_{og}\nn\\
&&+{\bigg{(}\sqrt{v_1v_2v_3}-\epsilon_1\sqrt{\frac{v_1}{v_2v_3}}+
\epsilon_2\sqrt{\frac{v_2}{v_1v_3}}-
\epsilon_3\sqrt{\frac{v_3}{v_1v_2}}\bigg{)}}^2\tau_{of}\nn\\
&&+{\bigg{(}\sqrt{v_1v_2v_3}-\epsilon_1\sqrt{\frac{v_1}{v_2v_3}}-
\epsilon_2\sqrt{\frac{v_2}{v_1v_3}}+
\epsilon_3\sqrt{\frac{v_3}{v_1v_2}}\bigg{)}}^2\tau_{oh}\bigg{\}}.
\end{eqnarray}
In the above expression, it is seen that the charges for the
orientifold planes can be changed in accordance to particular
model classes of the parameter (\ref{eqn:Epsilon}).


\subsection{Open Decedents}\label{sec:OpenSpectrum}


By incorporating the signs $\epsilon_j$ into the annulus, one can
introduce brane supersymmetry breaking.  As such, it is convenient
to use the compact notation
\begin{eqnarray}\label{eqn:charsusybreak}
\tilde{T}_{nm}^{(\epsilon_i)}=T_{nm}^{NS}-\epsilon_i T_{nm}^{R}.
\end{eqnarray}
Strings that couple to brane--antibrane pairs provide character
sets that now differ from the usual supersymmetric ones
(\ref{eqn:char}). Under $S$ transformation, characters of the form
$\tilde{T}_{nm}^{(-1)}$ go to $T^{(-1)}_{nm}$ which are the same
as in (\ref{eqn:char}) except for the changes of
$O_2\leftrightarrow V_2$ and $S_2\leftrightarrow C_2$ in the last
three factors. These sets are shown in equation
(\ref{eqn:NonSUSYChar}). The characters that correspond to
$\tilde{T}_{nm}^{(+1)}$ are simply denoted $T_{nm}$.
\begin{eqnarray}\label{eqn:NonSUSYChar}
\tau_{oo}^{(-1)}&=&O_2O_2O_2O_2+V_2V_2V_2V_2-C_2S_2S_2S_2-S_2C_2C_2C_2 \nn\\
\tau_{og}^{(-1)}&=&V_2V_2O_2O_2+O_2O_2V_2V_2-S_2C_2S_2S_2-C_2S_2C_2C_2 \nn\\
\tau_{oh}^{(-1)}&=&V_2O_2O_2V_2+O_2V_2V_2O_2-S_2S_2S_2C_2-C_2C_2C_2S_2 \nn\\
\tau_{of}^{(-1)}&=&V_2O_2V_2O_2+O_2V_2O_2V_2-S_2S_2C_2S_2-C_2C_2S_2C_2 \nn\\
\nn\\
\tau_{go}^{(-1)}&=&O_2O_2S_2C_2+V_2V_2C_2S_2-C_2S_2V_2O_2-S_2C_2O_2V_2 \nn\\
\tau_{gg}^{(-1)}&=&V_2V_2S_2C_2+O_2O_2C_2S_2-C_2S_2O_2V_2-S_2C_2V_2O_2 \nn\\
\tau_{gh}^{(-1)}&=&V_2O_2S_2S_2+O_2V_2C_2C_2-S_2S_2V_2V_2-C_2C_2O_2O_2 \nn\\
\tau_{gf}^{(-1)}&=&V_2O_2C_2C_2+O_2V_2S_2S_2-C_2C_2V_2V_2-S_2S_2O_2O_2 \nn\\
\nn\\
\tau_{ho}^{(-1)}&=&O_2S_2C_2O_2+V_2C_2S_2V_2-S_2O_2V_2C_2-C_2V_2O_2S_2 \nn\\
\tau_{hg}^{(-1)}&=&V_2C_2C_2O_2+O_2S_2S_2V_2-S_2O_2O_2S_2-C_2V_2V_2C_2 \nn\\
\tau_{hh}^{(-1)}&=&V_2S_2C_2V_2+O_2C_2S_2O_2-C_2O_2V_2S_2-S_2V_2O_2C_2 \nn\\
\tau_{hf}^{(-1)}&=&V_2S_2S_2O_2+O_2C_2C_2V_2-S_2V_2V_2S_2-C_2O_2O_2C_2 \nn\\
\nn\\
\tau_{fo}^{(-1)}&=&O_2S_2O_2C_2+V_2C_2V_2S_2-C_2V_2S_2O_2-S_2O_2C_2V_2 \nn\\
\tau_{fg}^{(-1)}&=&V_2C_2O_2C_2+O_2S_2V_2S_2-S_2O_2S_2O_2-C_2V_2C_2V_2 \nn\\
\tau_{fh}^{(-1)}&=&V_2S_2O_2S_2+O_2C_2V_2C_2-S_2V_2S_2V_2-C_2O_2C_2O_2 \nn\\
\tau_{ff}^{(-1)}&=&V_2S_2V_2C_2+O_2C_2O_2S_2-S_2V_2C_2O_2-C_2O_2S_2V_2. \nn\\
\end{eqnarray}

Moreover, as will be seen, this arrangement defines a perfect
square structure for the annulus $R$--$R$ sector that agrees with
that of the Klein amplitude, and will thus facilitate consistent
$R$--$R$ tadpole cancellation, as is required.

The differences in the closed amplitudes associated with the sign
freedom $\epsilon=\pm 1$ have been demonstrated.  The different
choices lead to quite distinct open amplitudes with very different
phenomenological characteristics.

In the transverse channel, the annulus is interpreted as closed
string states propagating between $D$--branes, that in this case
are either $D9$ or $D5_k$ branes.  As was done in the construction
of the annulus of the previous case
($\mathbb{Z}_2\times\mathbb{Z}_2^\text{s}$), one obtains the
transverse annulus from the torus partition function.  The
previous case consisted of characters of $SO(8)\rightarrow
SO(4)^2$ which from equations (\ref{eqn:ThetaS}) and
(\ref{eqn:ThetaC}) can be seen to be real. With the breaking
$SO(8)\rightarrow SO(2)^4$ the corresponding characters $S_2$ and
$C_2$ are complex.  The transverse annulus states are of the form
\begin{eqnarray}\label{eqn:TimeOp}
\langle{\cal T}(S)|q^{L_0}|S\rangle
\end{eqnarray}
(where $\cal T$ is a time reversal operator).  The operator $\cal
T$ maps complex values to their conjugates, thus interchanging
$S_2$ and $C_2$.

It is obvious from (\ref{eqn:TorusZero}) and (\ref{eqn:char}) that
for $\epsilon=-1$, such terms which are of the form
(\ref{eqn:TimeOp}) exist for both the untwisted and twisted
sectors of the torus.  For the choice $\epsilon=+1$, it is seen
that one can only retain such characters in the untwisted sector.
Consequently, the open descendants of the two models $\epsilon=\pm
1$ have very different structures.


\subsection{Models Without Discrete Torsion
$(\epsilon=+1)$}\label{sec:MWDT}


The subclass of models is generated by $\epsilon=(+,+,+)$, where
$\epsilon_k=+1$, and $(+,-,-)$ with two additional permutations
$(-,+,-)$ and $(-,-,+)$.  As has been shown in
(\ref{eqn:charsusybreak}), the presence of any $\epsilon_k=-1$
breaks supersymmetry.

The transverse annulus amplitude is defined by
\begin{eqnarray}\label{eqn:TransverseAnnulusWT}
\tilde{{\cal
A}}=&\frac{2^{-5}}{16}&\bigg{\{}\bigg{(}N^2v_1v_2v_3(W^1W^2W^3+
W^1_{n+\frac{1}{2}}W^2_{n+\frac{1}{2}}W^3_{n+\frac{1}{2}})\nn\\
&&+D^2_{ko}\frac{v_k}{2v_lv_s}W^kP^lP^s\frac{\big{(}1+(-1)^{m_l+m_s}\big{)}}{2}
\bigg{)}T_{oo}\nn\\
&&+2D_{ko}Nv_kW^k\bigg{(}\frac{2\eta}{\theta_2}\bigg{)}^2\tilde{T}_{ok}^{(\epsilon_k)}\nn\\
&&+D_{ko}D_{lo}\frac{1}{v_s}\frac{P^s(1+(-1)^s)}{2}\bigg{(}\frac{2\eta}{\theta_2}\bigg{)}^2\tilde{T}_{os}^{(\epsilon_k
\epsilon_l )}
\end{eqnarray}
where the construction follows from that done in the
$\mathbb{Z}_2\times\mathbb{Z}_2^\text{s}$ orientifold case, with the
addition of strings stretched between the sets of $D5_k-D5_l$
branes.  The brane configuration is of course similar to that of
figure \ref{dia:Z2xZ2branes} (after relevant $T$--dualities) where
one now has instead three species of $D5$ branes.

The numerical coefficients are constrained by the requiring that
(\ref{eqn:AnnulusOrigin1}) is obeyed. The origin of the lattice
towers shows the perfect square structure as
\begin{eqnarray}
\tilde{{\cal
A}}_{o}&=&\frac{2^{-5}}{16}\bigg{\{}{\bigg{(}N\sqrt{v_1v_2v_3}+
D_{go}\sqrt{\frac{v_1}{v_2v_3}}+D_{fo}\sqrt{\frac{v_2}{v_lv_3}}+
D_{ho}\sqrt{\frac{v_3}{v_1v_2}}\bigg{)}}^2\tau_{oo}^{NS}\nn\\ \nn\\
&&-{\bigg{(}N\sqrt{v_1v_2v_3}+\epsilon_1D_{go}\sqrt{\frac{v_1}{v_2v_3}}
+\epsilon_2D_{fo}\sqrt{\frac{v_2}{v_1v_3}}+\epsilon_3D_{ho}
\sqrt{\frac{v_3}{v_1v_2}}\bigg{)}}^2\tau_{oo}^{R}\nn\\ \nn\\
&&+{\bigg{(}N\sqrt{v_1v_2v_3}+D_{go}\sqrt{\frac{v_1}{v_2v_3}}
-D_{fo}\sqrt{\frac{v_2}{v_1v_3}}-D_{ho}\sqrt{\frac{v_3}{v_1v_2}}
\bigg{)}}^2\tau_{og}^{NS}\nn
\end{eqnarray}
\begin{eqnarray}\label{eqn:AnnulusOrigin1}
&&-{\bigg{(}N\sqrt{v_1v_2v_3}+\epsilon_1D_{go}\sqrt{\frac{v_1}{v_2v_3}}
-\epsilon_2D_{fo}\sqrt{\frac{v_2}{v_1v_3}}-\epsilon_3D_{ho}
\sqrt{\frac{v_3}{v_1v_2}}\bigg{)}}^2\tau_{og}^{R}\nn\\ \nn\\
&&+{\bigg{(}N\sqrt{v_1v_2v_3}-D_{go}\sqrt{\frac{v_1}{v_2v_3}}
+D_{fo}\sqrt{\frac{v_2}{v_1v_3}}-D_{ho}\sqrt{\frac{v_3}{v_1v_2}}
\bigg{)}}^2\tau_{of}^{NS}\nn\\ \nn\\
&&-{\bigg{(}N\sqrt{v_1v_2v_3}-\epsilon_1D_{go}\sqrt{\frac{v_1}{v_2v_3}}
+\epsilon_2D_{fo}\sqrt{\frac{v_2}{v_1v_3}}-\epsilon_3D_{ho}
\sqrt{\frac{v_3}{v_1v_2}}\bigg{)}}^2\tau_{of}^{R}\nn\\ \nn\\
&&+{\bigg{(}N\sqrt{v_1v_2v_3}-D_{go}\sqrt{\frac{v_1}{v_2v_3}}-
D_{fo}\sqrt{\frac{v_2}{v_1v_3}}+D_{ho}\sqrt{\frac{v_3}{v_1v_2}}
\bigg{)}}^2\tau_{oh}^{NS}\nn\\ \nn\\
&&-{\bigg{(}N\sqrt{v_1v_2v_3}-\epsilon_1D_{go}\sqrt{\frac{v_1}{v_2v_3}}
-\epsilon_2D_{fo}\sqrt{\frac{v_2}{v_1v_3}}+\epsilon_3D_{ho}\sqrt{\frac{v_3}
{v_1v_2}}\bigg{)}}^2\tau_{oh}^{R}\bigg{\}}.\nn\\
\end{eqnarray}
An $S$ transform shows the direct channel amplitude to be
\begin{eqnarray}\label{eqn:DirectA}
{\cal A}&=&\frac{1}{16} \bigg\{\bigg(N^2P_1P_2P_3\big(1+(-1)^{m_1+m_2+m_3}\big)\nn\\
&&+\frac{1}{4}D^2_{ko}P^k\big(W^lW^s+W^l_{n+\frac{1}{2}}W^s_{n+\frac{1}{2}}\big)\bigg)T_{oo}\nn\\
&&+2D_{ko}NP^k\bigg{(}\frac{\eta}{\theta_4}\bigg{)}^2T_{ko}^{(\epsilon_k)}\nn\\
&&\frac{1}{2}D_{ko}D_{lo}\big(W^s+W^s_{n+\frac{1}{2}}\big)\bigg{(}
\frac{\eta}{\theta_4}\bigg{)}^2T_{so}^{(\epsilon_k \epsilon_l
)}\bigg\}.
\end{eqnarray}
\begin{table}
\begin{center}
\begin{tabular}{|c|c|}\hline
Plane diagrams & Volumes \\ \hline $D9-D9$, $D9-O9$,
$O9-O9$ & $v_1v_2v_3$ \\
$D5_k-D5_k$, $D5_k-O5_k$, $O5_k-O5_k$ & $\frac{v_k}{v_lv_m}$ \\
$D5_k-D5_l$, $D5_k-O5_l$, $O5_k-O5_l$ & $\frac{1}{v_m}$ \\
$D9-D5_k$, $D9-O5_k$, $D5_k-O9$ & $v_k$ \\ \hline \hline
$\tilde{\cal A}$ and $\tilde{\cal K}$ Plane Diagrams & Lattice
Couplings
\\\hline
$D9-D9$ & $W^1W^2W^3+W^1_{n+\frac{1}{2}}W^2_{n+\frac{1}{2}}
W^3_{n+\frac{1}{2}}$
\\
$D5_k-D5_k$ & $W^kP^lP^m\big(1+(-1)^{m_l+m_m}\big)$ \\
$O9-O9$ & $W^1_eW^2_eW^3_e+W^1_oW^2_oW^3_o$ \\
$O5_k-O5_k$ & $W^kP^l_eP^m_e$ \\ \hline \hline $\tilde{\cal M}$
plane diagrams & Lattice Couplings
\\\hline
$D9-O9$ & $W^1_eW^2_eW^3_e+W^1_oW^2_oW^3_o$ \\
$D9-O5_k$ & $W^k$ \\
$D5_k-O9$ & $W^k_e$ \\
$D5_k-O5_k$ & $W^k_eP^l_eP^m_e+(-1)^{m_l+m_m}W^k_oP^l_eP^m_e$ \\
$D5_k-O5_l$ & $P^m_e$ \\\hline
\end{tabular}
\end{center}
\caption{Lattice restrictions}\label{tab:cc}
\end{table}
Having fixed the relevant factors in the annulus, the Mobius can
now be constructed.  The Mobius is dressed with lattice states in
the same manner as that of the previous model.  However, with the
shift acting in all tori, the phase that was introduced to give
correct direct channel annulus-Mobius symmetrization is extended
accordingly.  The appropriate Mobius term is shown in table
\ref{tab:cc}.

With the introduction of the signs $\epsilon_j$ into the annulus
and Klein amplitudes, the Mobius shows these at its lattice origin
accordingly:
\begin{eqnarray}\label{eqn:mobzm}
\tilde{{\cal
M}}_o&=&\pm\frac{1}{8}\bigg{\{}Nv_1v_2v_3\hat{T}_{oo}+\epsilon_k
D_{ko}\frac{v_k}{2v_lv_m}\hat{\tilde{T}}_{oo}^{(\epsilon_k)}\nn\\
&&+\epsilon_kNv_k\hat{T}_{ok}+D_{ko}v_k
\hat{\tilde{T}}_{ok}^{(\epsilon_k)}+\epsilon_mD_{lo}\frac{1}{v_k}
\hat{\tilde{T}}_{ok}^{(\epsilon_l)}\bigg{\}}.
\end{eqnarray}

With the correct lattice terms, one finds the full transverse
Mobius amplitude to be
\begin{eqnarray}
\tilde{{\cal
M}}&=&-\frac{1}{8}\bigg{\{}Nv_1v_2v_3(W^1_eW^2_eW^3_e+W^1_oW^2_oW^3_o)
\hat{T}_{oo}\nn\\
&&+\epsilon_kD_{ko}\frac{v_k}{2v_lv_s}\big(W^k_eP^l_eP^s_e+(-1)^{m_l+m_s}
W^k_oP^l_eP^s_e\big)\hat{\tilde{T}}_{oo}^{(\epsilon_k)}\nn
\end{eqnarray}
\begin{eqnarray}
&&+\big(\epsilon_kNv_kW^k\hat{T}_{ok}+D_{ko}v_kW^k_e
\hat{\tilde{T}}_{ok}^{(\epsilon_k)}\big){\bigg(\frac{2\hat{\eta}}
{\hat{\theta}_2}\bigg)}^2\nn\\
&&+\epsilon_mD_{lo}\frac{P^k_e}{v_k}\hat{\tilde{T}}_{ok}^{(\epsilon_l)}
{\bigg(\frac{2\hat{\eta}}{\hat{\theta}_2}\bigg)}^2\bigg{\}}
\end{eqnarray}
where the sign ambiguity is $-1$ for consistent tadpole
cancellation.

The corresponding direct channel is obtained by $P$
transformation. It is noted that while $P$ has non trivial effect
of the lattice modes, it leaves the characters unchanged with the
exception of a sign change for the orbifold sector. This can be
seen by the representation defined in (\ref{eqn:Pmatrix}).  As
such
\begin{eqnarray}\label{eqn:DirectM}
{\cal
M}&=&-\frac{1}{16}\bigg{\{}NP^1P^2P^3\big(1+(-1)^{m_1+m_2+m_3}\big)
\hat{T}_{oo}\nn\\
&&+\epsilon_k\frac{1}{2}D_{ko}\big(P^kW^lW^s+(-1)^{m_k}P^k
W^l_{n+\frac{1}{2}}W^s_{n+\frac{1}{2}}\big)\hat{\tilde{T}}_{oo}^{(\epsilon_k)}\nn\\
&&-\big(2\epsilon_kNP^k_e\hat{T}_{ok}+D_{ko}P^k
\hat{\tilde{T}}_{ok}^{(\epsilon_k)}\big){\bigg(\frac{2\hat{\eta}}{\hat{\theta}_2}
\bigg)}^2\nn\\
&&-\epsilon_mD_{lo}W^k\hat{\tilde{T}}_{ok}^{(\epsilon_l)}{\bigg(\frac{2\hat{\eta}}
{\hat{\theta}_2}\bigg)}^2\bigg{\}}.
\end{eqnarray}

The tadpole conditions for the $D9$ branes are
\begin{eqnarray}\label{eqn:Ntadpole}
\frac{2^5}{16}+\frac{2^{-5}}{16}N^2-\frac{N}{8}=0~\Rightarrow~N=32.
\end{eqnarray}
It is seen that the tadpole conditions for the $D5$ branes in the
$NS$ and $R$ sector cannot lead to mutual cancellation for cases
other than $(+,+,+)$. The tadpole for $N$, as seen by equation
(\ref{eqn:Ntadpole}), is unaffected by this. However, allowing the
cancellation of the $R$--$R$ sector forces a tree level dilaton
tadpole correlated with the creation of a potential for the
$NS$--$NS$ sector. This has an interpretation of increased vacuum
energy. The tadpoles arising form the $R$--$R$ sector must be
satisfied in order to allow the total $R$--$R$ charge to be
cancelled.

{From} the amplitudes, one finds
\begin{eqnarray}\label{eqn:Dtadpole}
D_{ko}^{(NS)}=\epsilon_k32{\rm ,}~D_{ko}^{(R)}=32.
\end{eqnarray}
%
%


\subsection{Model Classes of $\epsilon=+1$}


The direct channel amplitudes defined by equations
(\ref{eqn:DirectA}) and (\ref{eqn:DirectM}) require a rescaling of
$N\rightarrow 2n$ and $D_{ko}\rightarrow 4d_k$ to give proper
symmetrization. As such, one now has
\begin{eqnarray}\label{eqn:AnnulusRescaled}
{\cal A} &=&\frac{1}{4}\bigg\{\bigg(n^2P_1P_2P_3\big(1+(-1)^{m_1+m_2+m_3}\big)\nn\\
&&+d^2_{k}P^k\big(W^lW^s+W^l_{n+\frac{1}{2}}W^s_{n+\frac{1}{2}}\big)\bigg)T_{oo}\nn\\
&&+4d_knP^k\bigg{(}\frac{\eta}{\theta_4}\bigg{)}^2T_{ko}^{(\epsilon_k)}\nn\\
&&+4d_kd_l\frac{\big(W^s+W^s_{n+\frac{1}{2}}\big)}{2}
\bigg{(}\frac{\eta}{\theta_4}\bigg{)}^2T_{mo}^{(\epsilon_k\epsilon_l)}\bigg\}
\end{eqnarray}
and
\begin{eqnarray}\label{eqn:MobiusRescaled}
{\cal
M}&=&-\frac{1}{8}\bigg{\{}nP^1P^2P^3\big(1+(-1)^{m_1+m_2+m_3}\big)
\hat{T}_{oo}\nn\\
&&+\epsilon_kd_k\big(P^kW^lW^s+(-1)^{m_k}P^k
W^l_{n+\frac{1}{2}}W^s_{n+\frac{1}{2}}\big)\hat{\tilde{T}}_{oo}^{(\epsilon_k)}\nn\\
&&-\big(2\epsilon_knP^k_e\hat{T}_{ok}+2d_kP^k
\hat{\tilde{T}}_{ok}^{(\epsilon_k)}\big){\bigg(\frac{2\hat{\eta}}{\hat{\theta}_2}
\bigg)}^2\nn\\
&&-2\epsilon_md_lW^k\hat{\tilde{T}}_{ok}^{(\epsilon_l)}{\bigg(\frac{2\hat{\eta}}
{\hat{\theta}_2}\bigg)}^2\bigg{\}}.
\end{eqnarray}
Here, it appears that some massive modes, in particular the
$(-1)^{m_k}P^kW^l_{n+\frac{1}{2}}W^s_{n+\frac{1}{2}}$ towers do
not symmetrize with proper numerical factors. The annulus and
Mobius are required to symmetrize modulo two. In this case, one
has a multiplicity of the common states in the Mobius and annulus
as
\begin{eqnarray}
2^3\times\bigg(\frac{d_k^2}{4}-\frac{d_k}{8}\bigg)=3\frac{d_k(d_k-1)}{2}+\frac{d_k(d_k+1)}{2}
\end{eqnarray}
where the multiplicity of $2^3$ comes from the interchange of the
indices $l$ and $s$ and the degeneracy of massive states under an
orbifold element or $\Omega$ as $\alpha:n+\frac{1}{2}\rightarrow
-n-\frac{1}{2}$.  So one finds that the group interpretation is
preserved as the decomposition into three orthogonal copies and
one simplectic.

Firstly, I discuss the simplest and fully supersymmetric case of
$(+,+,+)$.  The expanded lattice origin form of the Mobius
amplitude is shown in appendix \ref{app:moborigin}. The massless
open spectra is given by
\begin{eqnarray}\label{eqn:AnnulusMobiusOrigin1}
{\cal A}_o+{\cal M}_o
=&&\bigg[\frac{n(n+1)}{2}+\frac{d_{g}(d_{g}+1)}{2}\nn\\
&&+\frac{d_{f}(d_{f}+1)}{2}+\frac{d_{h}(d_{h}+1)}{2}\bigg]\tau_{oo}\nn\\
&&+\bigg[\frac{n(n-1)}{2}+\frac{d_{g}(d_{g}-1)}{2}\nn\\
&&+\frac{d_f(d_f-1)}{2}+\frac{d_h(d_h-1)}{2}\bigg](\tau_{og}+\tau_{of}+\tau_{oh})\nn\\
&&+(nd_g+d_fd_h)(\tau_{go}+\tau_{gg}+\tau_{gf}+\tau_{gh})\nn\\
&&+(nd_f+d_gd_h)(\tau_{fo}+\tau_{fg}+\tau_{ff}+\tau_{fh})\nn\\
&&+(nd_h+d_gd_f)(\tau_{ho}+\tau_{hg}+\tau_{hf}+\tau_{hh}).
\end{eqnarray}
The vector multiplet, contained in $\tau_{oo}$, combined with the
tadpole conditions (\ref{eqn:Ntadpole}) and (\ref{eqn:Dtadpole}),
and with the rescaling $N=2n$ and $D_{ko}=4d_k$, shows the gauge
group to be $USp(16)_9\times USp(8)_{5_{\{1,2,3\}}}$.  The
suffixes refer to the groups of the $D9$ and three copies of $D5$.
{From} (\ref{eqn:char}), one can see that $N=1$ with chiral
multiplets that arise in the untwisted sector from $\tau_{ok}$ and
from the twisted sector in $\tau_{gf}$, $\tau_{hg}$ and
$\tau_{fg}$. This model therefore has chiral multiplets in the
representations described in table \ref{tab:ChiralReps1}.
\begin{table}[!ht]
\begin{center}
\begin{tabular}{|l|l|l|}\hline
Sector & Characters & Reps. in $(D9;D5_1,D5_2,D5_3)$ \\
\hline Twisted & $\tau_{gf}+\tau_{gh}$ & $(16;8,1,1)+(1;1,8,8)$
\\
& $\tau_{fg}+\tau_{fh}$ & $(16;1,8,1)+(1;8 ,1,8)$
\\
& $\tau_{hg}+\tau_{hf}$ & $(16;1,1,8)+(1;8,8,1)$
\\\hline
Untwisted & $\tau_{ok}$ & $(120;1,1,1)+(1;28,1,1)$ \\
& & $+(1;1,28,1)+(1;1,1,28)$
\\\hline
\end{tabular}
\end{center}
\caption{Chiral multiplet representations for
$\epsilon=(+,+,+)$}\label{tab:ChiralReps1}
\end{table}

The remaining cases break supersymmetry for states coupling to
$\bar{D}5_k$ branes, or antibranes that are aligned with the
directions that satisfy $\epsilon_k=-1$.  These are permutations
of $(+,-,-)$.

For $(+,-,-)$, one has low lying spectrum
\begin{eqnarray}
{\cal A}_o+{\cal M}_o
=&&\bigg[\frac{n(n-1)}{2}+\frac{d_g(d_g-1)}{2}\bigg](\tau_{oo}+\tau_{of}+\tau_{oh})\nn\\
&&+\bigg[\frac{n(n+1)}{2}+\frac{d_g(d_g+1)}{2}\bigg]\tau_{og}\nn\\
&&+\bigg[\frac{d_f(d_f-1)}{2}+\frac{d_h(d_h-1)}{2}\bigg]
\tau_{og}^{\rm NS}\nn\\
&&+\bigg[\frac{d_f(d_f+1)}{2}+\frac{d_h(d_h+1)}{2}\bigg]
\tau_{og}^{\rm R}\nn
\end{eqnarray}
\begin{eqnarray}\label{eqn:AnnulusMobiusOrigin2}
&&+\bigg[\frac{d_f(d_f+1)}{2}+\frac{d_h(d_h+1)}{2}\bigg]
(\tau_{oo}^{\rm NS}+\tau_{of}^{\rm NS}+\tau_{oh}^{\rm NS})\nn\\
&&+\bigg[\frac{d_f(d_f-1)}{2}+\frac{d_h(d_h-1)}{2}\bigg]
(\tau_{oo}^{\rm R}+\tau_{of}^{\rm R}+\tau_{oh}^{\rm R})\nn\\
&&+(nd_g+d_fd_h)(\tau_{go}+\tau_{gg}+\tau_{gf}+\tau_{gh})\nn\\
&&+(nd_f+d_gd_h)(\tau_{fo}^{(-)}+\tau_{fg}^{(-)}+\tau_{ff}^{(-)}+\tau_{fh}^{(-)})\nn\\
&&+(nd_h+d_gd_f)(\tau_{ho}^{(-)}+\tau_{hg}^{(-)}+\tau_{hf}^{(-)}+\tau_{hh}^{(-)}).
\end{eqnarray}
\begin{table}[!ht]
\begin{center}
\begin{tabular}{|l|l|l|}\hline
Sector & Characters & Reps. in $(D9;D5_1,D5_2,D5_3)$ \\
\hline Twisted & $\tau_{gf}+\tau_{gh}$ & $(16;8,1,1)+(1;1,8,8)$
\\\hline
Untwisted & $\tau_{og}$ & $(136;1,1,1)+(1;36,1,1)$
\\
& $\tau_{of}$ & $(120;1,1,1)+(1;28,1,1)$
\\
& $\tau_{oh}$ & $(120;1,1,1)+(1;28,1,1)$
\\
\hline
\end{tabular}
\end{center}
\caption{Chiral multiplet representations for
$\epsilon=(+,-,-)$}\label{tab:ChiralReps2}
\end{table}
Supersymmetry is seen to be broken in this expansion in a twofold
way.  Firstly, the representations for the Neveu-Schwarz and
Ramond terms are different in the untwisted sectors.  Secondly,
the presence of signs $\epsilon_k$ forces character sets to
transform differently under $S$, as can be seen by reference to
(\ref{eqn:NonSUSYChar}).  So the $f$ and $h$ twisted characters
are no longer supersymmetric.

In this case, the gauge group is $SO(16)_9\times SO(8)_{5_1}\times
USp(8)_{\bar{5}_{\{2,3\}}}$. The representations of the chiral
multiplets in this model are displayed in table
\ref{tab:ChiralReps2}.
\begin{table}[!ht]
\begin{center}
\begin{tabular}{|l|l|l|}\hline
Sector & Characters & Reps. in $(D9;D5_1,D5_2,D5_3)$ \\
\hline Twisted & $\tau_{hg}+\tau_{hf}$ & $(16;1,1,8)+(1;8,8,1)$
\\\hline
Untwisted & $\tau_{og}$ & $(120;1,1,1)+(1;1,1,28)$
\\
& $\tau_{of}$ & $(120;1,1,1)+(1;1,1,28)$
\\
& $\tau_{oh}$ & $(136;1,1,1)+(1;1,1,36)$
\\
\hline
\end{tabular}
\end{center}
\caption{Chiral multiplet representations for
$\epsilon=(-,-,+)$}\label{tab:ChiralReps4}
\end{table}
For $(-,+,-)$, the low lying modes correspond to
\begin{eqnarray}
{\cal A}_o+{\cal M}_o
=&&\bigg[\frac{n(n-1)}{2}+\frac{d_f(d_f-1)}{2}\bigg]
(\tau_{oo}+\tau_{og}+\tau_{oh})\nn\\
&&+\bigg[\frac{n(n+1)}{2}+\frac{d_f(d_f+1)}{2}\bigg]
\tau_{of}\nn\\
&&+\bigg[\frac{d_g(d_g-1)}{2}+\frac{d_h(d_h-1)}{2}\bigg]
\tau_{of}^{\text{NS}}\nn\\
&&+\bigg[\frac{d_g(d_g+1)}{2}+\frac{d_h(d_h+1)}{2}\bigg]
\tau_{of}^{\text{R}}\nn
\end{eqnarray}
\begin{eqnarray}
&&+\bigg[\frac{d_g(d_g+1)}{2}+\frac{d_h(d_h+1)}{2}\bigg]
(\tau_{oo}^{\text{NS}}+\tau_{og}^{\text{NS}}
+\tau_{oh}^{\text{NS}})\nn\\
&&+\bigg[\frac{d_g(d_g-1)}{2}+\frac{d_h(d_h-1)}{2}\bigg]
(\tau_{oo}^{\text{R}}+\tau_{og}^{\text{R}}
+\tau_{oh}^{\text{R}})\nn\\
&&+(nd_g+d_fd_h)(\tau_{go}^{(-)}+\tau_{gg}^{(-)}+\tau_{gf}^{(-)}+\tau_{gh}^{(-)})\nn\\
&&+(nd_f+d_gd_h)(\tau_{fo}+\tau_{fg}+\tau_{ff}+\tau_{fh})\nn\\
&&+(nd_h+d_gd_f)(\tau_{ho}^{(-)}+\tau_{hg}^{(-)}+\tau_{hf}^{(-)}+\tau_{hh}^{(-)})
\end{eqnarray}
\begin{table}[!ht]
\begin{center}
\begin{tabular}{|l|l|l|}\hline
Sector & Characters & Reps. in $(D9;D5_1,D5_2,D5_3)$ \\
\hline Twisted & $\tau_{fg}+\tau_{fh}$ & $(16;1,8,1)+(1;8,1,8)$
\\\hline
Untwisted & $\tau_{og}$ & $(120;1,1,1)+(1;1,28,1)$
\\
& $\tau_{of}$ & $(136;1,1,1)+(1;1,36,1)$
\\
& $\tau_{oh}$ & $(120;1,1,1)+(1;1,28,1)$
\\
\hline
\end{tabular}
\end{center}
\caption{Chiral multiplet representations for
$\epsilon=(-,+,-)$}\label{tab:ChiralReps3}
\end{table}
with gauge group $SO(16)_9\times USp(8)_{\bar{5}_1}\times
SO(8)_{5_2}\times USp(8)_{\bar{5}_{3}}$ and representations for
the chiral multiplets are displayed in table
\ref{tab:ChiralReps3}.

Finally, the $(-,-,+)$ model gives rise to
\begin{eqnarray}
{\cal A}_o+{\cal M}_o
=&&\bigg[\frac{n(n-1)}{2}+\frac{d_h(d_h-1)}{2}\bigg]
(\tau_{oo}+\tau_{og}+\tau_{of})\nn\\
&&+\bigg[\frac{n(n+1)}{2}+\frac{d_h(d_h+1)}{2}\bigg]
\tau_{oh}\nn\\
&&+\bigg[\frac{d_g(d_g-1)}{2}+\frac{d_f(d_f-1)}{2}\bigg]
\tau_{oh}^{\text{NS}}\nn\\
&&+\bigg[\frac{d_g(d_g+1)}{2}+\frac{d_f(d_f+1)}{2}\bigg]
\tau_{oh}^{\text{R}}\nn\\
&&+\bigg[\frac{d_g(d_g+1)}{2}+\frac{d_f(d_f+1)}{2}\bigg]
(\tau_{oo}^{\text{NS}}+\tau_{og}^{\text{NS}}
+\tau_{of}^{\text{NS}})\nn\\
&&+\bigg[\frac{d_g(d_g-1)}{2}+\frac{d_f(d_f-1)}{2}\bigg]
(\tau_{oo}^{\text{R}}+\tau_{og}^{\text{R}}
+\tau_{of}^{\text{R}})\nn\\
&&+(nd_g+d_fd_h)(\tau_{go}^{(-)}+\tau_{gg}^{(-)}+\tau_{gf}^{(-)}+\tau_{gh}^{(-)})\nn\\
&&+(nd_f+d_gd_h)(\tau_{fo}^{(-)}+\tau_{fg}^{(-)}+\tau_{ff}^{(-)}+\tau_{fh}^{(-)})\nn\\
&&+(nd_h+d_gd_f)(\tau_{ho}+\tau_{hg}+\tau_{hf}+\tau_{hh})
\end{eqnarray}
with $SO(16)_9\times USp(8)_{\bar{5}_{\{1,2\}}}\times SO(8)_{5_3}$
gauge group.  Representations for the chiral multiplets are
displayed in table \ref{tab:ChiralReps4}.

This then exhausts all possible configurations of the
$\epsilon=+1$ models.  I now turn to those with discrete torsion.


\subsection{Models With Discrete Torsion $(\epsilon=-1)$}


The oriented open sector for this class of models is far richer
than those without discrete torsion.  By reference to
(\ref{eqn:TorusZero}), one has the existence of left moving states
coupled to their corresponding conjugates. In this case, in
addition to the states in (\ref{eqn:TransverseAnnulusWT}) there
are twisted sectors.

In addition to the transverse untwisted states defined by
(\ref{eqn:TransverseAnnulusWT}), one now has
\begin{eqnarray}
\tilde{{\cal
A}}&=&\frac{2^{-5}}{16}\bigg{\{}\bigg{(}N_o^2v_1v_2v_3(W^1W^2W^3+
W^1_{n+\frac{1}{2}}W^2_{n+\frac{1}{2}}W^3_{n+\frac{1}{2}})\nn\\
&&+\frac{v_k}{2v_lv_s}D^2_{ko}W^kP^lP^m\frac{\big{(}1+(-1)^{m_l+m_s}\big{)}}{2}
\bigg{)}T_{oo}\nn\\
&&+\bigg[M_1N_k^2v_k(W^k+W^k_{n+\frac{1}{2}})\nn\\
&&+M_2D_{kk}^2v_kW^k+M_3D^2_{lk}\frac{P^k}{v_k}\bigg]T_{ko}{\bigg{(}\frac{\eta}{\theta_4}\bigg{)}}^2\nn\\
&&+2N_oD_{ko}v_kW^k\tilde{T}_{ok}^{(\epsilon_k)}{\bigg{(}\frac{2\eta}
{\theta_2}\bigg{)}}^2\nn\\
&&+M_4N_kD_{kk}v_kW^k\tilde{T}_{kk}^{(\epsilon_k)}{\bigg{(}\frac{\eta}
{\theta_3}\bigg{)}}^2\nn\\
&&+M_5N_lD_{kl}\tilde{T}_{lk}^{(\epsilon_k)}\frac{8{\eta}^3}{\theta_2
\theta_3\theta_4}\nn\\
&&+D_{ko}D_{lo}\frac{P^s}{v_s}\frac{\big{(}1+(-1)^{m_s}\big{)}}{2}
\tilde{T}_{os}^{(\epsilon_k\epsilon_l)}{\bigg{(}\frac{2\eta}{\theta_2}
\bigg{)}}^2\nn\\
&&+M_{6}D_{km}D_{lm}\frac{P^m}{v_m}\tilde{T}_{mm}^{(\epsilon_k
\epsilon_l)}{\bigg{(}\frac{\eta}{\theta_3}\bigg{)}}^2\nn\\
&&+M_{7}D_{kk}D_{lk}\tilde{T}_{km}^{(\epsilon_k\epsilon_l)}
\frac{8{\eta}^3}{\theta_2\theta_3\theta_4}\bigg{\}}.
\end{eqnarray}
The coefficients $M_i$ are determined from the origin of the
twisted sector with the same arguments as used for the
$\mathbb{Z}_2\times\mathbb{Z}_2^\text{s}$ model.

The $N_g$ term fills all compact and non-compact dimensions and thus
has the coefficient $v_k$.  With the volume $v_k$ being provided by
the remaining compact directions that are not acted on by an
orbifold.  When considering the factors involved with terms like
$D_{kl}$, $l$ represents the fixed point configuration of $T^2_{45}
\times T^2_{67} \times T^2_{89}$ and $k$ represents whether the
brane is wrapped or transverse. For example, $D_{gf}$ has fixed
points in the first and third torus corresponding to $f$. The index
$g$ implies that the $D_{gf}$ brane is wrapped around the first tori
and is transverse to the second and third, consistent with the
representation $g=(+,-,-)$. Hence, it \textit{sees} four fixed
points.

Looking at the $g$-twisted sector of the $\epsilon=(+,+,-)$ model,
this has a total of sixteen fixed points, four located in each of
the second and third tori. Under the operation of the shift, half
are identified. The independent fixed points are as in table
\ref{tab:shiftedFixedPoints}.

For the $g$-twisted sector, one has terms in the annulus as
\begin{eqnarray}
\tilde{{\cal
A}}^g=&\frac{2^{-5}}{16}&\bigg{\{}\bigg[(M_1N_g^2+M_2D_{gg}^2)v_1+
M_3D^2_{lg}\frac{1}{v_1}\bigg]T_{go}\nn\\ \nn\\
&&+M_4N_gD_{gg}v_1\tilde{T}_{gg}^{(\epsilon_1)}\nn\\ \nn\\
&&+4M_5N_gD_{kg}\tilde{T}_{gk}^{(\epsilon_k)}\nn\\ \nn\\
&&+M_{6}D_{kg}D_{lg}\frac{1}{v_1}\tilde{T}_{gg}^{(\epsilon_k\epsilon_l)}
\nn\\ \nn\\
&&+4M_{7}D_{gg}D_{lg}\tilde{T}_{gm}^{(\epsilon_k\epsilon_l)}\bigg\}.
\end{eqnarray}

All brane types $N_g$, $D_{gg}$, $D_{fg}$ and $D_{hg}$ see the
fixed point $(0,0;0,0)$, and therefore arrange into a perfect
square with multiplicity 1.  The arguments set out in the simpler
$\mathbb{Z}_2\times \mathbb{Z}_2^\text{s}$ model with regard to
the wrapping of $D5$ branes is generalized here with the inclusion
of three distinct types. The counting of their fixed point
occupation is then a little more complicated. $N_g$ and $D_{hg}$
see the fixed points $(0,0,\frac{1}{2},0)$, $(0,0;0,\frac{1}{2})$
and $(0,0;\frac{1}{2},\frac{1}{2})$, which correspond to their own
perfect square with multiplicity 3.  The coefficients of the $N_g$
and $D_{hg}$ terms are 1 and 2 respectively, which can easily be
seen by reference to (\ref{eqn:FixedPointSummary}).  Similar holds
for the square of $N_g$ and $D_{fg}$. The remaining fixed points
are taken into account by $N_g$ alone.

The resulting perfect square structures for the $NS$--$NS$ and
$R$--$R$ portions of $\tau_{gl}$ are
\begin{eqnarray}
&& 2\times\frac{2^{-5}}{16}\bigg{\{}{\big{(}\sqrt{v_1}N_g-
4\sqrt{v_1}D_{gg}-2\frac{1}{\sqrt{v_1}}D_{fg}+2
\frac{1}{\sqrt{v_1}}D_{hg}\big{)}}^2\nn\\ \nn\\
&&+3{\big{(}\sqrt{v_1}N_g-2\frac{1}{\sqrt{v_1}}D_{fg}\big{)}}^2
+3{\big{(}\sqrt{v_1}N_g+2\frac{1}{\sqrt{v_1}}D_{hg}\big{)}}^2+
v_1N_g^2\bigg{\}},
\end{eqnarray}
and
\begin{eqnarray}
&&2\times\frac{2^{-5}}{16}\bigg{\{}{\big{(}\sqrt{v_1}N_g-
4\sqrt{v_1}D_{gg}-2\frac{1}{\sqrt{v_1}}D_{fg}-2
\frac{1}{\sqrt{v_1}}D_{hg}\big{)}}^2\nn\\ \nn\\
&&+3{\big{(}\sqrt{v_1}N_g-2\frac{1}{\sqrt{v_1}}D_{fg}\big{)}}^2
+3{\big{(}\sqrt{v_1}N_g-2\frac{1}{\sqrt{v_1}}D_{hg}\big{)}}^2+
v_1N_g^2\bigg{\}}
\end{eqnarray}
respectively.  Similar results follow for the same procedure in
the $f$ and $h$ twisted sectors.  The overall factor of 2 is to
account for the multiplicity of the shifted fixed points in the
same fashion as for the $\mathbb{Z}_2\times \mathbb{Z}_2^\text{s}$
case.

With the aid of the identity
\begin{eqnarray}
\theta_2 \theta_3 \theta_4 = 2\eta^3,\nn
\end{eqnarray}
the transverse annulus is now seen to be
\begin{eqnarray}
\tilde{{\cal
A}}=&\frac{2^{-5}}{16}&\bigg{\{}\bigg{(}N_o^2v_1v_2v_3
(W^1W^2W^3+W^1_{n+\frac{1}{2}}W^2_{n+\frac{1}{2}}
W^3_{n+\frac{1}{2}})\nn\\
&&+\frac{v_k}{2v_lv_s}D^2_{ko}W^kP^lP^s
\frac{\big{(}1+(-1)^{m_l+m_s}\big{)}}{2}\bigg{)}T_{oo}\nn\\
&&+2\times2\bigg{[}N_k^2v_k(W^k+W^k_{n+\frac{1}{2}})\nn\\
&&+2D_{kk}^2v_kW^k+2D^2_{lk}\frac{P^k}{v_k}\bigg{]}
T_{ko}{\bigg{(}\frac{2\eta}{\theta_4}\bigg{)}}^2\nn\\
&&+2N_oD_{ko}v_kW^k\tilde{T}_{ok}^{(\epsilon_k)}
{\bigg{(}\frac{2\eta}{\theta_2}\bigg{)}}^2\nn\\
&&+2\times2N_kD_{kk}v_kW_k\tilde{T}_{kk}^{(\epsilon_k)}
{\bigg{(}\frac{2\eta}{\theta_3}\bigg{)}}^2\nn\\
&&+2\times4N_lD_{kl}\tilde{T}_{lk}^{(\epsilon_k)}
\frac{8{\eta}^3}{\theta_2\theta_3\theta_4}\nn\\
&&+D_{ko}D_{lo}\frac{P^s}{v_s}\frac{\big{(}1+(-1)^{m_s}\big{)}}{2}
\tilde{T}_{os}^{(\epsilon_k\epsilon_l)}{\bigg{(}\frac{2\eta}{\theta_2}
\bigg{)}}^2\nn\\
&&+2D_{km}D_{lm}\frac{P^m}{v_m}\tilde{T}_{mm}^{(\epsilon_k\epsilon_l)}
{\bigg{(}\frac{2\eta}{\theta_3}\bigg{)}}^2\nn\\
&&+2\times4D_{kk}D_{lk}\tilde{T}_{km}^{(\epsilon_k\epsilon_l)}
\frac{8{\eta}^3}{\theta_2\theta_3\theta_4}\bigg{\}}.
\end{eqnarray}
With corresponding direct channel
\begin{eqnarray}
{\cal
A}=&\frac{1}{16}&\bigg{\{}\bigg{(}N_o^2P^1P^2P^3\big(1+(-1)^{m_1+m_2+m_3}\big)
\nn\\
&&+\frac{1}{2}\frac{D^2_{ko}}{2}P^k(W^lW^m+W^l_{n+\frac{1}{2}}
W^m_{n+\frac{1}{2}}\bigg{)}T_{oo}\nn\\
&&+\bigg{[}N_k^2P^k\big(1+(-1)^{m_k}\big)\nn\\
&&+2D_{kk}^2P^k+2D^2_{lk}W^k\bigg{]}T_{ok}{\bigg{(}\frac{2\eta}{\theta_2}
\bigg{)}}^2\nn\\
&&+2N_oD_{ko}P^k
T_{ko}^{(\epsilon_k)}{\bigg{(}\frac{\eta}{\theta_4}\bigg{)}}^2\nn\\
&&-2\times2N_kD_{kk}P^kT_{kk}^{(\epsilon_k)}{\bigg{(}\frac{\eta}{\theta_3}
\bigg{)}}^2\nn\\
&&+2\times2i(-1)^{k+l}N_lD_{kl}T_{kl}^{(\epsilon_k)}
\frac{2{\eta}^3}{\theta_2\theta_3\theta_4}\nn\\
&&+\frac{1}{2}D_{ko}D_{lo}(W^m+W^m_{n+\frac{1}{2}})
T_{mo}^{(\epsilon_k\epsilon_l)}{\bigg{(}\frac{\eta}{\theta_4}\bigg{)}}^2
\nn
\end{eqnarray}
\begin{eqnarray}\label{eqn:dirannorigin}
&&-2\times D_{km}D_{lm}W^mT_{mm}^{(\epsilon_k\epsilon_l)}{\bigg{(}
\frac{\eta}{\theta_3}\bigg{)}}^2\nn\\
&&+2\times2i(-1)^{m+k}D_{kk}D_{lk}T_{mk}^{(\epsilon_k\epsilon_l)}
\frac{2{\eta}^3}{\theta_2\theta_3\theta_4}\bigg{\}}.
\end{eqnarray}

It is here that consistent particle interpretation does not occur.
The inconsistency is generated in the $D5_i-D5_j$ (for $i\neq j$)
sector.  All other sectors give rise to the proper massless and
massive counting.  Moreover, the problem exists in the twisted
sector that has to symmetrize by itself, as the Mobius has only
untwisted sectors present.  The Mobius has the same form as for
the case without discrete torsion, up to signs $\epsilon_k$ that
in this case are required to give $\epsilon=-1$ (with reference to
equation (\ref{eqn:Epsilon})), and a different definition of the
Chan--Paton charges.

To begin with one must define the Chan-Paton charge
parameterization which is given in table \ref{tab:ModelCharges}
for the case in consideration of $\epsilon=(+,+,-)$. As with the
$\mathbb{Z}_2\times \mathbb{Z}_2^\text{s}$ case, the
parameterization involves factors of the form $n$, $\bar{n}$,
etc$\ldots$. These apply for branes that are aligned with
$\epsilon_k=+1$ (i.e. $D_{kl}$ and $N_{k}$ for $l=1,2,3$) for the
cases which are permutations of $(+,+,-)$.

As with the previous models, the factors of two for the terms in
each $D_{ko}$ are necessary and induced by the shift to give
proper integer particle interpretation.  All sectors that involve
a coupling to a $D9$ brane are consistent.

\begin{table}
\begin{center}
\begin{tabular}{|llllll|}\hline
$N_o$ & $=$ & $(n+m+\bar{n}+\bar{m})$, & $N_g$ & $=$ &
$i(n+m-\bar{n}-\bar{m})$
\\
$N_f$ & $=$ & $i(n-m-\bar{n}+\bar{m})$, & $N_h$ & $=$ &
$(n-m+\bar{n}-\bar{m})$\\
$D_{go}$ & $=$ & $2(o_1+g_1+\bar{o}_1+\bar{g}_1)$, & $D_{fo}$ &
$=$ &
$2(o_2+g_2+\bar{o}_2+\bar{g}_2)$\\
$D_{ho}$ & $=$ & $2(a+b+c+d)$, & $D_{gg}$ & $=$ &
$i(o_1+g_1-\bar{o}_1-\bar{g}_1)$\\
$D_{ff}$ & $=$ & $i(o_2+g_2-\bar{o}_2-\bar{g}_2)$, & $D_{hh}$ &
$=$ &
$a-b-c+d$\\
$D_{gf}$ & $=$ & $o_1-g_1+\bar{o}_1-\bar{g}_1$, & $D_{gh}$ & $=$ &
$-i(o_1-g_1-\bar{o}_1+\bar{g}_1)$\\
$D_{fg}$ & $=$ & $o_2-g_2+\bar{o}_2-\bar{g}_2$, & $D_{fh}$ & $=$ &
$i(o_2-g_2-\bar{o}_2+\bar{g}_2)$\\
$D_{hg}$ & $=$ & $a+b-c-d$, & $D_{hf}$ & $=$ & $a-b+c-d$\\\hline
\end{tabular}
\end{center}
\caption{$\epsilon=(1,1,-1)$ Model
Charges}\label{tab:ModelCharges}
\end{table}
With this parameterization, the untwisted sector provides
consistent massless spectrum as
\begin{eqnarray}\label{egn:AnplusMobzero}
{\cal A}_o+{\cal
M}_o&=&(n\bar{n}+m\bar{m}+g_1\bar{g}_1+o_1\bar{o}_1+o_2\bar{o}_2+g_2\bar{g}_2)
\tau_{oo}\nn\\
&&+(n\bar{m}+m\bar{n}+o_1\bar{g}_1+g_1\bar{o}_1+ab+cd)\tau_{og}\nn\\
&&+(nm+\bar{n}\bar{m}+o_2\bar{g}_2+g_2\bar{o}_2+ac+bd)\tau_{of}\nn\\
&&+(\bar{o}_1\bar{g}_1+o_1g_1+o_2g_2+\bar{o}_2\bar{g}_1+ad+bc)\tau_{oh}\nn\\
&&+\frac{\big(a(a+1)+b(b+1)+c(c+1)+d(d+1)\big)}{2}\tau_{oo}^{NS}\nn\\
&&+\frac{\big(a(a-1)+b(b-1)+c(c-1)+d(d-1)\big)}{2}\tau_{oo}^{R}\nn\\
&&+\frac{\big(o_2(o_2-1)+g_2(g_2-1)+\bar{o}_2(\bar{o}_2-1)+\bar{g}_2
(\bar{g}_2-1)\big)}{2}\tau_{og}\nn
\end{eqnarray}
\begin{eqnarray}
&&+\frac{\big(o_1(o_1-1)+g_1(g_1-1)+\bar{o}_1(\bar{o}_1-1)+\bar{g}_1
(\bar{g}_1-1)\big)}{2}\tau_{of}\nn\\
&&+\frac{\big(n(n-1)+m(m-1)+\bar{n}(\bar{n}-1)+
\bar{m}(\bar{m}-1)\big)}{2}\tau_{oh}.
\end{eqnarray}
Indeed, the twisted massless spectrum also gives rise to a
consistent particle interpretation in all sectors $g$, $f$ and
$h$.  The spectrum obtained from the twisted $D9$--$D5$ couplings
also allows consistent state counting for all mass levels.

Turning lastly to the $D5_i-D5_j$ sector, one has consistency for
massless and all integer massive levels.  However, for the
$n+\frac{1}{2}$ massive states one has the term
\begin{eqnarray}
\frac{1}{2}D_{fo}D_{lo}W^m_{n+\frac{1}{2}}
T_{mo}^{(\epsilon_k\epsilon_l)}{\bigg{(}\frac{\eta}{\theta_4}\bigg{)}}^2.
\end{eqnarray}
Any combination of the operators $g$, $f$, $h$ and $\Omega$ will
map $W_{n+\frac{1}{2}}$ to $W_{\pm(n+\frac{1}{2})}$, this winding
tower will then have a degeneracy of two.  Taking into account the
interchange counting $k\leftrightarrow l$ and the rescaling
defined in table \ref{tab:ModelCharges}, the end result is a state
with numerical coefficient of $\frac{1}{2}$.

The same term occurred in the model without discrete torsion in
equation (\ref{eqn:AnnulusRescaled}).  In that case, it did not
cause any inconsistency because of the generic rescaling
$N\rightarrow 2n$ and $D\rightarrow 2d$ in addition to the
rescaling induced by the freely acting shift.  In the models with
discrete torsion, such a rescaling is taken into account (for the
integer massive and massless levels) by the presence of the
breaking terms $N_k, D_{kk},\ldots$.

In the transverse annulus, one has the freedom to introduce Wilson
lines via phases of the form $e^{2\pi i \alpha}$, for $\alpha\in
(0,1)$. If such phases are introduced, the resulting amplitude
must respect symmetrization in the direct channel and the
corresponding terms in the transverse annulus must exist in the
torus. Introducing phases in the group of $D5_i-D5_j$ terms that
must symmetrize together as
\begin{eqnarray}\label{eqn:Problem}
&&\frac{1}{2}D_{ko}D_{lo}(W^m+W^m_{n+\frac{1}{2}})
T_{mo}^{(\epsilon_k\epsilon_l)}{\bigg{(}\frac{\eta}{\theta_4}\bigg{)}}^2
\nn\\
&&-2\times D_{km}D_{lm}W^mT_{mm}^{(\epsilon_k\epsilon_l)}{\bigg{(}
\frac{\eta}{\theta_3}\bigg{)}}^2\nn\\
&&+2\times2i(-1)^{m+k}D_{kk}D_{lk}T_{mk}^{(\epsilon_k\epsilon_l)}
\frac{2{\eta}^3}{\theta_2\theta_3\theta_4}
\end{eqnarray}
only leads to amplitudes that violate this requirement, or indeed
cause the same problem of incorrect counting at other mass levels.

To illustrate this point, I begin with a phase in the first term
as
\begin{eqnarray}
D_{ko}D_{lo}P^s\frac{\big(1+(-1)^{m_s^1+\beta m_s^2}\big)}{2}
\quad\overrightarrow{S}\quad
\frac{1}{2}D_{ko}D_{lo}\big(W^s+W_{n_1+\frac{1}{2},n_2+\frac{\beta}{2}}^s\big).
\end{eqnarray}
Where $m_s^1$ and $m_s^2$ are the momentum quantum numbers on the
first and second directions of the torus $T^s$. For $\beta=1$, the
momentum lattice in the transverse channel has expanded form
\begin{eqnarray}
P^s_{2m_1,2m_2}+P^s_{2m_1+1,2m_2+1}
\end{eqnarray}
which includes odd states that do not exist in the torus.  For
$\beta\neq 1$, the projection which leads to the counting of even
momentum states, as required by the torus, no longer exists unless
$\beta=0~({\rm mod}~2)$.  Therefore, the introduction of a phase
for this term has to be one which multiplies the whole lattice
expression.  Doing this will however lift the massless spectrum in
the direct channel amplitude of terms associated with
$D_{ko}D_{lo}$ couplings.

Allowing a phase for the second term in (\ref{eqn:Problem}), while
solving the $n+\frac{1}{2}$ state counting problem, would then
cause a similar effect of non--integer particle counting for the
massless states.

The third term does not carry any momenta or winding states, and
so any phase possibilities are only applicable to the first two.

I reiterate here that the spectrum is consistent at massive and
massless levels for all other couplings in this model, consistency
only breaks down for the $D5_{i}D5_{j}$ couplings.


\section{Discussion}


The spectra of freely acting orbifolds with non-freely acting
winding and or Kaluza Klein shifts have been exhaustively studied
in \cite{AADS} as (shift) orbifolds. The inclusion of shift
operators within the $\mathbb{Z}_2\times \mathbb{Z}_2$ orbifold
generators leads to (in the cases with two $D5$ branes, and some
models with only one $D5$ brane) richer geometries.  In
particular, such cases involve shifted fixed points, which give
rise to unique massive lattices of the form $W_{n+\frac{1}{4}}$ in
the direct channel annulus. In addition, the arrangement of shifts
within the $\mathbb{Z}_2\times \mathbb{Z}_2$ generators imposes
restrictions on the number of distinct $D5$ branes that can exist.
The models defined by $\mathbb{Z}_2\times \mathbb{Z}_2$ (shift)
projections that essentially exhaust all interesting
configurations are defined \cite{AADS} by
\begin{eqnarray}\label{eqn:(shift)orbifolds}
\sigma_1(\delta_1,\delta_2,\delta_3)= \left(\begin{array}{rrr}
\delta_1 & -\delta_2 & -1 \\ -1 & \delta_2 & -\delta_3 \\
-\delta_1 & -1 & \delta_3
\end{array}\right),\quad
\sigma_2(\delta_1,\delta_2,\delta_3)= \left(\begin{array}{rrr}
\delta_1 & -1 & -1 \\ -1 & \delta_2 & -\delta_3 \\ -\delta_1 &
-\delta_2 & \delta_3
\end{array}\right).
\end{eqnarray}
The parameters $\delta_i$ are winding or momentum shifts. Wherever
a $\delta$ operation exists in a column, the corresponding brane
is eliminated.

In the freely acting shift models, all $D5$ branes are allowed to
exist but have a more conventional geometry with regard to their
relative placements.

The closed string states in the $\mathbb{Z}_2\times \mathbb{Z}_2$
(shift) models are not as rich as those in the $\mathbb{Z}_2\times
\mathbb{Z}_2\times \mathbb{Z}_2^\text{s}$ models.  This is
especially evident in that such (shift) orbifolds do not allow
contributions from $\mathbb{Z}_2\times \mathbb{Z}_2$ orbits that
lie outside $S$ and $T$ transformations on the principle orbits
$(o,o),(o,g),(o,f)$ and $(o,h)$ (as illustrated in appendix
\ref{app:Boundaries}).

With the inclusion of independent orbits, one has a class of
models which exhibit possible scenarios of supersymmetry breaking
(according to the sign freedom associated with the independent
modular orbits). The cases without discrete torsion which include
$(+,+,+)$, $(+,-,-)$, $(-,+,-)$ and $(-,-,+)$ lead to fully
consistent amplitudes with ${N}=1$ supersymmetry in the open
sector with brane supersymmetry breaking associated with strings
attached to a $\bar{D}5_k$ antibrane (which are aligned with the
directions corresponding to $\epsilon_k=-1$).

There is however an unresolved problem of consistent particle
interpretation for cases with discrete torsion for $n+\frac{1}{2}$
massive modes stretched between distinct $D5$ branes.  The
spectral content is entirely consistent for the counting of string
states which do not include $D5_i$--$D5_j$ couplings.  However, at
the time of writing, a solution to this problem has not been
found.

The sign $\epsilon$ associated with the inclusion of the
additional independent orbits is a freedom.  There is no mechanism
outlined yet that guides the choice of which model is preferred,
other than perhaps phenomenological requirements. In contrast to
the (shift) orbifold models, although the closed spectrum is not
as rich in such cases, they do eliminate this freedom.

Unitary groups are obtained only from the models in the
$\epsilon=-1$ in using the underlying geometry
$T^6(\mathbb{Z}_2\times\mathbb{Z}_2)$. In many of the possible
arrangements of (shift) orbifold models, the corresponding torus
amplitude allows the propagation of twisted states that give three
different models with unitary gauge groups (for models with both
$D9$'s and $D5$'s present).  However, none of these has all three
twisted sectors contributing to the massless spectrum.  Use of the
freely acting shift allows one to keep the three twisted sectors
at massless level.

\setcounter{footnote}{0}

\chapter{Magnetic Deformation of the Type I
$\mathbb{Z}_2\times \mathbb{Z}_2$ Model with Discrete
Torsion}\label{ch:MagDeform}

In this last discussion of type I phenomenology, I look at the
interesting scenario of including background magnetic fields in
the generic $\mathbb{Z}_2\times\mathbb{Z}_2$ model (see appendix
\ref{app:Z2xZ2Generic}). In particular, it is the model with
discrete torsion ($\epsilon=-1$).  It will be shown that
magnetizing this model, which has twisted terms in the transverse
annulus, causes complications that force the model to be
inconsistent.

In addition, I will review the instabilities inherent with
magnetizing the $\mathbb{Z}_2\times\mathbb{Z}_2$ structure.

\section{Magnetic Background Construction for Bosons}

The effects of introducing background magnetic fields are
discussed in \cite{CIES}, \cite{CAAS} and in the bosonic case
\cite{ACNY}.  I review the basic concepts in the context of the
bosonic string here, so that the construction of the
$\mathbb{Z}_2\times\mathbb{Z}_2$ model is more transparent.

The terms in the standard bosonic action are supplemented with
additional terms that allow a gauge field to interact with the
string end points.  The full bosonic action is
\begin{eqnarray}\label{eqn:action}
S_{\text{bosonic}}&=&-\frac{1}{4\pi \alpha^{\prime}}\int
d^2\sigma\partial_\alpha X_\mu \partial^\alpha X^\mu\nn\\
&&-q_L\int d\tau A_\mu
\partial_{\tau}X^{\mu}\bigg{|}_{\sigma=0}
-q_R\int d\tau A_\mu
\partial_{\tau}X^{\mu}\bigg{|}_{\sigma=\pi}
\end{eqnarray}
with $A_\mu=-\frac{1}{2}F_{\mu \nu}X^\nu$ for a constant field
tensor $F_{\mu \nu}$. In addition to the well know wave equation
for the fields $X^\mu$, one also has boundary conditions that now
mix Dirichlet and Neumann conditions
\begin{eqnarray}\label{eqn:BoundaryConditions}
&&\partial_{\sigma}X^\mu-2\pi \alpha^{\prime}q_LF^\mu_\nu
\partial_{\tau}X^\nu=0,\quad\sigma=0,\nn\\
&&\partial_{\sigma}X^\mu+2\pi\alpha^{\prime}q_RF^\mu_\nu
\partial_{\tau}X^\nu=0,\quad\sigma=\pi.
\end{eqnarray}

The action (\ref{eqn:action}) describes the effect of open strings
coupling to boundaries that carry the source charge for such
magnetic fields.  As such, these boundaries are considered as
magnetic monopoles, where the left and right boundaries carry the
charges $q_L$ and $q_R$ respectively.

After the complex redefinitions $Z=\frac{1}{\sqrt{2}}(X^1+iX^2)$
and $\bar{Z}=\frac{1}{\sqrt{2}}(X^1-iX^2)$ in the $j^\text{th}$
torus, in the case that the total charge
\begin{eqnarray}\label{eqn:TotalCharge}
Q=q_L+q_R
\end{eqnarray}
is nonzero, the corresponding wave function has shifted mode
numbers. This can be seen from the wave function
\begin{eqnarray}\label{eqn:wavefunction}
\psi_n(\tau,\sigma)=\frac{1}{\sqrt{|n-\xi_j|}}cos[(n-\xi_j)\sigma+\gamma_j]
e^{-i(n-\xi_j)\tau},
\end{eqnarray}
with $\gamma_j=tan^{-1}(2\pi\alpha^{\prime}q_L H_j)$, of the
string coordinate
\begin{eqnarray}
Z=z+i\sqrt{2\alpha'}\left[\sum_{n=1}a_n
\psi_n(\tau,\sigma)-\sum_{m=0}b_m^\dag
\psi_{-n}(\tau,\sigma)\right].
\end{eqnarray}
Where $\xi$ is defined as
\begin{eqnarray}\label{eqn:ZetaFunction}
\xi_j=\frac{1}{\pi}(tan^{-1}(2\pi\alpha^{\prime}q_L
H_j)+tan^{-1}(2\pi\alpha^{\prime}q_R H_j)).
\end{eqnarray}

This will result in a modification to the components of the energy
momentum tensor aligned with the magnetized torus, which defines
the relevant parts of the zero Laurent mode as \cite{ACNY}
\begin{eqnarray}
L_0=\sum_{m=1}^{\infty}(m-\xi_j)a^{\dag}_m . a_m
+\sum_{m=0}^{\infty}(m+\xi_j)b^{\dag}_m . b_m.
\end{eqnarray}
The general modes will have a modified Virasoro algebra
\begin{eqnarray}
\big[L_n,L_m\big]=(n-m)L_{n+m}+\delta_{n+m,0}\big[\frac{c}{12}
(n^3-n)+n\xi_j(1-\xi_j)\big],
\end{eqnarray}
with an additional $c$ number piece.  The original algebra
\cite{ACNY} is recovered with the redefinition of the zero mode
$L_o'=L_o+\frac{1}{2}\xi_j(1-\xi_j)$.  Hence, the contribution to
states in the partition function is now
\begin{eqnarray}\label{eqn:ShifedModes}
\text{Tr}q^{L_0^{\prime}}=-i\left(q^{\frac{1}{24}}\right)^2
q^{-\frac{1}{2}\xi^2_j}
\bigg(\frac{k_j\eta}{\theta_1(\xi_j\tau|\tau)}\bigg),
\end{eqnarray}
Including the presence of an orbifold operation and twisted modes,
the results of (\ref{eqn:LatticeModesTheta's}) also deform in a
similar way to (\ref{eqn:ShifedModes}).

The theta function $\theta_1$ is defined by equations
(\ref{eqn:Theta's}) and (\ref{eqn:ThetaProExp}) with argument
$z=\xi\tau$ in the direct channel.

The numbers $k_j$ arise from the non--commutativity of the zero
modes and the quantization condition (\ref{eqn:QuantCond}) that
allows a degeneracy of exactly $k_j$ for states in of the
$j^\text{th}$ torus.  Such numbers will only be present for the
bosonic modes that are untwisted and not accompanied by an
orbifold operation. Twisted strings will not have Landau levels as
they do not have zero modes.

In what follows, I will use the shorthand notation
\begin{eqnarray}\label{eqn:notation}
\alpha_j=2\pi\alpha'qH_j.
\end{eqnarray}
for the $j^\text{th}$ torus

In the case that the total charge is zero $q_L=-q_R$, which
represent a dipole string, the situation is a little more subtle.
The oscillator frequencies are no longer shifted as $\xi=0$,
although there is still a phase shift according to
(\ref{eqn:wavefunction}). As such, the spectrum for this solution
is that of a string in the absence of magnetic fields, up to
momenta that are now boosted as
\begin{eqnarray}
\tilde{p}=\frac{m}{R\sqrt{1+\alpha^2}}.
\end{eqnarray}
This rescaling ensures the consistency of the massless transverse
channel amplitude where contributions should form perfect squares.

As such, there will be momenta and winding lattices $\tilde{P}^j$
and $\tilde{W}^j$ that simply denote the conventional lattices
with correspondingly altered radii.  Under $S$ transformation, one
has
\begin{eqnarray}
S:\tilde{P}^j\rightarrow
\frac{v_j}{2}\big(1+\alpha^2_j\big)\tilde{W}^j
\end{eqnarray}

For untwisted modes, the change in mass caused by the magnetic
deformation in the direct channel relates to the field $H_j$ in
the following way
\begin{eqnarray}\label{eqn:MassShift}
\triangle M^2\sim\sum_{j=1,2,3}\big\{
(2n_j+1)|2\pi\alpha'(q_L+q_R)H_j|+4\pi\alpha'(q_L+q_R)\Sigma_j
H_j\big\}.
\end{eqnarray}
The first term originates from the Landau levels as can be seen
from taking into account the contributions from the deformed world
sheet bosons.  The second arises from the magnetic moments due to
spin $\Sigma_j$ which arise from the world sheet fermions.  In the
case of contributions from the twisted sector, the Landau terms
are absent.  As will be discussed, with the absence of the term
$(2n_j+1)|2\pi \alpha^\prime (q_L+q_R)H_j|$, problems with respect
to stability in $\mathbb{Z}_2\times \mathbb{Z}_2$ models will
arise. For such cases, tachyonic modes will appear when the second
term in $\triangle M^2$ is negative.

\subsection{Magnetically Deformed Fermions}

I review the details of the action on fermionic states by magnetic
fields \cite{B}.

In addition to the bosonic action defined in (\ref{eqn:action}),
fermions behave, with the introduction of magnetic fields,
according to
\begin{eqnarray}
S_{\text{fermionic}}&=&\frac{i}{4\pi\alpha'}\int d^2\sigma
\bar{\psi}^\mu\gamma^\beta\partial_\beta \psi_\mu
\nn\\
&&+\frac{iq_L}{2}\int d\tau F_{\nu \mu}\bar{\psi}^\nu \gamma^0
\psi^\mu \bigg|_{\sigma=0}+\frac{iq_R}{2}\int d\tau F_{\nu
\mu}\bar{\psi}^\nu \gamma^0 \psi^\mu \bigg|_{\sigma=\pi}.
\end{eqnarray}
Where the matrices $\gamma^\beta$ ($\beta=0,1$) satisfy the relation
$\{\gamma_\alpha,\gamma_\beta\}=2h_{\alpha\beta}$.  In particular,
one has
\begin{equation}
\gamma^0=\left(\begin{array}{rr} 1 & 0 \\
0 & 1
\end{array}\right),\quad
\gamma^1=\left(\begin{array}{rr} 1 & 0 \\
0 & -1
\end{array}\right).
\end{equation}

The variation of $S_{\text{fermionic}}$ with respect to the fields
gives the following relations between the left and right moving
parts as
\begin{eqnarray}\label{eqn:MagFermBC}
\psi^\mu_L+(-1)^a \psi^\mu_R&=&-\pi\alpha'q_R
F_\nu^\mu(\psi^\nu_R-(-1)^a \psi^\nu_L)\quad(\sigma=\pi),\nn\\
\psi^\mu_L-\psi^\mu_R&=&\pi\alpha'q_L F_\nu^\mu(\psi^\nu_R+
\psi^\nu_L)\quad\quad\quad~~~~(\sigma=0).
\end{eqnarray}
For a fermion written as $\psi=(\psi_R,\psi_L)^\text{T}$ with NS
or R boundary condition $a=0,1$ respectively.  The coordinates,
when made complex using the two coordinates of a torus
$\psi_{L,R}^{\pm}=\frac{1}{\sqrt{2}}(\psi_{L,R}\pm i\psi_{L,R})$,
reduce (\ref{eqn:MagFermBC}) to
\begin{eqnarray}\label{eqn:MagFermBCR}
-(-1)^a\psi_R^{\pm}&=&\frac{(1\mp i\alpha_R)}{(1\pm
i\alpha_R)}\psi^{\pm}_L\quad~(\sigma=\pi),
\nn\\
\psi_R^{\pm}&=&\frac{(1\pm i\alpha_L)}{(1\mp
i\alpha_L)}\psi^{\pm}_L \quad~(\sigma=0),
\end{eqnarray}
using the notation (\ref{eqn:notation}) with charges $q_L$ and
$q_R$.  The wave functions that satisfy these conditions are
\begin{eqnarray}\label{eqn:Wavfunctions}
\chi_{R,n}^{\pm}&=&\frac{1}{\sqrt{2}}exp\left[-i(n\pm
\xi)(\tau-\sigma_2)\pm itan^{-1}\alpha_L\right],\nn\\
\chi_{L,n}^{\pm}&=&\frac{1}{\sqrt{2}}exp\left[-i(n\pm
\xi)(\tau+\sigma_2)\mp itan^{-1}\alpha_L\right],
\end{eqnarray}
for the fermionic coordinates
\begin{eqnarray}
\psi_{L,R}^{\pm}(\sigma,\tau)&=&\sum_nd_n^{\pm}\chi^{\pm}_{(n)L,R}(\sigma,\tau).
\end{eqnarray}
The resulting $L_0$ mode is \cite{B}
\begin{eqnarray}
L_0=-\sum_{n\in\mathbb{Z}+\nu}(n+\xi):d^+_{-n}d^-_n:+\Delta
\end{eqnarray}
with normal ordering constants $\Delta=\frac{\xi^2}{2}$ (NS) and
$\Delta=\frac{1}{8}-\frac{\xi}{2}(1-\xi)$ (R).

The characters $O_{2}$, $V_{2}$, $S_{2}$ and $C_{2}$ are then
deformed in the direct channel \cite{CAAS} according to
\begin{eqnarray}\label{eqn:ThetaBreakings}
O_{2}(\xi)&=&\frac{q^{\frac{1}{2}\xi^2}}{2\eta(\tau)}
\big[\theta_3(\xi\tau|\tau)+\theta_4(\xi\tau|\tau)\big],\nn\\
V_{2}(\xi)&=&\frac{q^{\frac{1}{2}\xi^2}}{2\eta(\tau)}
\big[\theta_3(\xi\tau|\tau)-\theta_4(\xi\tau|\tau)\big],\nn\\
S_{2}(\xi)&=&\frac{q^{\frac{1}{2}\xi^2}}{2\eta(\tau)}
\big[\theta_2(\xi\tau|\tau)-i\theta_1(\xi\tau|\tau)\big],\nn\\
C_{2}(\xi)&=&\frac{q^{\frac{1}{2}\xi^2}}{2\eta(\tau)}
\big[\theta_2(\xi\tau|\tau)+i\theta_1(\xi\tau|\tau)\big].
\end{eqnarray}

The factor $\xi$ contains the charge term which is associated to
the magnetic charge of the $D9$ brane, to which the deformed open
string is attached.  In the case where an open string is attached
to two branes of like charge, this term appears in the theta
functions as $\pm 2\xi$.  Similarly, for strings stretched between
a neutral and charged brane, the contributions are $\pm \xi$.

In studying the tadpole terms, it is only necessary to look at the
low lying contributions associated with the deformed theta
functions. Such modes, which here are written in the transverse
channel, are given as
\begin{eqnarray}\label{eqn:ThetaDeformations}
\theta_1 (\pm n\xi_j|\tau)&=& \mp 2sin(n\pi
\xi_j)q^{\frac{1}{8}}\big(1-2qcos(2n\pi \xi_j)
+\ldots\big)\nn\\
\theta_2 (\pm n\xi_j|\tau)&=& 2cos(n\pi
\xi_j)q^{\frac{1}{8}}\big(1+2qcos(2n\pi \xi_j)
+\ldots\big)\nn\\
\theta_3 (\pm n\xi_j|\tau)&=& 1+2q^{\frac{1}{2}}cos(2n\pi \xi_j)+\ldots\nn\\
\theta_4 (\pm n\xi_j|\tau)&=& 1-2q^{\frac{1}{2}}cos(2n\pi
\xi_j)+\ldots
\end{eqnarray}
where $n$ takes the value of $1$ or $2$, in accordance with the
charged ends of the string.  Here $\xi_j=2\pi \alpha' qH_j$ (for
small $H_j$) and the charge $q$ is the absolute value of the
charge associated with a brane $m$ or $\bar{m}$.

The low lying character modes (\ref{eqn:ThetaBreakings}) in the
transverse channel reduce to
\begin{eqnarray}\label{eqn:CharDeformations1}
O_2 (\pm \xi_j) &\sim& O_2 (\pm 2\xi_j) \sim \frac{1}{\eta},\nn\\
V_2 (\pm \xi_j) &\sim&
\frac{q^{\frac{1}{2}}}{\eta}\bigg(\frac{1-\alpha^2_j}{1+\alpha^2_j}
\bigg),\nn\\
V_2 (\pm 2\xi_j) &\sim& \frac{q^{\frac{1}{2}}}{\eta}\bigg(
\frac{1-6\alpha_j^2+\alpha_j^4}{(1+\alpha_j^2)^2}\bigg),\nn\\
S_2 (\pm n\xi_j) &\sim& \frac{q^{\frac{1}{8}}}{\eta}\bigg(
\frac{1+(1-n)\alpha_j^2\pm in\alpha_j}{(1+\alpha_j^2)^\frac{n}{2}}\bigg),\nn\\
C_2 (\pm n\xi_j) &\sim& \frac{q^{\frac{1}{8}}}{\eta}\bigg(
\frac{1+(1-n)\alpha_j^2\mp
in\alpha_j}{(1+\alpha_j^2)^\frac{n}{2}}\bigg).
\end{eqnarray}
These relations illustrate how the tadpoles will be influenced by
particular couplings to magnetically charged branes.

Now that the details for the building of magnetically deformed
models have been discussed, I now turn to the case of the
$\mathbb{Z}_2\times\mathbb{Z}_2$ model.


\section{$(T^6/\mathbb{Z}_2\times \mathbb{Z}_2)(\mathbf{H})$ Structure}


The structure of the generators of the $\mathbb{Z}_2\times
\mathbb{Z}_2$ allows a rather rich spectral content in the
presence of background magnetic fields.  However, the breaking
terms that follow from the choice of models with discrete torsion
($\epsilon=-1$) provides inconsistencies. The second class of
models belonging to $\epsilon=+1$ has been reported in
\cite{MLGP}, which does not experience such problems, as will be
explained in more detail later.

To illustrate this, I only consider the case with two magnetized
tori. As such, the field tensor $F_{\mu \nu}$ has the block form
$(\mu=4,5,6,7)$
\begin{eqnarray}\label{eqn:background1}
\left( \begin{array}{cccc} 0 & H_1 & 0 & 0 \\
-H_1 & 0 & 0 & 0 \\
0 & 0 & 0 & H_2 \\
0 & 0 & -H_2 & 0 \\
\end{array}\right).
\end{eqnarray}

The model that is considered here is one with $\epsilon=(+,+,-)$.
As such, the Chan-Paton charges corresponding to $D9$ branes,
which have been defined for the model with freely acting shifts in
table \ref{tab:ModelCharges}, break according to
\begin{eqnarray}\label{eqn:breaking}
&&N_o'=n'+s+\bar{n}'+\bar{s}\rightarrow N_o+m+\bar{m}\nn\\
&&N_g'=i(n'+s-\bar{n}'-\bar{s})\rightarrow N_g+i(m-\bar{m})\nn\\
&&N_f'=i(n'-s-\bar{n}'+\bar{s})\rightarrow N_f+i(m-\bar{m})\nn\\
&&N_h'=n'-s+\bar{n}'-\bar{s}\rightarrow N_h+m+\bar{m}.
\end{eqnarray}
Where $n'=n+m$, $\bar{n}'=\bar{n}+\bar{m}$, the prime therefore
refers the original non--deformed stack of branes.

One begins with a stack of $l$ parent $D9$ branes, $2p$ of them
are then charged magnetically, $p$ with charge $q$ and the same
number of charge $-q$.  This allows the compact tori to support
fields from monopoles to have vanishing total charge. Any factors
assigned to the parent Chan-Paton charges are inherited by the
magnetically charged stacks, as shown in (\ref{eqn:breaking}).

The direct channel annulus that results from such breaking's of
the $D9$'s separates into three pieces. Firstly, the uncharged
sectors, or $Q=0$, are given by
\begin{eqnarray}
{\cal A}_{(Q=0)}&=&\frac{1}{8}\bigg\{\bigg(N_o^2P^1P^2P^3
+2m\bar{m}\tilde{P}_1\tilde{P}_2P_3\bigg)T_{oo}(0,0)
+\frac{D^2_{ko}}{2}P^kW^lW^sT_{oo}\nn\\
&&+\bigg(N_g^2P^1+2m\bar{m}\tilde{P}^1\bigg)T_{og}(0,0){\bigg{(}\frac{2\eta}
{\theta_2(0)}\bigg{)}}^2\nn\\
&&+\bigg(N_f^2P^2+2m\bar{m}\tilde{P}^2\bigg)T_{of}(0,0){\bigg{(}\frac{2\eta}
{\theta_2(0)}\bigg{)}}^2\nn\\
&&+\bigg(N_h^2P^3+2m\bar{m}P^3\bigg)T_{oh}(0,0){\bigg{(}\frac{2\eta}
{\theta_2(0)}\bigg{)}}^2\nn\\
&&+\bigg(D_{kk}^2P^k+D^2_{lk}W^k\bigg)T_{ok}{\bigg{(}\frac{2\eta}{\theta_2
}\bigg{)}}^2\nn\\
&&+2N_oD_{ko}P^kT_{ko}^{(\epsilon_k)}(0,0){\bigg{(}
\frac{\eta}{\theta_4}\bigg{)}}^2\nn\\
&&-2N_kD_{kk}P_kT_{kk}^{(\epsilon_k)}(0,0){\bigg{(}
\frac{\eta}{\theta_3}\bigg{)}}^2\nn\\
&&+2i(-1)^{k+l}N_lD_{kl}T_{kl}^{(\epsilon_k)}(0,0)
\frac{2{\eta}^3}{\theta_2\theta_3\theta_4}\nn\
\end{eqnarray}
\begin{eqnarray}\label{eqn:UnchargedAnnulus}
&&+D_{ko}D_{lo}W^sT_{so}^{(\epsilon_k\epsilon_l)}
{\bigg(\frac{\eta}{\theta_4}\bigg)}^2\nn\\
&&-D_{ks}D_{ls}W^sT_{ss}^{(\epsilon_k\epsilon_l)}
{\bigg{(}\frac{\eta}{\theta_3}\bigg{)}}^2\nn\\
&&+2i(-1)^{m+k}D_{kk}D_{lk}T_{mk}^{(\epsilon_k\epsilon_l)}\frac{2{\eta}^3}
{\theta_2\theta_3\theta_4}\bigg{\}}.
\end{eqnarray}
Then, the $|Q|=1$ terms are
\begin{eqnarray}\label{eqn:ChargedAnnulusQ1}
{\cal
A}_{(|Q|=1)}&=&\frac{1}{8}\bigg\{-2N_o(m+\bar{m})T_{oo}(\xi_1\tau,\xi_2\tau)
{\bigg{(}\frac{k_1\eta}{\theta_1(\xi_1\tau|\tau)}\bigg{)}}{\bigg{(}
\frac{k_2\eta}{\theta_1(\xi_2\tau|\tau)}\bigg{)}}P^3 \nn\\
&&+2N_g(m-\bar{m})T_{og}(\xi_1\tau,\xi_2\tau)
{\bigg(\frac{k_1\eta}{\theta_1(\xi_1\tau|\tau)}\bigg)}{\bigg(\frac{2\eta}
{\theta_2(\xi_2\tau|\tau)}\bigg)}
{\bigg{(}\frac{2\eta}{\theta_2(0)}\bigg{)}}\nn\\
&&+2N_f(m-\bar{m})T_{of}(\xi_1\tau,\xi_2\tau)
{\bigg(\frac{2\eta}{\theta_2(\xi_1\tau|\tau)}\bigg)}{\bigg(\frac{k_2\eta}
{\theta_1(\xi_2\tau|\tau)}\bigg)}
{\bigg{(}\frac{2\eta}{\theta_2(0)}\bigg{)}}\nn\\
&&+2N_h(m+\bar{m})T_{oh}(\xi_1\tau,\xi_2\tau){\bigg(\frac{2\eta}{\theta_2
(\xi_1\tau|\tau)}\bigg)}
{\bigg(\frac{2\eta}{\theta_2(\xi_2\tau|\tau)}\bigg)} P^3\nn\\
&&-2i(m+\bar{m})D_{go}T_{go}^{(\epsilon_1)}(\xi_1\tau,\xi_2\tau)
{\bigg(\frac{k_1\eta}{\theta_1(\xi_1\tau|\tau)}\bigg)}
{\bigg(\frac{\eta}{\theta_4(\xi_2\tau|\tau)}\bigg)}{\bigg(\frac{\eta}
{\theta_4(0)}\bigg)}\nn\\
&&-2i(m+\bar{m})D_{fo}T_{fo}^{(\epsilon_2)}(\xi_1\tau,\xi_2\tau)
{\bigg(\frac{\eta}{\theta_4(\xi_1\tau|\tau)}\bigg)}
{\bigg(\frac{k_2\eta}{\theta_1(\xi_2\tau|\tau)}\bigg)}{\bigg(\frac{\eta}
{\theta_4(0)}\bigg)}\nn\\
&&+2(m+\bar{m})D_{ho}T_{ho}^{(\epsilon_3)}(\xi_1\tau,\xi_2\tau)
{\bigg(\frac{\eta}{\theta_4(\xi_1\tau|\tau)}\bigg)}
{\bigg(\frac{\eta}{\theta_4(\xi_2\tau|\tau)}\bigg)} P_3\nn\\
&&-2(m-\bar{m})D_{gg}T_{gg}^{(\epsilon_1)}(\xi_1\tau,\xi_2\tau)
{\bigg(\frac{k_1\eta}{\theta_1(\xi_1\tau|\tau)}\bigg)}
{\bigg(\frac{\eta}{\theta_3(\xi_2\tau|\tau)}\bigg)}
{\bigg{(}\frac{\eta}{\theta_3(0)}\bigg{)}}\nn\\
&&-2(m-\bar{m})D_{ff}T_{ff}^{(\epsilon_2)}(\xi_1\tau,\xi_2\tau)
{\bigg(\frac{\eta}{\theta_3(\xi_1\tau|\tau)}\bigg)}
{\bigg(\frac{k_2\eta}{\theta_1(\xi_2\tau|\tau)}\bigg)}
{\bigg{(}\frac{\eta}{\theta_3(0)}\bigg{)}}\nn\\
&&-2(m+\bar{m})D_{hh}T_{hh}^{(\epsilon_3)}(\xi_1\tau,\xi_2\tau)
{\bigg(\frac{\eta}{\theta_3(\xi_1\tau|\tau)}\bigg)}
{\bigg(\frac{\eta}{\theta_3(\xi_2\tau|\tau)}\bigg)}P_3\nn\\
&&+2(m-\bar{m})D_{fg}T_{fg}^{(\epsilon_2)}
\frac{2\eta^3}{\theta_4(\xi_1\tau|\tau)\theta_2(\xi_2\tau|\tau)\theta_3(0)}
\nn\\
&&-2(m-\bar{m})D_{hg}T_{hg}^{(\epsilon_3)}
\frac{2\eta^3}{\theta_4(\xi_1\tau|\tau)\theta_3(\xi_2\tau|\tau)\theta_2(0)}
\nn\\
&&+2(m-\bar{m})D_{gf}T_{gf}^{(\epsilon_1)}
\frac{2\eta^3}{\theta_2(\xi_1\tau|\tau)\theta_4(\xi_2\tau|\tau)\theta_3(0)}
\nn\\
&&+2(m-\bar{m})D_{hf}T_{hf}^{(\epsilon_3)}
\frac{2\eta^3}{\theta_3(\xi_1\tau|\tau)\theta_4(\xi_2\tau|\tau)\theta_2(0)}
\nn\\
&&+2i(m+\bar{m})D_{gh}T_{gh}^{(\epsilon_1)}
\frac{2\eta^3}{\theta_2(\xi_1\tau|\tau)\theta_3(\xi_2\tau|\tau)\theta_4(0)}
\nn\\
&&-2i(m+\bar{m})D_{fh}T_{fh}^{(\epsilon_2)}
\frac{2\eta^3}{\theta_3(\xi_1\tau|\tau)\theta_2(\xi_2\tau|\tau)\theta_4(0)}
\bigg\}.
\end{eqnarray}
And finally, the $|Q|=2$ terms are
\begin{eqnarray}\label{eqn:ChargedAnnulusQ2}
{\cal
A}_{(|Q|=2)}&=&\frac{1}{8}\bigg\{-(m^2+\bar{m}^2)T_{oo}(2\xi_1\tau,2\xi_2\tau)
{\bigg{(}\frac{2k_1\eta}{\theta_1(2\xi_1\tau|\tau)}\bigg{)}}
{\bigg{(}\frac{2k_2\eta}{\theta_1(2\xi_2\tau|\tau)}\bigg{)}}P^3\nn\\
&&+i(m^2+\bar{m}^2)T_{og}(2\xi_1\tau,2\xi_2\tau)
{\bigg(\frac{2k_1\eta}{\theta_1(2\xi_1\tau|\tau)}\bigg)}
{\bigg{(}\frac{2\eta}{\theta_2(2\xi_2\tau|\tau)}\bigg{)}}
{\bigg{(}\frac{2\eta}{\theta_2(0)}\bigg{)}}\nn\\
&&+i(m^2+\bar{m}^2)T_{of}(2\xi_1\tau,2\xi_2\tau)
{\bigg{(}\frac{2\eta}{\theta_2(2\xi_1\tau|\tau)}\bigg{)}}
{\bigg(\frac{2k_2\eta}{\theta_1(2\xi_2\tau|\tau)}\bigg)}
{\bigg(\frac{2\eta}{\theta_2(0)}\bigg)}\nn\\
&&+(m^2+\bar{m}^2)T_{oh}(2\xi_1\tau,2\xi_2\tau)
{\bigg{(}\frac{2\eta}{\theta_2(2\xi_1\tau|\tau)}\bigg{)}}
{\bigg{(}\frac{2\eta}{\theta_2(2\xi_2\tau|\tau)}\bigg{)}} P^3.
\end{eqnarray}
The amplitude stated above is contracted for the sake of brevity,
the couplings to the different boundaries $m$ and $\bar{m}$ will
of course effect the parameter $\xi_j$ through a change of charge
sign.

For comparison, this amplitude is the breaking of the
$\mathbb{Z}_2\times \mathbb{Z}_2$ direct channel annulus shown in
equation (\ref{eqn:GenericZ2xZ2-1case}).

With the direct channel amplitude fully specified, it is then a
simple matter to perform an $S$ transformation to obtain the
transverse channel.


\subsection{Tadpole Conditions}


One finds that the untwisted sectors from the breaking's
(\ref{eqn:UnchargedAnnulus}), (\ref{eqn:ChargedAnnulusQ1}) and
(\ref{eqn:ChargedAnnulusQ2}) give rise to tadpoles that cancel
consistently.  These do present some subtleties that are similar
in nature to those discussed in \cite{CAAS} for the simpler six
dimensional model $T^4/\mathbb{Z}_2(H_1,H_2)$. However, these are
made to reduce to consistent tadpole cancellation conditions with
the constraints $m=\bar{m}$ and $H_1=-H_2$ as was the case for the
$T^4/\mathbb{Z}_2(H_1,H_2)$ model.  It is noted that the choice
for the fields that minimizes tachyonic excitations (which will be
discussed thoroughly later) is $H_1=-H_2$.

The twisted sector is more complicated. To illustrate this, I
first consider the $g$-twisted sector.  The $R$--$R$ contributions
from $\tau_{gh}$ and $\tau_{gf}$ reduce to their generic fixed
point arrangements with an additional term which is proportional
to $\alpha_1$.

In the case of $\tau_{gh}$, with the deformations generated by
(\ref{eqn:breaking}), and the constraint $H_1=-H_2$, one has the
$R$--$R$ contribution as
\begin{eqnarray}\label{eqn:TwistedTadpole}
\tilde{{\cal A}}^g_{o}&=&\frac{2^{-5}}{8}\bigg{\{}\bigg(\sqrt{v_1}
\big[N_g+i(m-\bar{m})+(m+\bar{m})\alpha_1\big]\nn\\
&&-4\sqrt{v_1}D_{gg}-2\frac{1}{\sqrt{v_1}}D_{fg}+
2\frac{1}{\sqrt{v_1}}D_{hg}\bigg)^2 \nn\\
&&+3{\bigg(\sqrt{v_1}\big[N_g+i(m-\bar{m})+(m+\bar{m})\alpha_1\big]
-2\frac{1}{\sqrt{v_1}}D_{f;g}\bigg)}^2\nn\\
&&+3{\bigg(\sqrt{v_1}\big[N_g+i(m-\bar{m})+(m+\bar{m})\alpha_1\big]
+2\frac{1}{\sqrt{v_1}}D_{h;g}\bigg)}^2\nn\\
&&+v_1\bigg(N_g+i(m-\bar{m})+(m+\bar{m})\alpha_1\bigg)^2\bigg{\}}.
\end{eqnarray}
This is just the fixed point structure of the generic
$\mathbb{Z}_2\times\mathbb{Z}_2$ for $\tau_{gh}$ after the
substitution $N_g\rightarrow
N_g+i(m-\bar{m})+(m+\bar{m})\alpha_1$. So, after identification of
conjugate multiplicities $(m-\bar{m})=N_g=0$, the expression
(\ref{eqn:TwistedTadpole}) reduces to
\begin{eqnarray}
2^{-5}(m+\bar{m})^2\alpha_1^2.
\end{eqnarray}
A similar case arises for the $NS-NS$ contributions, which after
the identification $H_1=-H_2$ and $m=\bar{m}$ has a residual term
proportional to $\alpha_2^2$. In a reversed sense, those from the
$f$--twisted sector are opposite, such that $R-R$ sector carries a
residual contribution from $\alpha_2^2$, and the $NS-NS$ similarly
has remaining term proportional to $\alpha_1^2$. For terms in the
$h$--twisted sector, the resulting spectrum is inherently free of
such residual terms, and so behaves as the tadpole term of the
non--deformed case.

In the transverse channel, the constraint on the twisted sector is
that it should cancel by itself.  For the untwisted terms, the
Klein bottle and Mobius contributions allow a constraint on the
Chan-Paton factors other than zero. The Klein and the Mobius
amplitude do not contain twisted terms with which to help
cancellation of twisted terms in the annulus. In table
\ref{tab:TadCond}, such conditions are shown as $N_j=0$ ($j\neq
0$).

In chapter \ref{ch:OpenDescendents} it was seen, in relation to
the cancellation of the $D5$ tadpoles of the $\epsilon=+1$ cases,
that the $NS$--$NS$ and $R$--$R$ sectors do not mutually cancel
for models other than $(+,+,+)$. However, one could still obtain
an spectrum that was anomaly free with respect to the cancellation
of the $R$--$R$ charges of the $D5$'s and $O5$'s by ensuring that
the $R$--$R$ conditions where satisfied.  Non--zero $NS$--$NS$
tadpole terms have an interpretation as a correction to the vacuum
energy through the low--energy effective action \cite{CAAS}. In
the above case, the residual term is present for both the $R$--$R$
and $NS$--$NS$ sectors.  Therefore, the model is not capable of
cancelling the net $R$--$R$ charge.

Moreover, the form of the twisted tadpole terms are representative
of the correct counting of the occupation of fixed points by the
various brane types.  Under magnetic deformation, equation
(\ref{eqn:TwistedTadpole}) shows that the residual term is not
consistent with this interpretation. The fixed point occupation
counting for the magnetically charged branes are taken in to
account by $i(m-\bar{m})$ and those of the uncharged ones by
$N_g$. By arguments above, this will also be the case for the
$f$--twisted sector.

These types of residual terms in the twisted sector are an
artifact of the action on the world sheet bosonic and fermionic
modes by the $\mathbb{Z}_2\times \mathbb{Z}_2$ group generators.
The six dimensional case considered \cite{CAAS} was an orbifold
breaking of the form $T^4/\mathbb{Z}_2(H_1,H_2)\equiv
\big(T^2(H_1)\times T^2(H_2)\big)/\mathbb{Z}_2$. The action of the
orbifold on the bosonic coordinates would then act in all internal
directions, thereby allowing the same mode deformations, caused by
the magnetic fields in both tori. In the case of the
$\mathbb{Z}_2\times \mathbb{Z}_2$, the set of generators are
permutations of the operator $(+1,-1,-1)$ that act on
$T^2_{45}\times T^2_{67}\times T^2_{89}$. Consequently, it is only
possible to align the orbifold directions with those of the
charged branes (in this model with $H_1, H_2\neq 0$ and $H_3=0$)
consistent with the $h=(-1,-1,+1)$ generator. This alignment
effectively allows the $h$--twisted sector to be free of this
complication.

Alternatively, for the choice of $\epsilon=+1$ (without discrete
torsion) as discussed in \cite{MLGP} and section
\ref{sec:Z2xZ2xZ2Model1}, the parent torus amplitude leaves such
twisted terms absent in the resulting annulus amplitude. However,
this removes the breaking terms that could otherwise define
unitary representations for the $D9$ (that are not magnetically
charged) and $D5$ branes. For the case ($\epsilon=+1$), one has
either orthogonal or simplectic groups, and so the resulting
models are less appealing phenomenologically, although the branes
that carry magnetic charge are still unitary.

The method for keeping groups of unitary type while magnetizing
two tori is discussed in chapter \ref{ch:Conclusions}, this method
was reported in \cite{MLGP}.


\subsection{Mass Shifting and Tachyonic Instabilities}


In the presence of magnetized tori, tachyonic modes are present by
virtue of the shift in mass that a magnetic deformation causes
(\ref{eqn:MassShift}). This shifting occurs in the direct channel,
since the deforming parameter $\xi$ is now accompanied by the
measure parameter $\tau$ as $\xi\tau$. These additional parts now
provide a change in the mass squared as they can be written as
powers of $q$.

I briefly review the respective contributions of the world sheet
bosons and fermions to the mass shift (\ref{eqn:MassShift}). The
world sheet fermion deformations are given by
\begin{eqnarray}
&&O_2(m\xi_j\tau)\sim_z \frac{1}{\eta},~V_2(m\xi_j\tau)\sim_z
\frac{q^\frac{1}{2}}{\eta}(q^{m\xi_j}+q^{-m\xi_j})\nn\\
&&C_2(m\xi_j\tau)\sim_z
\frac{q^{\frac{1}{8}}}{\eta}q^{-\frac{m\xi_j}{2}},
S_2(m\xi_j\tau)\sim_z
\frac{q^{\frac{1}{8}}}{\eta}q^{\frac{m\xi_j}{2}}.\nn
\end{eqnarray}

In the case of magnetically deformed world sheet bosonic modes,
with and without an orbifold projection, the low lying modes are
respectively
\begin{eqnarray}\label{eqn:LowlyingThetaModes}
&&\frac{mk_j\eta}{\theta_1(m\xi_j\tau|\tau)}\sim_z
\frac{-mk_jq^{-\frac{1}{12}}}{sin(m\pi\xi_j\tau)}
=im|k_j|q^{-\frac{1}{12}}\sum_{n_j=0}^\infty
q^{\frac{1}{2}|m\xi_j|(2n_j+1)},
\nn\\
&&\frac{\eta}{\theta_2(m\xi_j\tau|\tau)}\sim_z
\frac{q^{-\frac{1}{12}}}{cos(m\pi\xi_j\tau)}=q^{-\frac{1}{12}}\sum_{n_j=0
}^\infty (-1)^{n_j}q^{\frac{1}{2}|m\xi_j|(2n_j+1)}.
\end{eqnarray}
Where the index $j$ labels the two directions of the $j^\text{th}$
torus. The bosonic coordinates are not effected by the change in
charge from say an $m^2$ to $\bar{m}^2$ coupling. For the
deformation of the world sheet bosonic modes, the Landau number
$k_j$ is odd under $\xi_j\rightarrow -\xi_j$, as is the zero mode
and thus the deformed lattice as a whole is even. This is
reflected by the appearance of $|\xi_j|$ and $|k_j|$.

The dependence of $k_j$ on the charge and field values can be seen
by the Dirac quantization condition
\begin{eqnarray}\label{eqn:QuantCond}
k_j=2\pi \alpha^\prime q H_j v_j,
\end{eqnarray}
for the $j^{\rm th}$ torus.

Firstly, I discuss the instabilities that arise in the untwisted
sector.  In the case of the spacetime vector $V_2O_2O_2O_2$ and
the scalar $O_2O_2O_2V_2$ the mass shift $\triangle M^2$ is
proportional to $2|H|$ for both $\bar{m}^2$ and $m^2$ couplings.
The remaining untwisted scalars have internal vectors that lie in
the tori with magnetic fields and so $\triangle M^2$ has
contributions from the magnetic moment terms.  For $O_2V_2O_2O_2$
one finds that the shift in mass is $3|H|$ and $|H|$ for $m^2$ and
$\bar{m}^2$ couplings respectively.  In a reversed sense (because
$H_1=-H_2$) the relevant terms are $|H|$ and $3|H|$ for $m^2$ and
$\bar{m}^2$ respectively for $O_2O_2V_2O_2$.

In the Ramond untwisted sector, for terms of the form $S_2S_2$
lying in the first two tori, the mass shift is $4|H|$ for $m^2$
couplings and zero for the $\bar{m}^2$ ones.  Similarly, those of
$C_2C_2$ have the same values for opposite couplings.  The terms
with $C_2S_2$ or $S_2C_2$ aligned with the first two tori don't
have contributions from the magnetic moment couplings and so give
$\triangle M^2$ proportional to $2|H|$.

In conclusion, strings from the untwisted sector that arise from
the couplings $m^2$ and $\bar{m}^2$ give either massless or
massive states and so tachyonic excitations are absent.  States
from terms involving $\bar{m}m$ coupings remain massless since
$q_L+q_R=0$.

The terms that contribute massless modes in the
$\mathbb{Z}_2\times \mathbb{Z}_2$ twisted sector without magnetic
fields are illustrated in table \ref{tab:shifts1} (in the second
column) alongside their mass shifts under magnetic deformation
(the $\pm$ refers to the net charge sign of the branes in the
coupling, and is correlated for the two terms in a given
character).

There are six distinct choices for the fields $H_j$ that will
minimize instabilities (for two magnetized tori)
\begin{eqnarray}
&&H_2=-H_3, H_1=0, H_3>0~{\rm or }~H_3<0,\nn\\
&&H_1=-H_2, H_3=0, H_2>0~{\rm or }~H_2<0,\nn\\
&&H_1=-H_3, H_2=0, H_1>0~{\rm or }~H_1<0.\nn
\end{eqnarray}

In the case at hand I consider only $H_1$ and $H_2$ to be nonzero.
With this configuration, characters in the $h$--twisted sector are
exactly massless. The characters of the remaining sectors contain
tachyons. So again one sees that the $h$--sector is free of
possible complications from the presence of magnetic fields.

It is clear that it is not possible to remove these instabilities
by choice of fields due to the absence of Landau contributions and
the sensitivity to charge sign from the magnetic moment parts.
\begin{table}
\begin{center}
\begin{tabular}{|c|c|l|}\hline
Character & Diagram & Mass shift \\ \hline
$\tau_{gh}$ & $O_2O_2S_2S_2$ & $\Delta M^2\sim \pm(H_2+H_3)$ \\
& $S_2C_2O_2O_2$ & $\Delta M^2\sim \mp H_1$ \\
\hline \hline
$\tau_{gf}$ & $O_2O_2C_2C_2$ & $\Delta M^2\sim \pm(H_2+H_3)$ \\
& $C_2S_2O_2O_2$ & $\Delta M^2\sim \mp H_1$ \\
\hline \hline
$\tau_{hg}$ & $O_2C_2C_2O_2$ & $\Delta M^2\sim \pm(H_1+H_2)$ \\
& $C_2O_2O_2S_2$ & $\Delta M^2\sim \mp H_3$ \\
\hline\hline
$\tau_{hf}$ & $O_2S_2S_2O_2$ & $\Delta M^2\sim \pm(H_1+H_2)$ \\
& $S_2O_2O_2C_2$ & $\Delta M^2\sim \mp H_3$ \\
\hline \hline
$\tau_{fg}$ & $O_2C_2O_2C_2$ & $\Delta M^2\sim \pm(H_1+H_3)$ \\
& $C_2O_2S_2O_2$ & $\Delta M^2\sim \mp H_2$ \\
\hline \hline
$\tau_{fh}$ & $O_2S_2O_2S_2$ & $\Delta M^2\sim \pm(H_1+H_3)$ \\
& $S_2O_2C_2O_2$ & $\Delta M^2\sim \mp H_2$ \\
\hline
\end{tabular}
\end{center}
\caption{Twisted sector mass shifts for $\mathbb{Z}_2\times
\mathbb{Z}_2$ model }\label{tab:shifts1}
\end{table}
%
%


\section{Discussion}


It has been shown in the previous section that the form of the
$\mathbb{Z}_2\times \mathbb{Z}_2$ ($\epsilon=-1$) with two
background magnetic fields, gives low lying states in the
transverse amplitude that provide inconsistencies through residual
terms proportional to the magnetic fields. In the conventional
$\mathbb{Z}_2\times \mathbb{Z}_2$ model (without magnetized
$D9$'s), the tadpole conditions for the terms $N_g$, $N_f$ and
$N_h$ must be (and are) identically zero.

There are no twisted sectors in the transverse Klein or Mobius
with which to assist with cancellation of these terms. Therefore,
the annulus amplitude defined with the presence of discrete
torsion, cannot be magnetized to give a consistent spectrum.  This
is true for the case of choosing either one or two fields to be
nonzero. What form these inconsistencies might have for the case
of three internal fields is not clear, though it is expected that
similar problems will arise.

In order to include magnetic backgrounds in models with transverse
twisted sectors, one must have some control over what form these
states have, as has been done in \cite{MLGP}. This will be
discussed more concisely in chapter \ref{ch:Conclusions}.

As an additional point, the aim of implementing background fields
in this way was to obtain a solution of the form $U(8-m)_9\times
U(m)_9\times U(8)_5$. Here the original stack of $D9$'s could be
broken to $U(6)_9\times U(2)_9\times U(8)_5$ using the solution of
$m=2$. In order to reduce the rank of all families of branes, one
could then allow the existence of a background antisymmetric
tensor \cite{CAAS}, \cite{C}.

The $B_{ab}$ field is introduced as a generalization of the
momenta in the compact directions
\begin{eqnarray}
&&p_{L,a}=m_a+\frac{1}{\alpha^\prime}(g_{ab}-B_{ab})n^b,\nn\\
&&p_{R,a}=m_a-\frac{1}{\alpha^\prime}(g_{ab}+B_{ab})n^b,\nn
\end{eqnarray}
for momenta and winding quantum numbers $m_a$ and $n^b$. Retaining
the symmetry of world sheet parity under $\Omega$, the $B_{ab}$
field takes on quantized values.  The projector that is necessary
in the trace of the partition function carries a normalization
factor which is dependant on the rank of this field. This factor
forces the tadpole conditions to be sensitive to the value of the
rank $r$ of the $B_{ab}$ field. Thus the rank of the gauge group
can be reduced accordingly.

The next step would be to allow the presence of discrete Wilson
lines that would allow a reduction by half of the $D5$ stacks,
thus obtaining a group of the form $U(3)_9\times U(2)_5 \times
U(1)_9$ (after using the antisymmetric field $B_{ab}$ of
appropriate rank).

The method for doing this kind of background setting is also
reviewed in chapter \ref{ch:Conclusions}, where I discuss work
similar to the intended idea presented here that was implemented
in \cite{MLGP}.

\setcounter{footnote}{0}

\chapter{Conclusions}\label{ch:Conclusions}

The three chapters that have been presented detail particular
phenomenologically motivated aspects of the free fermionic
heterotic and type I string limits.

The heterotic model described in the context of this work shows a
construction based on the NAHE set $\{1,S,b_1,b_2,b_3\}$ and the
additional vectors $\{\alpha,\beta,\gamma\}$.  The particular
choice of phase
\begin{eqnarray}\label{eqn:Phase11}
C\left(\begin{array}{c} b_3 \\ \beta
\end{array}\right)=-1\nn
\end{eqnarray}
is made. It was noted in contrast to model 3, which chooses this
term as $+1$, that this causes an enhancement to the observable
and hidden sector gauge groups.  In particular, they are enhanced
from $SU(3)$ to $SU(4)$.

The trace of the anomalous charge $Q_A$ of this model is
$\text{Tr}Q_A=-72$. It was shown that the spectrum does not
contain Abelian VEVs that could cancel this contribution to allow
$D$--flatness.  The only VEVs that could be utilized to cancel the
contribution of $-72$ arose from non--Abelian fields.

This is the first instance where the use of non--Abelian VEVs has
been enforced in order for $D$--flatness, as opposed to other
possible phenomenological reasons.

However, it was outlined that this particular model does not
pertain to a realistic model. And so this study is discussed as an
instance of an extended NAHE set based toy model which requires
non--Abelian VEVs to satisfy the flat direction constraints. In
examples such as that presented here, one can gain some intuition
for phenomenology that arises from basis vectors that give rise to
a large spectral content.  While efforts have been made to
quantify the effects that certain vectors and phases have on a
models content, there is no general set of instructions to
determine the precise phenomenology desired when attempting to
design appropriate basis vectors and GSO phases.  Therefore, this
is yet another model which will help to highlight the relationship
between the physics of models and their basis vectors/GSO phases.
In particular, an enhancement to the observable and hodden sector
gauge groups and more importantly, the need for non--Abelian VEVs
to retain supersymmetry at the FI scale.

Models based on the NAHE set have provided a basis on which a
great deal of study has been focused.  The NAHE set contains the
basis vectors $\{b_1,b_2,b_3\}$ which provide the
$\mathbb{Z}_2\times\mathbb{Z}_2$ structure of the one--loop
partition function.  The gauge group after this set is
$SO(10)\times E_8 \times SO(6)^3$, which as stated before contains
three $\mathbf{16}$'s of $SO(10)$, one from each of the vectors
$b_j$.

Any study of string theory is motivated by the correspondence it
has with the real world.  Agreement with low energy physics, or
the phenomenology of the standard model, is a basis on which to
guide the efforts of research.  The NAHE set has shown to be an
underlying set with which to build and explore phenomenologically
viable models within the heterotic limit.

In the type I limit, the models studied are freely acting shift
$\mathbb{Z}_2^\text{s}$ modulations of the $\mathbb{Z}_2$ and
$\mathbb{Z}_2\times \mathbb{Z}_2$ orbifolds. In contrast to the
model that includes the $\mathbb{Z}_2\times \mathbb{Z}_2$ orbifold
geometry, the toy model is a pseudo $\mathbb{Z}_2\times
\mathbb{Z}_2$ geometry with only orbifold actions in the last two
tori.  This leads to a simplified spectrum with an absence of
those terms corresponding to independent orbits (see appendix
\ref{app:Boundaries}), since the other $\mathbb{Z}_2\times
\mathbb{Z}_2$ generator acts only to shift the coordinates by
$\frac{\pi R}{2}$.

The spectrum of the open sector admitted propagation of twisted
sectors in the transverse channel. In turn, this allowed
Chan-Paton factors to be interpreted as unitary terms.  The gauge
group therefore became unitary.

Adding in an additional orbifold generator to the trace of this
model, and allowing the shift to act in all directions, one then
has the next model based on the $\mathbb{Z}_2\times \mathbb{Z}_2$
orbifold. This model proved to be more subtle both in its
orientifold structure and its open descendants. To begin with, the
torus partition function by itself defines an enhancement of the
generic $\mathbb{Z}_2\times \mathbb{Z}_2$ torus with additional
$n+\frac{1}{2}$ massive states in the twisted sector. While the
untwisted sector also has these types of states, it acquires a
projection of the form $(1+(-1)^{m_1+m_2+m_3})$. The low lying
states in the torus shows that the number of fixed points,
originally 16 for the generic $\mathbb{Z}_2\times \mathbb{Z}_2$
model, are halved.  Similar arguments are valid for the toy model.

In taking the orientifold projection of the torus amplitude, one
finds a subtlety.  Use of a generic orientifold gave rise to the
counting of states in the Klein amplitude that give an
inconsistent perfect square structure.  The perfect square in the
Klein amplitude represents the reflection coefficients of a closed
string propagating between two $O$--plane's.  The requirement that
the coefficients of propagating closed string states between
boundaries be perfect squares is condition that must be satisfied.

In this respect, a naive $\Omega$ projection gave rise to massive
states of the form $W_{n+\frac{1}{2}}$ in the $O5_i$--$O5_j$
($i\neq j$) couplings.  As it would stand, this term maps under
$S$ to give rise to contributions that don't admit a perfect
square structure. In this case, the eigenvalues of this coupling
where forced to acquire different signs, so that the total was
zero.  As such, this term no longer appears in the counting and
was then made so as to not interfere with the perfect squares.

A similar case was shown to arise in a $T^4/\mathbb{Z}_2$ model.
However, this involved an eigenvalue reassignment (or sign
redefinition of $\Omega$) when using the unusual global form of
$\Omega$ with the phase $(-1)^m$ as a model freedom.  As such, a
reassignment of eigenvalues was forced by taking a different
choice of the parity operator.  In the model discussed here, one
begins with a conventional $\Omega$ which leads to the same
reassignment for similar consistency conditions.

The additional modulation of a $\mathbb{Z}_2$ shift on the
underlying $\mathbb{Z}_2\times \mathbb{Z}_2$ spectrum allows this
model to have orbits that are not related to the principle ones
$\{(o,o),(o,g),(o,f),(o,h)\}$.  As such there are two classes in a
total of eight models that can be defined.  The two classes are
defined by the choices of
$\epsilon=\epsilon_1\epsilon_2\epsilon_3=\pm 1$ which is a sign
freedom attached to the terms corresponding to the independent
orbits. The spectral content is quite different between the two
classes, while those models within each class differ with respect
to the types of $O$--plane and $D$--brane present (which in turn
effects a models supersymmetry). The only model that is fully
supersymmetric is that defined by $\epsilon_j=+1$.

Models with $\epsilon=+1$ define the simplest of the two cases. It
was shown in equation (\ref{eqn:TorusZero}) that $\epsilon=+1$
choice does not admit the appropriate form of the twisted states
so as to have twisted sectors in the transverse annulus. The open
descendants are thus limited to untwisted states in the transverse
channel.

The annulus amplitude of the $\mathbb{Z}_2\times \mathbb{Z}_2$
($\epsilon=+1$) models required a rescaling of the Chan-Paton
charges, since there are no breaking terms present.   The
inclusion of the $\mathbb{Z}_2$ shift required an additional
rescaling of the $D5$ charges, thus halving the corresponding
group size.  The gauge groups display different behavior in the
case (+,+,+) from the others in the same class.  In this case the
groups corresponding to all branes where unitary simplectic.  In
the cases that where permutations of $(+,-,-)$, one sees a change
for the $D9$--branes as orthogonal. The $D5$--branes where
orthogonal, whereas the $\bar{D}5$--branes where unitary
simplectic.

While the class of models belonging to $\epsilon=+1$ allow a rich
spectrum with rank reduction of the $D5$ groups, the breaking
terms that come from the twisted sectors in the transverse channel
are absent.  As such, unitary groups, and hence a more
phenomenologically interesting spectrum (in the motivation to look
at matter coupled to unitary groups) could not be obtained.

In the second class of models corresponding to $\epsilon=-1$, one
has a vastly richer spectrum of states.  In this case, it was seen
that the torus admits the flow of twisted sectors in the
transverse annulus.  {From} this, one obtains breaking terms that
allow the Chan-Paton charges as unitary for $D9$ and $D5$
couplings and simplectic for $\bar{D}5$--branes.

However, the complexity of this amplitude causes a problem in the
twisted sector of the direct channel $D5_i$--$D5_j$ ($i\neq j$)
couplings. In this case, the massless spectrum is entirely
consistent with a gauge group of $U(8)_9\times U(4)_{5_{\{1,2\}}}$
for the $D9$ and two $D5$'s and $USp(4)^4$ for the $\bar{D}5$ with
the required halving of the Chan-Paton charges. The
$n+\frac{1}{2}$ massive states for the $D5_i$--$D5_j$ couplings
does not give a proper particle interpretation.  After taking into
account all necessary factors, the coefficients of these terms are
equal to $\frac{1}{2}$.

While a solution was not found that gives a proper particle
interpretation at all lattice levels, it is regarded that such a
solution should exist.  The complexity of this model makes it very
difficult to implement possible solutions.  For example, one could
not impose a rescaling of say the $D5$ branes alone, this would
cause inconsistent particle counting in both the $D5$--$D5$ and
$N$--$D5$ sectors.  The presence of the twisted terms in the
transverse channel allows the Chan-Paton charges to be broken in
such a way that the rescaling that was done for the $\epsilon=+1$
is taken into account by this breaking.

Various possibilities for the introduction of auxiliary Wilson
lines where discussed.  The term which causes the inconsistency
will only admit a global phase for the particular lattice state
that the $D5_i$--$D5_j$ term has.  Global in this context is
simply meant that the phase must act outside the lattice state
projector $(1+(-1)^m_k)$, which would otherwise not be consistent
with the corresponding torus amplitude states.

In trying to gain freedom in controlling phenomenologically
interesting aspects of these models, such as gauge group rank and
type, spectral content (such as matter families and supersymmetry)
and the study of backgrounds has proved very forthcoming. The work
that has been presented in chapter \ref{ch:OpenDescendents} looks
at the effect of discrete Wilson lines in the form of freely
acting shifts.  The aim was to obtain models that might be used in
conjunction with other backgrounds, to yield a four dimensional
gauge group of the form $SU(3)\times SU(2)\times U(1)$.  However,
the class of $\epsilon$ that is consistent with model building
requirements only admits matter coupled to orthogonal or
simplectic groups.

The motivation for using shifts in this manner was to set the
foundation that would lead to a three generation model with a
realistic gauge group.  One could begin with the
$\mathbb{Z}_2\times \mathbb{Z}_2$ spectra for the case of
$(+,+,-)$, which has a unitary configuration of $U(8)\times U(8)$
for the $D9$ and $D5$'s in the $1^\text{st}$ and $2^\text{nd}$
tori.  The freely acting shift allows one to retain the full
$\mathbb{Z}_2\times \mathbb{Z}_2$ characters, with various changes
in lattice contributions.  This would then yield a low lying
spectrum with gauge group $U(8)\times U(4)$.  By use of an
additional breaking mechanism (such as background magnetic
fields), one could realize a group of the form $U(8-m)\times
U(m)\times U(4)$.  Switching on the background antisymmetric
tensor, it would be possible to further control the rank of these
groups to get close to or reproduce a standard model like
structure.  However, as has been shown, the $\epsilon=-1$ model
cannot be straight forwardly magnetized.

The work done on freely acting shifts is an extension of the study
discussed in \cite{AADS}.  Here, the shifts where used as integral
parts of the $\mathbb{Z}_2\times \mathbb{Z}_2$ generators as
\begin{eqnarray}\label{eqn:Gen'sOfNonFreelyActing1}
\sigma_1(\delta_1,\delta_2,\delta_3)= \left(\begin{array}{rrr}
\delta_1 & -\delta_2 & -1 \\ -1 & \delta_2 & -\delta_3 \\
-\delta_1 & -1 & \delta_3
\end{array}\right),~
\sigma_2(\delta_1,\delta_2,\delta_3)= \left(\begin{array}{rrr}
\delta_1 & -1 & -1 \\ -1 & \delta_2 & -\delta_3 \\ -\delta_1 &
-\delta_2 & \delta_3
\end{array}\right)\nn\\
\end{eqnarray}
where $\delta$ is a shift operator.  In all cases of these models,
the terms corresponding to independent orbits are absent.  As a
consequence of defining shifts and orbifold operations as
(\ref{eqn:Gen'sOfNonFreelyActing1}) the spectrum is projected by
operators of the form $-\delta$. On the complex coordinates of the
torus, this composite operation has the effect of providing
shifted fixed points, as one can see from
\begin{eqnarray}
z_i\sim -z_i+\frac{1}{2}
\end{eqnarray}
and so the fixed points coordinate set is
$z^f_i=\{\frac{1}{4},\frac{3}{4}\}$.  One then sees in contrast to
the freely acting case, that there no fixed points at the origin.
Those of the freely acting shift have the standard fixed points of
$\{0,\frac{1}{2}\}$. In addition, of the cases arising from
(\ref{eqn:Gen'sOfNonFreelyActing1}), most of the models are
capable of providing matter coupled to unitary gauge groups.

The discussion on background magnetic deformations in chapter
\ref{ch:MagDeform} looks at the effect of magnetizing the last two
tori in the configuration $T^2_{45}\times T^2_{67}\times T^2_{89}$
of the $\epsilon=-1$ case of the $\mathbb{Z}_2\times \mathbb{Z}_2$
open descendants.

Magnetic deformations are a very good way of breaking more
precisely the gauge group rank of the $D9$ branes.  The presence
of such fields introduces additional massive states in the
spectrum that correspond to lattices that depend on radii that are
scaled according to $R_j\rightarrow R_j(1+(2\pi\alpha'q
H_j))^{\frac{1}{2}}$.  Moreover, the Landau degeneracies that
exist allow one some freedom (with the guidelines of spectral
consistency, i.e. particle interpretation and symmetrization) to
control certain matter contributions in the spectrum.

The magnetic deformation of the open descendants that correspond
to $\epsilon_j=+1$ have been demonstrated in \cite{MLGP}. In this
case, the group structure is a composite of unitary and orthogonal
groups.  The magnetic charges force the $D9$ stacks to split from
the parent $D9$'s as $N\rightarrow N+m+\bar{m}$.  As can be seen
from this, the gauge groups corresponding to charged branes are
unitary.

The deformations presented in this study show the effect on the
low lying spectrum and structure of the annulus.  The tadpoles in
the untwisted sector present the usual subtleties where the
reflection coefficients are no longer perfect squares. This is due
to the structure of the transverse annulus (\ref{eqn:TimeOp}) and
that the boundary terms at either ends of the string are not
invariant under the operation $\cal T$. However, after
identification of the multiplicities $m=\bar{m}$, the invariance
is restored and the tadpoles behave in manner similar to the
$T^2(H)\times T^2(H)/\mathbb{Z}_2$ model (with the exception of
two additional $D5$ sectors and the signs $\epsilon_j$).

Therefore, the untwisted sector is well mannered with respect to
cancellation of the tadpoles. The usual disagreement between the
$NS$--$NS$ and $R$--$R$ sector appear where a negative sign is
used for one of the $\epsilon_j$ factors.

The tadpoles of the twisted sector however present a deviation
from the behavior of those of the untwisted ones.  It was shown
that after identification of conjugate multiplicities and
$H_1=-H_2$ (as demanded by the cancellation of the untwisted
tadpoles), one is left with contributions depending on the
magnetic fields.

These types of terms can be avoided in the $\mathbb{Z}_2\times
\mathbb{Z}_2$ formalism by the use of shifts integrated in the
$\mathbb{Z}_2\times \mathbb{Z}_2$ generators.  Such work has been
shown in \cite{MLGP} using projection operators of the form of
equation (\ref{eqn:Gen'sOfNonFreelyActing1}), which was published
independently of the finding in this work.

In particular, they consider a case with a winding shift acting in
the second torus and a momentum shift acting in the third.  The
generic model defined by this arrangement defines a twisted sector
which includes only a twist of the form $(+,-,-)$, which acts in
the second and third torus. As such, they align two magnetic
fields in the same directions as the twist and shift operators.
This allows the model to be consistent in terms of symmetrization
and free of problematic terms on the transverse channel.

The type I $\mathbb{Z}_2\times \mathbb{Z}_2$ vacuum generically
becomes unstable.  Background magnetic fields introduce tachyonic
modes in the direct channel.  In the untwisted sector, the $NS$
terms for two background fields can at least compensate for any
possible instabilities (coming from the magnetic field/spin
coupling) with contributions from the magnetically deformed
bosons, as can be seen from equation (\ref{eqn:MassShift}). In the
case of three fields, the magnetic/spin coupling terms are
overwhelmed, and thus these contribute massive states.

The $R$ untwisted terms are either at least stable (for two
fields) or massive (for three fields) due to the presence of only
spin-$\frac{1}{2}$ characters. The nature of the twisted
characters of the $\mathbb{Z}_2\times \mathbb{Z}_2$ inherently
allows instabilities, which is demonstrated in table
\ref{tab:shifts1}. At the time of writing there is yet no
mechanism for allowing a stable type I $\mathbb{Z}_2\times
\mathbb{Z}_2$ vacuum.  Such a mechanism would have to include a
way of lifting the problematic sectors in mass.

While it is possible to discuss the stability in the case of three
background fields, what conditions are forced upon such a
configuration are not known.  Indeed, what problems might occur
(as did with the two field case) are not known. For the two field
case, the condition $H_1=-H_2$ was a necessary requirement to
achieve the untwisted tadpole conditions of the form (as well as
partial stability in the twisted sector)
\begin{eqnarray}\label{eqn:TadpoleConditions}
N_o+m+\bar{m} &=& 32\nn\\
D_{go} &=& 32\nn\\
D_{fo} &=& 32\nn\\
D_{ho}+\epsilon_3(m+\bar{m})k_1k_2 &=& 32.
\end{eqnarray}
Such constraints in the three field case are not known, although I
would expect them to follow similar lines.

The study of three fields would be a necessary next step in
understanding the $\mathbb{Z}_2\times \mathbb{Z}_2$ with magnetic
fields.  Magnetizing all three tori would also mean that there is
no directional preference given. In the one field case, all
twisted sectors carry instabilities and the internal vectors of
the untwisted sector also develop tachyon's.

The six dimensional case that involves background magnetization of
$T^4/\mathbb{Z}_2$ that was studied in \cite{CAAS} is completely
stable and is free of tachyonic instabilities and inconsistencies
in all sectors.

In the interest of studying type I vacua with geometry of the type
$\mathbb{Z}_2\times \mathbb{Z}_2$, it is clear that the current
approaches with magnetizing tori is incomplete if one is to assert
stability in the vacuum. Continued research in the framework of
this geometry with background magnetic fields would ideally seek
to achieve the following points:
\begin{itemize}

\item Vacuum stability

\item Standard model like components

\item A way of ``freezing" the additional moduli

\end{itemize}
The last point refers to factors one finds in the spectrum that
are related to the Landau--level degeneracy, which manifest
themselves as the numbers $k_j$, as can be seen in the amplitudes
(\ref{eqn:ChargedAnnulusQ1}) and (\ref{eqn:ChargedAnnulusQ2}).

As motivated in the introduction, the study of different string
limits related to the $\mathbb{Z}_2\times \mathbb{Z}_2$ geometry
is well founded.  The heterotic limit in the free fermionic
formalism based on the NAHE set (which contains the
$\mathbb{Z}_2\times \mathbb{Z}_2$ vectors $b_1$, $b_2$, $b_3$)
harbors the $SO(10)$ group in the observable sector. Supplemented
by additional vectors, this has been shown to give rise to models
that reproduce the Minimal Supersymmetric Standard Model in the
effective low energy theory.

A mechanism that would provide insight into vacuum selection and a
framework with which to describe strings in a more dynamical sense
are two very important questions that need to be answered. In
addressing these points, it is prudent to continue the study of
the different limits and their phenomenological aspects.  This is
the best guide for determining wether they apply to the real
world.  This also helps to highlight possible structures that
promote new ideas for the resolution of aspects that are not yet
fully understood.  This is particularly relevant with respect to a
mechanism of vacuum selection.

It may well be the case that the hidden vacuum selection mechanism
will begin to show once a certain level of intuition is gained in
how to interpret the relative limits. However, string
phenomenology research has and should continue to proved important
guidelines in how the theory of strings is interpreted and applied
to the real world.

\appendix

\chapter{General Mobius Origin for the $A_1$
Shift}\label{app:moborigin}

This is the Mobius origin with Chan-Paton charge parameterizations
according to the class of models belonging to
$\epsilon=\epsilon_1\epsilon_2\epsilon_3=+1$. It is provided as a
reference to show the explicitly lattice origin structure of the
Mobius.  And in particular, how the choice of different signs
results in the change of group structure.
\begin{eqnarray}
{\cal
M}_o&=&-\frac{1}{4}\bigg{\{}\bigg(n(1-\epsilon_1-\epsilon_2-\epsilon_3)-
d_{go}(1-\epsilon_1+\epsilon_2+\epsilon_3)\nn\\
&&-d_{fo}(1+\epsilon_1-\epsilon_2+\epsilon_3)-d_{ho}(1+\epsilon_1+
\epsilon_2-\epsilon_3)\bigg)\hat{\tau}_{oo}^{NS}\nn\\
&&-\bigg(n(1-\epsilon_1-\epsilon_2-\epsilon_3)-d_{go}\epsilon_1(1-
\epsilon_1+\epsilon_2+\epsilon_3)\nn\\
&&-d_{fo}\epsilon_2(1+\epsilon_1-\epsilon_2+\epsilon_3)-
d_{ho}\epsilon_3(1+\epsilon_1+\epsilon_2-\epsilon_3)\bigg)
\hat{\tau}_{oo}^{R}\nn\\
&&+\bigg(n(1-\epsilon_1+\epsilon_2+\epsilon_3)-
d_{go}(1-\epsilon_1-\epsilon_2-\epsilon_3)\nn\\
&&-d_{fo}(-1-\epsilon_1-\epsilon_2+\epsilon_3)-
d_{ho}(-1-\epsilon_1+\epsilon_2-\epsilon_3)\bigg)\hat{\tau}_{og}^{NS}\nn\\
&&-\bigg(n(1-\epsilon_1+\epsilon_2+\epsilon_3)-
d_{go}\epsilon_1(1-\epsilon_1-\epsilon_2-\epsilon_3)\nn\\
&&-d_{fo}\epsilon_2(-1-\epsilon_1-\epsilon_2+\epsilon_3)-
d_{ho}\epsilon_3(-1-\epsilon_1+\epsilon_2-\epsilon_3)\bigg)
\hat{\tau}_{og}^{R}\nn\\
&&+\bigg(n(1+\epsilon_1-\epsilon_2+\epsilon_3)-
d_{go}(-1-\epsilon_1-\epsilon_2+\epsilon_3)\nn\\
&&-d_{fo}(1-\epsilon_1-\epsilon_2-\epsilon_3)-
d_{ho}(-1+\epsilon_1-\epsilon_2-\epsilon_3)\bigg)
\hat{\tau}_{of}^{NS}\nn\\
&&-\bigg(n(1+\epsilon_1-\epsilon_2+\epsilon_3)-
d_{go}\epsilon_1(-1-\epsilon_1-\epsilon_2+\epsilon_3)\nn\\
&&-d_{fo}\epsilon_2(1-\epsilon_1-\epsilon_2-\epsilon_3)-
d_{ho}\epsilon_3(-1+\epsilon_1-\epsilon_2-\epsilon_3)\bigg)
\hat{\tau}_{of}^{R}\nn
\end{eqnarray}
\begin{eqnarray}
&&+\bigg(n(1+\epsilon_1+\epsilon_2-\epsilon_3)-
d_{go}(-1-\epsilon_1+\epsilon_2-\epsilon_3)\nn\\
&&-d_{fo}(-1+\epsilon_1-\epsilon_2-\epsilon_3)-
d_{ho}(1-\epsilon_1-\epsilon_2-\epsilon_3)\bigg)
\hat{\tau}_{oh}^{NS}\nn\\
&&-\bigg(n(1+\epsilon_1+\epsilon_2-\epsilon_3)-
d_{go}\epsilon_1(-1-\epsilon_1+\epsilon_2-\epsilon_3)\nn\\
&&-d_{fo}\epsilon_2(-1+\epsilon_1-\epsilon_2-\epsilon_3)-
d_{ho}\epsilon_3(1-\epsilon_1-\epsilon_2-\epsilon_3)\bigg)
\hat{\tau}_{oh}^{R}\bigg\}.
\end{eqnarray}
%
%


\chapter{$\mathbb{Z}_2\times \mathbb{Z}_2$ Boundary Operators}
\label{app:Boundaries}

\begin{figure}[!h]
\begin{center}
\includegraphics[scale=0.7]{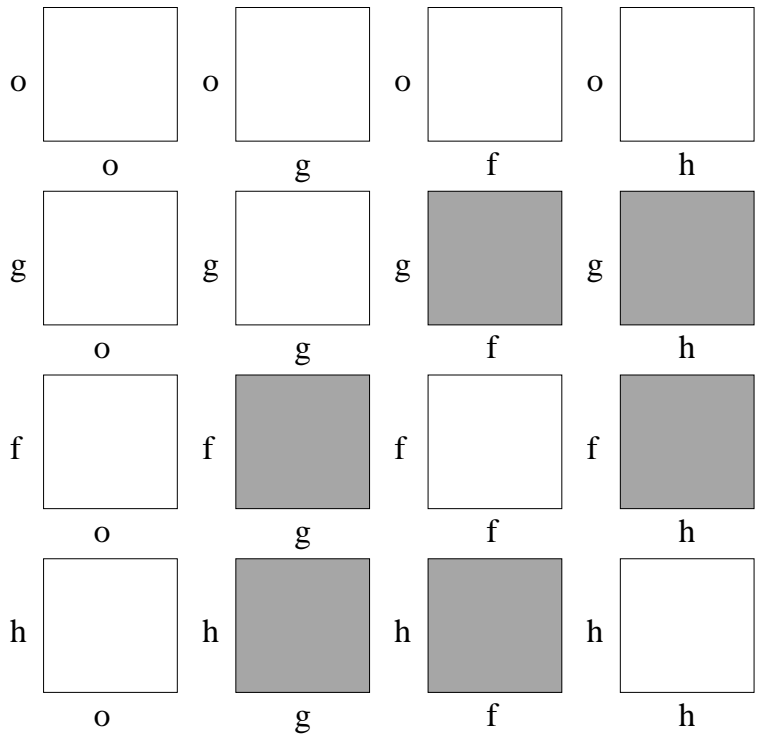}
\caption{Distinct boundary sets in the $\mathbb{Z}_2\times
\mathbb{Z}_2$}\label{app:Blocks}
\end{center}
\end{figure}

The $\mathbb{Z}_2\times \mathbb{Z}_2$ generators including the
identity lead to $16=4\times 4$ distinct boundary conditions on
the two dimensional sheet.  These are portrayed in fig.
\ref{app:Blocks}. The shaded blocks represent those which are not
connected to the unshaded ones by the modular invariance group
$SL(2,\mathbb{Z})/\mathbb{Z}_2$ which is generated by $S$ and $T$
transforms.

\chapter{Tadpole Diagrams}\label{app:TransverseTadpoleDiagrams}

\begin{figure}[!h]\label{app:TransDiagrams}
\begin{center}
\includegraphics[scale=0.5]{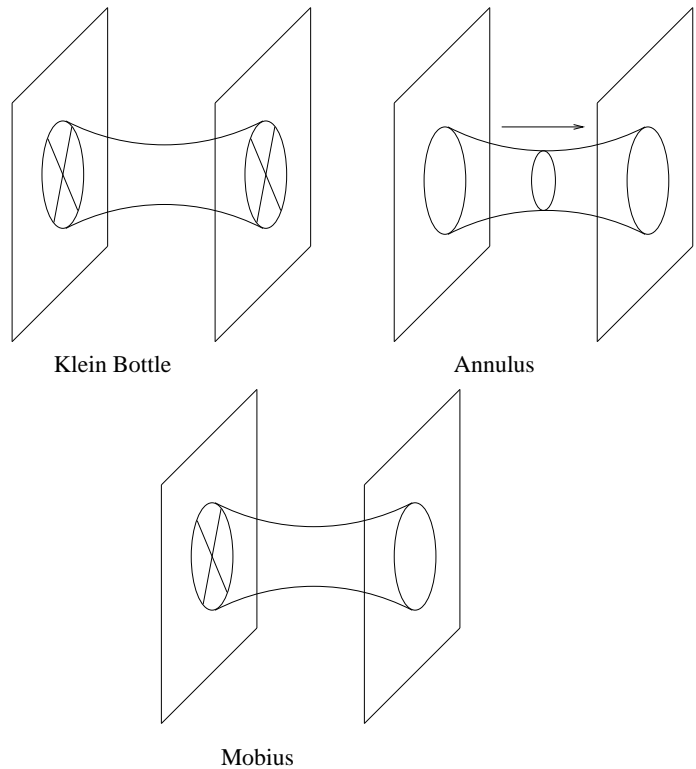}
\caption{Distinct boundary sets in the $\mathbb{Z}_2\times
\mathbb{Z}_2$}
\end{center}
\end{figure}

In the transverse (or tube channel) the open strings attached to
boundaries are interpreted in terms of closed strings propagating
between them.  In which case, the two types of boundaries
($O$-planes and branes) allow the topology of the tubes to be
illustrated as in diagram \ref{app:TransDiagrams}.  At massless
level, these are the tadpole diagrams for the Klein bottle (tube
with crosscaps), the annulus (simple tube) and the Mobius strip
(tube with one crosscap).

\chapter{The Torus Amplitude from its Corresponding Action}
\label{app:TorusFromAction}

Here, a simple case of the one-loop vacuum amplitude from the
action of a simple scalar mode with mass $M$ which exists in $D$
dimensions \cite{CAAS} is shown. The action of this field is
described by
\begin{eqnarray}
S=\int d^D x\frac{1}{2}\bigg[\partial_\mu \phi
\partial^\mu \phi -M^2\phi^2 \bigg].
\end{eqnarray}
In the Euclidean basis, after integrating the fields over all
paths, one has the vacuum energy $\Gamma$ defined as
\begin{eqnarray}
e^{-\Gamma}\sim {\rm det}^{-\frac{1}{2}}(-\Delta+M^2)
\end{eqnarray}
and using the identity
\begin{eqnarray}
{\rm log}({\rm det}(Y))=-\int_\Lambda^\infty\frac{dt}{t}{\rm
tr}(e^{-tY})
\end{eqnarray}
($\Lambda$ is an ultraviolet cutoff and $t$ is the Schwinger
parameter) $\Gamma$ can be written as
\begin{eqnarray}
\Gamma=-\frac{V}{2}\int_\Lambda^\infty \frac{dt}{t}e^{-tM^2}\int
\frac{d^D p}{(2\pi)^D}e^{-tp^2}.
\end{eqnarray}
Which leads to two results after performing the Gaussian integral
over momentum, one for the scalar field and the second for a Dirac
fermion for similar mass as
\begin{eqnarray}
\Gamma=-\frac{V}{2(4\pi)^{\frac{D}{2}}}\int_\Lambda^\infty
\frac{dt}{t^{\frac{D}{2}+1}}e^{-tM^2}
\end{eqnarray}
and
\begin{eqnarray}
\Gamma=\frac{2^{\frac{D}{2}}V}{2(4\pi)^{\frac{D}{2}}}\int_\Lambda^\infty
\frac{dt}{t^{\frac{D}{2}+1}}e^{-tM^2}.
\end{eqnarray}
respectively.  So the total contribution can be defined in terms
of a super trace which counts the signed multiplicities of the
above which are extended to generic Bose or Fermi fields, since
there physics modes determine $\Gamma$.  In addition, $\Gamma$ is
proportional to their number, and so
\begin{eqnarray}\label{eqn:TenDimBasic}
\Gamma_{\rm tot
}=-\frac{V}{2(4\pi)^{\frac{D}{2}}}\int_\Lambda^\infty
\frac{dt}{t^{\frac{D}{2}+1}}{\rm Str}\big(e^{-tM^2}\big).
\end{eqnarray}
In moving now to describe the equivalent function in string
theory, one uses the mass squared operator in the above
expression. The mass levels of the modes are determined by
\begin{eqnarray}
M^2=\frac{2}{\alpha^\prime}(L_o+\bar{L}_o),
\end{eqnarray}
where $L_o$ and $\bar{L}_o$ are the left and right moving
Hamiltonian's of a closed string, and are given by equation
(\ref{eqn:LaurantModes}).

For the superstring, the critical dimension is $D=10$. So, in
conjunction with an insertion of a representation of the Dirac
delta function which enforces the level matching condition
$L_o=\bar{L}_o$ one has
\begin{eqnarray}
\Gamma_{\rm tot
}=-\frac{V}{2(4\pi)^{5}}\int_{-\frac{1}{2}}^{\frac{1}{2}}ds\int_\Lambda^\infty
\frac{dt}{t^{6}}{\rm
tr}\big(e^{-\frac{2}{\alpha^\prime}(L_o+\bar{L}_o)t+2\pi i
(L_o-\bar{L}_o)s}\big).
\end{eqnarray}
All calculations in the context of this paper are determined by
the one loop amplitude, or torus.  The Teichm\"{u}ller $\tau_2$,
is naturally identified with the complex Schwinger parameter. So,
for the definition of the complex Schwinger parameter
$\tau=\tau_1+i\tau_2=s+i\frac{t}{\alpha^\prime\pi}$ with
$q=e^{2\pi i \tau}$ and $\bar{q}=e^{-2\pi i \bar{\tau}}$, results
in the torus amplitude
\begin{eqnarray}
{\cal T} =\int_{\cal F}\frac{d^2\tau}{\tau^2_2}
\frac{1}{\tau^{4}_2}{\rm tr}q^{L_o}\bar{q}^{\bar{L}_o}.
\end{eqnarray}
The fundamental region ${\cal
F}=\{-\frac{1}{2}<\tau_1\leq\frac{1}{2},|\tau|\geq1\}$ defines the
region which covers the counting of distinct tori.

It is of some assistance to illustrate the role of the parameter
$\tau_2$ in its effect when $S$ transforming amplitudes.  The $S$
transform of the Klein and annulus amplitudes will lead to factors
that appear in the transverse channels.

Now, using the Gaussian integration of
\begin{eqnarray}\label{eqn:Guass}
{\alpha'}^\frac{1}{2}\int
exp\left(-\pi\alpha'\tau_2p^2\right)=\frac{1}{\sqrt{\tau_2}}
\end{eqnarray}
the measure that begins as
\begin{eqnarray}
\int_{\cal F}\frac{d^2\tau}{\tau_2^2}\frac{1}{\tau_2^4}
\end{eqnarray}
can thus be rewritten as (under compactification to $T^2\times
T^2\times T^2$)
\begin{eqnarray}\label{eqn:measure}
\int_{\cal
F}\frac{d^2\tau}{\tau_2^2}\frac{1}{\tau_2}\sum_{\{m,n\}_i}
q^{\frac{\alpha'}{4}p^2_{L,i}}\bar{q}^{\frac{\alpha'}{4}p^2_{R,i}}.
\end{eqnarray}
Where $i=\{1,\ldots,6\}$ and the continuum of momentum states has
been replaced by the discrete summation over momenta and winding
states.

The Klein is based on the measure of the form $2i\tau_2$ in the
direct channel, as can be seen by the $\Omega$ identification of
the torus.  The direct annulus has the measure
$\frac{1}{2}i\tau_2$.  As such, in $S$ transforming to their
respective transverse channels, one makes the substitutions of
$2\tau_2=\frac{1}{l}$ for the Klein and
$\frac{\tau_2}{2}=\frac{1}{l}$ for the annulus.  This allows the
measure dependency of the Klein, annulus and Mobius amplitudes as
$il$. The appropriate rescaling's show that the measure, as
defined in equation (\ref{eqn:measure}), gives a factor of $2^2$
for the transformation to the transverse Klein amplitude, and
$2^{-2}$ for the annulus.  The Poisson resummation formula of
equation (\ref{eqn:Resummation}) then gives the additional factors
of $2^3$ and $2^{-3}$ for the Klein and annulus lattices.


\chapter{Effect of Non-trivial Wilson Lines on Branes}\label{app:WilsonLines}

This short discussion on the effect of a simple $U(1)$ gauge field
on the branes connected by an open string follows that which was
made in \cite{Pol}.

In the case of the 26 dimensional bosonic theory, one can have a
simple and constant gauge field
\begin{eqnarray}
A_{25}(x^{D})=-\frac{\gamma}{2\pi
R}=-i\frac{1}{\Lambda}\frac{\partial \Lambda}{\partial
x^{25}}\text{, } \Lambda(x^{25})=exp{\left(-\frac{i\gamma
x^{25}}{2\pi R}\right)}
\end{eqnarray}
with constant $\gamma$, and $x^{25}$ is the coordinate for the
compact space with topology $S^1$. The measure of the gauge field
around the compact transverse space picks up a phase according to
\begin{eqnarray}
W_Q=exp\left(iQ\oint A_{25}dx^{25}\right)=exp(-iQ\gamma).
\end{eqnarray}
The action of the bosonic open string which interacts with the
gauge field is given by
\begin{eqnarray}
\int d\tau \left(\frac{1}{2}\dot{X}^\mu\dot{X}_\mu-\frac{M^2}{2}
-iQ_iA_\mu\dot{X}^\mu\bigg{|}_{\sigma=0}-iQ_jA_\mu\dot{X}^\mu
\bigg{|}_{\sigma=\pi}\right)
\end{eqnarray}
for Chan-Paton charges $|ij>$.  The gauge group in this case is
$U(1)^n$ parameterized by $A_{25}=-\frac{1}{2\pi
R}diag(\gamma_1,\ldots,\gamma_n)$.  The state $|ij>$ is charged as
$Q_i$ under $U(1)_i$ and $U(1)_j$ and is neutral under the others.
It can than be seen that the string mass is shift according to
\begin{eqnarray}
M^2=\frac{(Q_i\gamma_i+Q_j\gamma_j)}{(2\pi R)^2}+\ldots.
\end{eqnarray}
By performing a $T$ duality along the $25^\text{th}$ coordinate,
difference in string end points is described by
\begin{eqnarray}
\Delta
X'^{25}=X'^{25}(0)-X'^{25}(\pi)=(Q_i\gamma_i+Q_j\gamma_j)R'+\ldots.
\end{eqnarray}
The distance between branes are therefore moved apart by $\Delta
X'^{25}$. Correspondingly, the mass is altered.


\chapter{Lattice Simplifications}

I present here some simplifications of the lattice definition
(\ref{eqn:compact}) which will be helpful in looking at the
momentum towers that appear in all amplitudes.

A general lattice for the torus amplitude is defined as
\begin{eqnarray}
\Lambda_{m+a,n+b}=\frac{q^{\frac{\alpha\prime}{4}{\big{(}\frac{(m+a)}
{R}+\frac{(n+b)R}{\alpha\prime}\big{)}}^2}\bar{q}^{\frac{\alpha\prime}{4}
{\big{(}\frac{(m+a)}{R}-\frac{(n+b)R}{\alpha\prime}\big{)}}^2}}{\eta(q)
\eta(\bar{q})}
\end{eqnarray}
for $q=exp(2\pi i \tau)$ and $\tau=\tau_1+i\tau$. This carries the
quantum numbers $|m_i,n_i>$ for a circle $S^1_{i}$.  The expanded
for of this is
\begin{eqnarray}\label{app:expL}
\Lambda_{m+a,n+b}=\frac{(-1)^{2
\tau_1(m+a)(n+b)}}{|\eta(\tau)|^2}exp{\left[-\pi\tau_2\alpha'\left
\{\left(\frac{(m+a)}
{R}\right)^2+\left(\frac{(n+b)R}{\alpha'}\right)^2\right\}\right]}
\end{eqnarray}
As such $\Lambda_{m+a,n+b}$ has the following expansions;
\begin{eqnarray}
\Lambda_{m+a,0;0,0}&=&\frac{1}{|\eta(\tau)|^2}
exp\left[-\pi\tau_2\alpha'\left(\frac{m+a}{R}\right)^2\right],
\end{eqnarray}
\begin{eqnarray}
\Lambda_{0,n+b;0,0}&=&\frac{1}{|\eta(\tau)|^2}
exp\left[-\pi\tau_2\alpha'\left(\frac{(n+a)R}{\alpha'}\right)^2\right],
\end{eqnarray}
\begin{eqnarray}
\Lambda_{m+a,n+b;0,0}&=&\frac{1}{|\eta(\tau)|^2}
exp\left[-\pi\tau_2\alpha'\left(\left(\frac{m+a}{R}\right)^2
+\left(\frac{(n+a)R}{\alpha'}\right)^2\right)\right].
\end{eqnarray}
Since the two directions of the torus are independent, one can
decompose in the following manner
\begin{eqnarray}
\Lambda_{m_1+a_1,n_1+b_1;m_2+a_2,n_2+b_2}=\Lambda_{m_1+a_1,n_1+b_1;0,0}
\Lambda_{0,0;m_2+a_2,n_2+b_2}
\end{eqnarray}
for all $\{m_i,n_i\}\in\mathbb{Z}$.


\chapter{Amplitudes of the $\mathbb{Z}_2\times \mathbb{Z}_2$ Partition
Function}\label{app:Z2xZ2Generic}

The sections of this work that deal with type I string mainly
involve discussions of the phenomenology arising from models based
on the $\mathbb{Z}_2\times \mathbb{Z}_2$ model. As such, it is of
some assistance to review the parent torus amplitude and the open
descendants that follow in the cases of $\epsilon=\pm 1$. The
following are simply a review of those defined and discussed in
detail in \cite{CAAS}.

The parent torus for this model is defined as
\begin{eqnarray}
{\cal T}&=&\frac{1}{4}\bigg\{
|T_{oo}|^2\Lambda_1\Lambda_2\Lambda_3
+|T_{ok}|^2\Lambda_k\left|\frac{2\eta}{\theta_2}\right|^4
+|T_{ko}|^2\Lambda_k\left|\frac{2\eta}{\theta_4}\right|^4
+|T_{kk}|^2\Lambda_k\left|\frac{2\eta}{\theta_3}\right|^4\nn\\
&&+\epsilon\big(|T_{gh}|^2+|T_{gf}|^2+|T_{fg}|^2+|T_{fh}|^2+|T_{hg}|^2+
|T_{hf}|^2\big)
\left|\frac{8\eta}{\theta_2\theta_3\theta_4}\right|^2\bigg\}.
\end{eqnarray}
The direct channel Klein follows as
\begin{eqnarray}
{\cal
K}&=&\frac{1}{8}\bigg\{\big(P_1P_2P_3+\frac{1}{2}P_kW_lW_s\big)T_{oo}
+2\times 16\epsilon_k\big(P_k+\epsilon W_k\big)
T_{ko}\left(\frac{\eta}{\theta_4}\right)^2\bigg\},
\end{eqnarray}

The direct channel open descendants take the forms
\begin{eqnarray}
{\cal
A}_{(-1)}&=&\frac{1}{8}\bigg\{\bigg(N_o^2P_1P_2P_3+\frac{D_{ko}^2}{2}
P_kW_lW_s\bigg)T_{oo}\nn\\
&&\bigg[(N^2_k+D^2_{kk})P_k+D^2_{lk}W_k\bigg]T_{ok}\left(\frac{2\eta}
{\theta_2}\right)^2\nn\\
&&+2N_oD_{ko}P_kT_{ko}^{(\epsilon_k)}\left(\frac{\eta}{\theta_4}\right)^2\nn\\
&&-2N_kD_{kk}P_kT_{kk}^{(\epsilon_k)}\left(\frac{\eta}{\theta_3}\right)^2\nn\\
&&+2i(-1)^{k+l}N_lD_{kl}T_{kl}^{(\epsilon_k)}\frac{2\eta^3}{\theta_2\theta_3
\theta_4}\nn
\end{eqnarray}
\begin{eqnarray}\label{eqn:GenericZ2xZ2-1case}
&&+D_{ko}D_{lo}W_mT_{mo}^{(\epsilon_k\epsilon_l)}\left(\frac{\eta}{\theta_4}
\right)^2\nn\\
&&-D_{km}D_{lm}W_mT_{mm}^{(\epsilon_k\epsilon_l)}\left(\frac{\eta}{\theta_3}
\right)^2\nn\\
&&+2i(-1)^{m+k}D_{kk}D_{lk}T_{mk}^{(\epsilon_k\epsilon_l)}\frac{2\eta^3}
{\theta_2\theta_3\theta_4}\bigg\}
\end{eqnarray}
for the $\epsilon=-1$ annulus, and
\begin{eqnarray}
{\cal
A}_{(+1)}&=&\frac{1}{2}\bigg\{\bigg(N^2P_1P_2P_3+\frac{D_k^2}{2}P_kW_lW_s\bigg)
T_{oo}\nn\\
&&+2ND_kP_kT_{ko}^{(\epsilon_k)}\left(\frac{\eta}{\theta_4}\right)^2\nn\\
&&+D_kD_lW_mT_{mo}^{(\epsilon_k\epsilon_l)}\left(\frac{\eta}{\theta_4}\right)^2
\bigg\}
\end{eqnarray}
for the $\epsilon=+1$ case.

The oriented amplitudes are
\begin{eqnarray}
{\cal M}&=&-\frac{1}{8}\bigg\{\bigg(N_oP_1P_2P_3\hat{T}_{oo}
+\frac{D_{ko}}{2}\epsilon_kP_kW_lW_s\hat{\tilde{T}}_{oo}^{(\epsilon_k)}\nn\\
&&-N_oP_k\epsilon_k\hat{T}_{ok}\left(\frac{2\hat{\eta}}{\hat{\theta_2}}\right)^2
\nn\\
&&-\bigg(D_{lo}\epsilon_kW_m\hat{\tilde{T}}_{om}^{(\epsilon_l)}
+D_{ko}P_m\hat{\tilde{T}}_{ok}^{(\epsilon_k)}\bigg)
\left(\frac{2\hat{\eta}}{\hat{\theta_2}}\right)^2\bigg\}
\end{eqnarray}
for the $\epsilon=-1$ model, and
\begin{eqnarray}
{\cal M}&=&-\frac{1}{4}\bigg\{\bigg(NP_1P_2P_3\hat{T}_{oo}
+\frac{D_k}{2}\epsilon_kP_kW_lW_s\hat{\tilde{T}}_{oo}^{(\epsilon_k)}\nn\\
&&-NP_k\epsilon_k\hat{T}_{ok}\left(\frac{2\hat{\eta}}{\hat{\theta_2}}\right)^2
\nn\\
&&-\bigg(D_l\epsilon_kW_m\hat{\tilde{T}}_{om}^{(\epsilon_l)}
+D_kP_m\hat{\tilde{T}}_{ok}^{(\epsilon_k)}\bigg)
\left(\frac{2\hat{\eta}}{\hat{\theta_2}}\right)^2\bigg\}.
\end{eqnarray}
The tadpole conditions are summarized in table \ref{tab:TadCond}.
\begin{table}[!ht]
\begin{center}
\begin{tabular}{|ll||l|}\hline
$\epsilon=-1$ Model & & $\epsilon=+1$ Model \\\hline $N_o=32$, &
$N_g=N_f=N_h=0$ & $N=16$\\
$D_{ko}=32$, & $D_{kg}=D_{kf}=D_{kh}=0$ & $D_k=16$
\\\hline
\end{tabular}
\end{center}
\caption{Tadpole conditions}\label{tab:TadCond}
\end{table}

This concludes all necessary detail for comparison to the
$\mathbb{Z}_2\times \mathbb{Z}_2$ models discussed in this study.


\chapter{Singular Valued Decomposition}\label{app:SVD}

This an approach to finding a solution to the $D$--flat
constraints that are imposed by equations (\ref{da}) and
(\ref{dalpha}).  This method was first applied to the resolution
of $D$--flat constraints by G. Cleaver.

The form of the these constraint equations becomes a set of linear
equations of the form
\begin{eqnarray}
\mathbf{D}.\mathbf{x}=\mathbf{b}.
\end{eqnarray}
For an $M\times N$ matrix $\mathbf{D}$. A matrix element is
written in terms of the charge $Q^{(i)}_j$ with
$i=1,\ldots,A,\ldots,N$ as $D_{ij}=Q^{(i)}_j$.  The row $i=A$,
corresponds to the constraint (\ref{da}), and otherwise
corresponds to (\ref{dalpha}).  The $j^\text{th}$ element of
$\mathbf{x}$ is the norm squared of the VEV for a field $\phi_j$.
The vector $\mathbf{b}$ contains one non--zero entry for $i=A$,
which has value $-\xi$, as appreciated by (\ref{da}).

The singular value decomposition provides the matrix $\mathbf{D}$
in the form
\begin{eqnarray}\label{eqn:decomp1}
\mathbf{D}_{M\times N}=\mathbf{U}_{M\times N}.\mathbf{
W}^{\text{diag}}_{N\times N}.\mathbf{V}^{\text{T}}_{N\times N}
\end{eqnarray}
for $M\geq N$.  Here, $\mathbf{U}$ is an $M\times N$ column
orthogonal matrix, $\mathbf{W}$ is an $N\times N$ diagonal matrix
containing only semi--positive--definite elements and
$\mathbf{V}^\text{T}$ is the transpose of an $N\times N$
orthogonal matrix.

This decomposition is unique up to;
\begin{itemize}

\item
forming linear combinations of any columns of $\mathbf{U}$ and
$\mathbf{V}$ whose corresponding elements of $\mathbf{W}$ are
degenerate

\item
making the same permutation of the columns of $\mathbf{U}$,
diagonal elements of $\mathbf{W}$ and columns of $\mathbf{V}$

\end{itemize}
In this method, if $M<N$, then one appends an $(M-N)\times N$ zero
matrix to $\mathbf{D}$ so that $M\geq N$.  When $\mathbf{D}$ is
singular, corresponding to the $M$ constraints not all being
linearly independent, there is a subspace of $\mathbf{x}$ termed
the \emph{nullspace} that is mapped to $\mathbf{0}$ in
$\mathbf{b}$ space by $\mathbf{D}$.  The subspace of $\mathbf{b}$
that can be reached by the matrix $\mathbf{D}$ is the range of
$\mathbf{D}$, the dimension of which is rank($\mathbf{D}$).  With
the dimension of the nullspace as nullity($\mathbf{D}$), one has
\begin{eqnarray}
\text{rank}(\mathbf{D})+\text{nullity}(\mathbf{D})=N
\end{eqnarray}
and rank($\mathbf{D}$) is the number of independent constraint
equations.

In the decomposition (\ref{eqn:decomp1}), the set of the
$l^\text{th}$ columns of $\mathbf{U}$ corresponding to the
$l^\text{th}$ non--zero diagonal components of $\mathbf{W}$ form
an orthonormal set of basis vectors that span the range of
$\mathbf{D}$.  The columns of $\mathbf{V}$ whose corresponding
diagonal components of $\mathbf{W}$ are zero form an orthonormal
basis for the nullspace.

One begins with the full matrix $D_{ij}=Q^{(i)}_j$ for the
$i^\text{th}$ charges of the $j^\text{th}$ field $\phi_j$.  Then
one decomposes this to the form (\ref{eqn:decomp1}).

The matrix used in the calculation of the flat directions is
$\mathbf{D}'$ which is the original $\mathbf{D}$ with the $i=A$
row removed. This enables one to extract a complete set of
directions that are solutions of the form
\begin{eqnarray}
\mathbf{D'}.\mathbf{x}_p=\mathbf{0}.
\end{eqnarray}
This then gives a set of independent solutions $\mathbf{x}_p$ (for
$p$ running over the number of zero components of $\mathbf{W}$) to
(\ref{da}). The next step, which helps to simplify the necessary
analysis of the $D$--flat directions, involves rotating the bases
$\mathbf{x}$ so that these contain a set of common fields and at
least one unique field. These rotated combinations can then be
linearly combined to exact a solution of the constraint equation
(\ref{dalpha}).


\begin{table}
{\rm \large\bf Left--Right~Symmetric~Model~1~Fields}
\begin{eqnarray*}
\begin{tabular}{|c|c|c|rrrrrrrr|c|}
\hline

  $F$ & SEC & $(C;L;R)$
   & $Q_A$ & $Q_{1^\prime}$ & $Q_{2^\prime}$ & $Q_{3^\prime}$ &
   $Q_{4^\prime}$ &
             $Q_{5^\prime}$ & $Q_{7^{\prime\prime}}$ & $Q_{8^\prime}$
   & $SU(4)_{H_{1;2}}$ \\

\hline

  $Q_{L_1}$ &                          & $(4,2,1)$
     & -2  &  2  &  2  &   4  & -2  & -2 &  8  &  0  & $(1,1)$\\
  $Q_{R_1}$ & $\mb_1\oplus$ & $({\bar4},1,2)$
     & -2  & -2  & -2  &  -4  & -2  & -2 &  -8  & 0 &  $(1,1)$\\
  $L_{L_1}$ &  $\mb_1+\mzeta+2\mgamma$ & $(1,2,1)$
     & -2  &  2  &  2  & -20  & -2  & -2 &  0  &  0 & $(1,1)$ \\
  $L_{R_1}$ & & $(1,1,2)$
     & -2  & -2  & -2  &  20  & -2  & -2 &  0  &  0 & $(1,1)$ \\
  $\cL{L_{1}}$  &         & $(1,2,1)$
     &  2 &   2  &  2  &   4  & 2  &  2 &  -32  & 0 & $(1,1)$ \\
  $\cL{R_{1}}$  &         & $(1,1,2)$
     &  2 &  -2  & -2  &  -4  &  2  &  2 &  32  & 0 & $(1,1)$ \\

\hline

  $Q_{L_2}$ &                   & $(4,2,1)$
     & -2 & -2 &  2 &   4 &  2 & -2 &  8 &  0 & $(1,1)$   \\
  $Q_{R_2}$ & $\mb_2\oplus$ & $({\bar4},1,2)$
     & -2 &  2 & -2 &  -4 &  2 & -2 &  -8 & 0 & $(1,1)$   \\
  $L_{L_2}$ & $\mb_2+\mzeta +2\mgamma$ & $(1,2,1)$
     & -2 & -2 &  2 & -20 &  2 & -2 &  0 &  0 & $(1,1)$   \\
  $L_{R_2}$ &         & $(1,1,2)$
     & -2 &  2 & -2 &  20 &  2 & -2 &  0 &  0 & $(1,1)$   \\
  $\cL{L_{2}}$  &        & $(1,2,1)$
     &  2 & -2 &  2 &   4 & -2 &  2 &  -32 & 0 & $(1,1)$   \\
  $\cL{R_{2}}$  &        & $(1,1,2)$
     &  2 &  2 & -2 &  -4 &  -2 &  2 &  32 & 0 & $(1,1)$   \\

\hline

  $Q_{L_3}$ &  & $(4,2,1)$
     &  -2 &  0 &  -4 &  4 &  0 &  4 &  8 &  0 & $(1,1)$     \\
  $Q_{R_3}$ & $\mb_3\oplus$ & $({\bar4},1,2)$
     &  -2 &  0 &   4 & -4 &  0 &  4 &  -8 & 0 & $(1,1)$     \\
  $L_{L_3}$ & $\mb_3+\mzeta +2\mgamma$ & $(1,2,1)$
     &  -2 &  0 &  -4 &-20 &  0 &  4 &  0 &  0 & $(1,1)$     \\
  $L_{R_3}$ & & $(1,1,2)$
     &  -2 &  0 &   4 & 20 &  0 &  4 &  0 &  0 & $(1,1)$     \\
  $\cL{L_{3}}$  &         & $(1,2,1)$
     &   2 &  0 &  -4 &  4 &  0 & -4 &  -32 & 0 & $(1,1)$    \\
  $\cL{R_{3}}$  &         & $(1,1,2)$
     &   2 &  0 &   4 & -4 &  0 & -4 &  32 & 0 & $(1,1)$    \\

\hline

  $\Phi_1$          &         & $(1,1,1)$
     &  0 &  0 &  0 &  0 &  0 &  0 &  0 &  0 & $(1,1)$        \\
  $\Phi_2$          &         & $(1,1,1)$
     &  0 &  0 &  0 &  0 &  0 &  0 &  0 &  0 & $(1,1)$        \\
  $\Phi_3$          &         & $(1,1,1)$
     &  0 &  0 &  0 &  0 &  0 &  0 &  0 &  0 & $(1,1)$        \\
  $\Phi_{12}$       & Neveu-  & $(1,1,1)$
     &  0 & -8 &  0 &  0 &  0 &  0 &  0 &  0 & $(1,1)$        \\
  ${\bar\Phi}_{12}$ & Schwarz & $(1,1,1)$
     &  0 &  8 &  0 &  0 &  0 &  0 &  0 &  0 & $(1,1)$        \\
  $\Phi_{23}$       &         & $(1,1,1)$
     &  0 &  4 &-12 &  0 &  0 &  0 &  0 &  0 & $(1,1)$        \\
  ${\bar\Phi}_{23}$ &         & $(1,1,1)$
     &  0 & -4 & 12 &  0 &  0 &  0 &  0 &  0 & $(1,1)$        \\
  $\Phi_{31}$       &         & $(1,1,1)$
     &  0 & -4 & -12&  0 &  0 &  0 &  0 &  0 & $(1,1)$        \\
  ${\bar\Phi}_{31}$ &         & $(1,1,1)$
     &  0 &  4 &  12&  0 &  0 &  0 &  0 &  0 & $(1,1)$        \\

\hline

$\D{3}$& & $(4,1,1)$
     & 0 &  0 &  4 & 8 &  0 &  0 &  -24 & 0 & $(1,1)$        \\

$\Db{3}$& & $({\bar 4},1,1)$
     & 0 &  0 &  -4 & -8 &  0 &  0 &  24 & 0 & $(1,1)$   \\

${\phi}_{\alpha\beta}$& & $(1,1,1)$
     & 0 &  0 & -12 & 0  & 0 &  0 &  0 & 0 & $(1,1)$        \\

${\bar \phi}_{\alpha\beta}$& $\xi\equiv\mS+\mb_1+\mb_2+$  &
$(1,1,1)$
     & 0 &  0 &  12 & 0  & 0 &  0 &  0 & 0 & $(1,1)$        \\

${\phi}_{1}$& $\malpha+\mbeta$ & $(1,1,1)$
     &  0 & 4 &  0 & 0 & 0 &  0 &  0 &  0 & $(1,1)$        \\

${\bar \phi}_{1}$& $\oplus$ & $(1,1,1)$
     &  0 & -4 &  0 & 0 & 0 &  0 &  0 &  0 & $(1,1)$        \\

${\phi}_{2}$& $\xi+\zeta$ & $(1,1,1)$
     &  0 & 4 &  0 & 0 &  0 &  0 &  0 &  0 & $(1,1)$        \\

${\bar \phi}_{2}$ &  & $(1,1,1)$
     &  0 &  -4 &  0 & 0 &  0 &  0 &  0 &  0 & $(1,1)$        \\

$\S{8}$ &  & $(1,1,1)$
     & 0 &  0 & 4 &  -16 &  0 &  0 &  -32 &  0 & $(1,1)$        \\

$\bS{8}$ & & $(1,1,1)$
     &  0 & 0 &  -4 & 16 &  0 &  0 & 32 &  0 & $(1,1)$        \\

\hline
\end{tabular}
\nolabel
\end{eqnarray*}
Table 1: {\it Model 1 fields}.
\end{table}

\begin{flushleft}
\begin{table}
{\rm \large\bf Left--Right~Symmetric~Model~1~Fields~Continued}
\begin{eqnarray*}
\begin{tabular}{|c|c|c|rrrrrrrr|c|}
\hline

  $F$ & SEC & $(C;L;R)$
   & $Q_A$ & $Q_{1^\prime}$ & $Q_{2^\prime}$ & $Q_{3^\prime}$ &
   $Q_{4^\prime}$ &
             $Q_{5^\prime}$ & $Q_{7^{\prime\prime}}$ & $Q_{8^\prime}$
   & $SU(4)_{H_{1;2}}$ \\

\hline

  $\D{1}$ & & $ (4,1,1)$
     &  0 &   2  &  2 &   4 &  0 &  0 &  8 &  -8 & $(1,1)$     \\

  $\Db{1}$        &                      &   $ ({\bar 4},1,1)$
     &  0 &  -2 &  -2 &  -4 &  0 &  0 &   -8 & 8 & $(1,1)$    \\

  $\S{1}$        & $\xi\equiv\mS+ \mb_2+\mb_3+$        &   $(1,1,1)$
     &  0 &  -2 &  6 &  -12 &  0 &  0 &  16 & 8 & $(1,1)$        \\

  $\bS{1}$ &  $~\mbeta+\mgamma~$ &   $(1,1,1)$
     &  0 & 2 &  -6 &  12 &  0 &  0 &  -16 & -8 & $(1,1)$         \\

  $\S{2}$ &       $\oplus$          &   $(1,1,1)$
     &  0 &  2 &  -6 &  -12 &  0 &  0 &  16 & 8 & $(1,1)$        \\

  $\bS{2}$ & $\xi+\zeta+2\gamma$ &  $(1,1,1)$
     &  0 &  -2 & 6 & 12 &  0 &  0 &  -16 & -8 & $(1,1)$        \\

  ${\cH{1}}$ &                             &   $(1,1,1)$
     &  0 & 2 &  2 &  -8 &  0 &  0 &  -16 & 0 & $(1,6)$         \\

 ${\cHb{1}}$ &                      &   $(1,1,1)$
     &  0 &  -2 & -2 & 8 &  0 &  0 &  16 & 0 & $(1,6)$ \\

\hline

$\D{2}$ & & $(4,1,1)$
     &  0 &  -2 & 2 & 4 & 0 &  0 &  8 & -8 & $(1,1)$        \\

$\Db{2}$ &  & $({\bar 4},1,1)$
     &  0 & 2 &  -2 & -4 &  0 & 0 & -8  & 8 & $(1,1)$        \\

$\S{3}$ &  $\xi\equiv\mS+\mb_1+\mb_3+$ & $(1,1,1)$
     &  0 &   2 &   6 &  -12 &  0 &  0 &   16 &  8 & $(1,1)$        \\

$\bS{3}$ & $\malpha+\mgamma$     & $(1,1,1)$
     &  0 &  -2 &  -6 &   12 &  0 &  0 &  -16 &  -8 & $(1,1)$        \\

$\S{4}$ &  $\oplus$  & $(1,1,1)$
     &  0 &  -2 &  -6 &  -12 &  0 &  0 &   16 & 8 & $(1,1)$        \\

${\bar S}_{4}$ & $\xi+\zeta+2\mgamma$ & $(1,1,1)$
     &  0 &  2 &  6 &  12 &  0 &  0 &  -16 &  -8 & $(1,1)$         \\

${\cH{2}}$ & & $(1,1,1)$
     &  0 &  -2 &  2 &  -8 &  0 & 0 &  -16 &  0 & $(1,6)$        \\

${\cHb{2}}$ & & $(1,1,1)$
     &  0 &  2 &  -2 &  8 & 0 &  0 &  16 &  0 & $(1,6)$         \\

\hline

$H_{1}$ & $\xi\equiv\mS+\mb_2+\mb_3+$ & $(1,1,1)$
     &  -4 & -2 & -2 &   2 &  2 &   2 & -16 & -4 & $(1,4)$        \\

${\bar H}_{1'}$ & $\malpha+2\mgamma$ & $(1,1,1)$
     &  -4 &  2 &  2 &  -2 &  2 &   2 &  16 &  4 & $(1,{\bar4})$        \\

$ H_{2}$ & $\oplus$ & $(1,1,1)$
     &  -4 & -2 & -2 &   2 &  2 &   2 &  -16 & 4 & $(4,1)$        \\

${\bar H}_{2'}$ & $\xi+\zeta$ & $(1,1,1)$
     &  -4 &  2 &  2 &  -2 &  2 &   2 & 16 &  -4 & $({\bar 4},1)$        \\

\hline

$H_{3}$& $\xi\equiv\mS+\mb_1+\mb_3+$ & $(1,1,1)$
     & -4 &  2 & -2 &  2 &  -2 &  2 &  -16 &  -4 & $(1,4)$        \\

${\bar H}_{3'}$ &  $\malpha+2\mgamma$ & $(1,1,1)$
     & -4 &  -2 &  2 &  -2 &  -2 &  2 &  16 &  4 & $(1,{\bar4})$        \\

$H_{4}$ & $\oplus$& $(1,1,1)$
     &  -4 &  2 &  -2 & 2 &  -2 &  2 &  -16 &  4 & $(4,1)$        \\

${\bar H}_{4'}$ &$\xi+\zeta$ & $(1,1,1)$
     &  -4 & -2 &  2 & -2 &  -2 &  2 &  16 & -4 & $({\bar 4},1)$        \\

\hline

$H_{5}$& $\xi\equiv\mS+\mb_1+\mb_2+$ & $(1,1,1)$
     & -4 &  0 &   4 &  2 &  0 & -4 & -16 & -4 & $(1,4)$        \\

${\bar H}_{5'}$ &  $\malpha+2\mgamma$ & $(1,1,1)$
     & -4 &  0 &  -4 & -2 &  0 & -4 &  16 &  4 & $(1,{\bar4})$        \\

$H_{6}$ & $\oplus$ & $(1,1,1)$
     &  -4 & 0 &   4 &  2 &  0 & -4 &  -16 & 4 & $(4,1)$        \\

${\bar H}_{6'}$ &  $\xi+\zeta+2\mgamma$ & $(1,1,1)$
     &  -4 &  0 & -4 & -2 &  0 & -4 & 16 &  -4 & $({\bar4},1)$        \\

\hline

${S}_{5}$ &  & $(1,1,1)$
     & 0 &  0 & -8 & -16  & 0 &  0 &  -32 &  0 & $(1,1)$        \\

${\bar S}_{5}$ & & $(1,1,1)$
     &  0 & 0 & 8 & 16 & 0 &  0 &  32 &  0 & $(1,1)$        \\

${S}_{6}$ & $\mS+\zeta+2\mgamma$ & $(1,1,1)$
     &  0 & -4 &  4 & -16 & 0 &  0 &  -32 &  0 & $(1,1)$        \\

${\bar S}_{6}$ & & $(1,1,1)$
     &  0 &  4 &  -4 & 16 & 0 &  0 &  32 & 0 & $(1,1)$        \\

${S}_{7}$ & & $(1,1,1)$
     &  0 &  4 &   4 & -16 & 0 &  0 &  -32 &  0 & $(1,1)$        \\

${\bar S}_{7}$ & & $(1,1,1)$
     &  0 & -4 &  -4 &  16 & 0 &  0 &  32 &   0 & $(1,1)$        \\

\hline

\end{tabular}
\nolabel
\end{eqnarray*}
{Table 1: \it Model 1 fields continued}.
\end{table}
\end{flushleft}


\begin{table}[!ht]
\begin{flushleft}
\begin{tabular}{|l|r|rrrrrrrr|r|}
\hline Basis  &$\QA$& $\S{5}/\bS{5}$
                   & $\Db{3}/\D{3}$
                         & $\Db{2}/\D{2}$
                               & $\cH{2}/\cHb{2}$
                                     & $\cH{1}/\cHb{1}$
                                           & $\bH{4'}$
                                                & $\bH{2'}$
                                                     &    &     \\
\hline
$\mathbf{x}_{ 1}$& 12&   0 & -1  & -1  & -2  &  3  &  0 & -2 &  2 &
$\cL{R1}$ \\
$\mathbf{x}_{ 2}$& 12&  -1 &  3  & -1  &  0  &  1  &  0 & -2 &  2 &
$\cL{L1}$ \\
$\mathbf{x}_{ 3}$& 12&  -1 &  3  & -1  &  0  &  1  & -2 &  0 &  2 &
$\cL{L2}$ \\
$\mathbf{x}_{ 4}$& 12&   0 & -1  & -1  &  2  & -1  & -2 &  0 &  2 &
$\cL{R2}$ \\
$\mathbf{x}_{ 7}$& 24&  -1 &  4  & -4  & -2  &  2  & -4 & -4 &  4 &
$\Q{R3}$ \\
$\mathbf{x}_{ 5}$& 24&  -3 &  4  & -4  &  2  &  6  & -4 & -4 &  4 &
$\Q{L3}$ \\
$\mathbf{x}_{ 6}$& 24&  -1 &  8  & -4  &  2  &  6  & -4 & -4 &  4 &
$\L{R3}$ \\
$\mathbf{x}_{ 8}$& 24&  -3 &  0  & -4  & -2  &  2  & -4 & -4 &  4 &
$\L{L3}$ \\
\hline
$\mathbf{x}_{ 9}$&-12&   0 & -1  &  1  &  0  & -1  &  2 &  0 &  2 &
$\Q{R2}$ \\
$\mathbf{x}_{10}$&-12&   1 & -3  &  1  & -4  & -1  &  2 &  0 &  2 &
$\L{L2}$ \\
$\mathbf{x}_{11}$&-12&   1 & -1  &  1  & -2  &  1  &  2 &  0 &  2 &
$\Q{L2}$ \\
$\mathbf{x}_{13}$&-12&   0 &  1  &  1  &  2  &  1  &  2 &  0 &  2 &
$\L{R2}$ \\
$\mathbf{x}_{12}$&-12&   0 &  1  &  1  &  2  &  1  &  0 &  2 &  2 &
$\L{R1}$ \\
$\mathbf{x}_{14}$&-12&   1 & -3  &  1  &  0  & -5  &  0 &  2 &  2 &
$\L{L1}$ \\
$\mathbf{x}_{15}$&-12&   0 & -1  &  1  &  0  & -1  &  0 &  2 &  2 &
$\Q{R1}$ \\
$\mathbf{x}_{16}$&-12&   1 & -1  &  1  &  2  & -3  &  0 &  2 &  2 &
$\Q{L1}$ \\
$\mathbf{x}_{17}$&-24&   1 & -3  &  1  &  0  & -1  &  2 &  2 &  2 &
$\bH{5'}$ \\
$\mathbf{x}_{18}$&-24&   1 & -1  &  3  &  0  & -3  &  2 &  2 &  2 &
$\H{5}$ \\
$\mathbf{x}_{19}$&-24&  -1 &  0  &  4  &  2  & -2  &  4 &  4 &  4 &
$\cL{L3}$ \\
$\mathbf{x}_{20}$&-24&   5 & -8  &  4  & -2  & -6  &  4 &  4 &  4 &
$\cL{R3}$ \\
$\mathbf{x}_{21}$&-48&   1 & -6  &  6  &  2  & -4  &  4 &  4 &  4 &
$\bH{6'}$ \\
$\mathbf{x}_{22}$&-48&   3 & -2  &  2  & -2  & -4  &  4 &  4 &  4 &
$\H{6}$ \\
\hline
$\mathbf{x}_{40}$&  0&  -3 &  4  &  0  & -2  &  6  &  0 & -4 &  4 &
$\H{1}$ \\
$\mathbf{x}_{37}$&  0&   1 &  0  & -4  & -2  &  2  &  0 & -4 &  4 &
$\bH{1'}$ \\
$\mathbf{x}_{41}$&  0&  -3 &  4  &  0  &  6  & -2  & -4 &  0 &  4 &
$\H{3}$ \\
$\mathbf{x}_{29}$&  0&   1 &  0  & -4  & -2  &  2  & -4 &  0 &  4 &
$\bH{3'}$ \\
$\mathbf{x}_{34}$&  0&  -1 &  2  & -2  & -2  &  4  &  0 & -2 &  2 &
$\H{2}$ \\
$\mathbf{x}_{39}$&  0&  -1 &  2  & -2  &  2  &  0  &  -2 & 0 &  2
&
$\H{4}$ \\
\hline
$\mathbf{x}^v_{32}$& 0&   0 &  0  &  0  & -2  &  2  &  0 &  0 &  1 &
$\P{12}$  \\
$\mathbf{x}^v_{35}$& 0&  -1 &  0  &  0  &  2  &  0  &  0 &  0 &  1 &
$\P{23}$  \\
$\mathbf{x}^v_{28}$& 0&  -1 &  0  &  0  &  0  &  2  &  0 &  0 &  1 &
$\P{31}$  \\
$\mathbf{x}^v_{31}$& 0&  -1 &  0  &  0  &  1  &  1  &  0 &  0 &  1 &
$\S{8}$ \\
$\mathbf{x}^v_{25}$& 0&   0 &  0  &  0  & -1  & -1  &  0 &  0 &  1 &
$\p{\alpha\beta}$\\
$\mathbf{x}^v_{33}$& 0&   0 &  0  &  0  & -1  &  1  &  0 &  0 &  1 &
$\pb{1,2}$\\
$\mathbf{x}^v_{30}$& 0&   0 & -1  & -1  & -1  &  1  &  0 &  0 &  1 &
$\S{4}$ \\
$\mathbf{x}^v_{27}$& 0&   0 & -1  & -1  &  0  &  0  &  0 &  0 &  1 &
$\S{2}$ \\
$\mathbf{x}^v_{36}$& 0&   1 & -1  & -1  & -1  & -1  &  0 &  0 &  1 &
$\S{3}$\\
$\mathbf{x}^v_{24}$& 0&   1 & -1  & -1  & -2  &  0  &  0 &  0 &  1 &
$\S{1}$ \\
$\mathbf{x}^v_{23}$& 0&   0 &  0  &  0  &  0  & -2  &  0 &  0 &  1 &
$\S{7}$ \\
$\mathbf{x}^v_{38}$& 0&   0 &  0  &  0  & -2  &  0  &  0 &  0 &  1
&
$\S{6}$ \\
$\mathbf{x}^v_{26}$& 0&   0 &  0  & -1  & -1  &  1  &  0 &  0 &  1 &
$\Db{1}$   \\
\hline \hline
\end{tabular}
\end{flushleft}
{Table 2: \it $D$--Flat Direction Basis Set for Model 3.}
\end{table}


\begin{thebibliography}{99}

\bibitem{AADS} I. Antoniadis, G. D'Appollonio, E. Dudas and A.
Sagnotti \NPB{565}{2000}{123-156}.

\bibitem{z2z2orient}  M. Berkooz and R.G. Leigh, \NPB{483}{1997}{187}.


\bibitem{z2z2orient3} J. Park, R. Rabadan and A.M. Uranga,
\NPB{570}{2000}{38}.

\bibitem{z2z2orient4} M. Bianchi, J.F. Morales and G. Pradisi,
\NPB{573}{2000}{314}.

\bibitem{fff} I. Antoniadis, C. Bachas, and C. Kounnas, \NPB{289}{1987}{87}.

\bibitem{ff1} H. Kawai, D.C. Lewellen, and S.H.H. Tye, \NPB{288}{1987}{1}.

\bibitem{nahe} A.E. Faraggi and D.V. Nanopoulos,
\PRD{48}{1993}{3288}.

\bibitem{nahe1} \AEF, \IJMP{14}{1999}{1663}.

\bibitem{vwaaf} C. Angelantonj, I. Antoniadis and K. F\"orger,
\NPB{555}{1999}{116}.

\bibitem{mshsm} G.B. Cleaver, A. E. Faraggi, D. V. Nanopoulos,
\IJMP{16}{2001}{425}.









\bibitem{lrsmodels} G.B. Cleaver, A.E. Faraggi and C. Savage,
\PRD{63}{2001}{066001}.


\bibitem{heterotic} D.J. Gross, J.A. Harvey, J.A. Martinec and R. Rohm,
                        \PRL{54}{1985}{502}; \NPB{256}{1986}{253}.

\bibitem{dsw1} M. Dine, N. Seiberg and E. Witten,
\NPB{289}{1987}{589}.

\bibitem{dsw2} J. Atick, L. Dixon and A. Sen, \NPB{292}{1987}{109}.

\bibitem{cff} D.J. Clements and A.E. Faraggi,
\PRD{65}{2002}{106003}.

\bibitem{cfo} D.J. Clements and A.E. Faraggi, hep-th/0302006.

\bibitem{systematic} G. Cleaver, M. Cvetic, J. Espinosa, L. Everett
and P. Langacker \NPB{545}{1999}{47}.

\bibitem{fh1} A.E. Faraggi and E. Halyo, \IJMP{11}{1996}{2357}.

\bibitem{kln} S. Kalara, J.L. Lopez and D.V. Nanopoulos,
\NPB{353}{1990}{650}.

\bibitem{ps} \AEF, \NPB{428}{1994}{111}.

\bibitem{masip} A.E. Faraggi and M. Masip, \PLB{388}{1996}{524}.

\bibitem{cf1} G.B. Cleaver and A.E. Faraggi, \IJMP{14}{1999}{2335}.

\bibitem{cfnw} G.B. Cleaver, A.E. Faraggi, D.V. Nanopoulos,
\NPB{620}{2002}{259-289}.

\bibitem{ACNY} A. Abouelsaood, C. G. Callan, C. R. Nappi and S. A.
Yost, \NPB{280}{1987}{599}.

\bibitem{CAAS} C Angelantonj and A Sagnotti, \PRT{371}{2002}{1-150}.

\bibitem{IAEDAS} I Antoniadis, E Dudas and A Sagnotti, \NPB{544}{1999}{469-502}.

\bibitem{C} C Angelantonj, \NPB{566}{2000}{126-150}.

\bibitem{CSU} Mirjam Cvetic, Gary Shiu and Angel M. Uranga, \NPB{615}{2001}{3-32}.

\bibitem{CIES} C. Angelantonj, I. Antoniadis, E. Dudas and A.
Sagnotti, \PLB{489}{2000}{223-232}.

\bibitem{MLGP} M. Larosa and G. Pradisi, \NPB{667}{2003}{261-309}.

\bibitem{Pol} J. Polchinski, see \emph{String Theory} Vol. 1,
p263-268.

\bibitem{FT} E. S. Franklin and A. A. Tseytlin
\PLB{163}{1985}{123}.

\bibitem{B} C. Bachas, \PLB{296}{1992}{77-84}.

\bibitem{KLT} Kawai H, Lewellen D C and Tye S-H
\NPB{288}{1987}{1}.

\end{thebibliography}
\end{document}